\documentclass[12pt,a4paper]{article}

\usepackage{url}
\usepackage{epsfig}

\newcommand{\ba}{\begin{array}}
\newcommand{\ea}{\end{array}}
\newcommand{\bd}{\begin{displaymath}}
\newcommand{\ed}{\end{displaymath}}
\newcommand{\be}{\begin{equation}}
\newcommand{\ee}{\end{equation}}
\newcommand{\bea}{\begin{eqnarray}}
\newcommand{\eea}{\end{eqnarray}}
\newcommand{\Dir}{\kern -6.4pt\Big{/}}
\newcommand{\Dirin}{\kern -10.4pt\Big{/}\kern 4.4pt}
\newcommand{\DDir}{\kern -10.6pt\Big{/}}
\newcommand{\DGir}{\kern -6.0pt\Big{/}}
\begin{document}
\
\def\bra{\langle}
\def\ket{\rangle}

\def\a{\alpha}
\def\as {\alpha_s}
\def\b{\beta}
\def\d{\delta}
\def\e{\epsilon}
\def\ve{\varepsilon}
\def\l{\lambda}
\def\m{\mu}
\def\n{\nu}
\def\G{\Gamma}
\def\D{\Delta}
\def\L{\Lambda}
\def\s{\sigma}
\def\p{\pi}

\def\etal{ {\em et al.}}
\def\mzs {M_Z^2}
\def\mws {M_W^2}
\def\q2 {q^2}
\def\sz {\sin^2\theta_W}
\def\cz {\cos^2\theta_W}
\def\lp{\lambda^{\prime}}
\def\lps{\lambda^{\prime *}}
\def\lpp{\lambda^{\prime\prime}}
\def\lpps{\lambda^{\prime\prime * }}

\def\bapp{b_1^{\prime\prime}}
\def\bbpp{b_2^{\prime\prime}}
\def\bcp{b_3^{\prime}}
\def\bdp{b_4^{\prime}}
\def\t {\times }
\def\slash {\!\!\!\!\!\!/}
\def\photino {\tilde\gamma}
\def\sel {\tilde{e}}
 \def\N10{\widetilde \chi_1^0}
                         \def\C1p{\widetilde \chi_1^+}
                         \def\C1m{\widetilde \chi_1^-}
                         \def\C1pm{\widetilde \chi_1^\pm}
 \def\Ntwo{\widetilde \chi_2^0}
                         \def\Ctwo{\widetilde \chi_2^\pm}
\def\lslep {{\tilde e}_L}
\def\rslep {{\tilde e}_R}
\def\sneu {\tilde \nu}
\def\msneu {M_\tilde \nu}
\def\mrslep {m_{\rslep}}
\def\mlslep {m_{\lslep}}
\def\mneu {m_{\neu}}
\def\mpT{p_T \hspace{-1em}/\;\:}
\def\mET{E_T \hspace{-1.1em}/\;\:}
\def\mE{E \hspace{-.7em}/\;\:}
\def\go{\rightarrow}
\def\beq{\begin{eqnarray}}
\def\Rp{R\!\!\!\!/}
\def\wrp {{\cal W}_{R\!\!\!\!/}}
\def\enq{\end{eqnarray}}
\def\goes{\longrightarrow}
\def\lsim{\:\raisebox{-0.5ex}{$\stackrel{\textstyle<}{\sim}$}\:}
\def\gsim{\:\raisebox{-0.5ex}{$\stackrel{\textstyle>}{\sim}$}\:}

 \begin{flushleft}
{SHEP-07-29} \hfill {~~~~~~~~~~~~~~} \\
{DFTT 39/2009} \hfill {\today}\\
 \end{flushleft}
\begin{center}
{\large\bf Production of Light Higgs Pairs
 in 2-Higgs Doublet Models}\\[2.5mm]
{\large\bf via the Higgs-strahlung Process at the LHC}\\[3mm]
{\large M. Moretti}\\
{\small\it Dipartimento di Fisica, Universit\`a di Ferrara and\\
INFN - Sezione di Ferrara, Via Paradiso 12, 44100 Ferrara, Italy
}\\[2mm]
{\large S. Moretti}\\
{\em School of Physics \& Astronomy, University of Southampton,\\
Highfield, Southampton SO17 1BJ, UK}
\\[2mm]
{\em and}
\\[2mm]
{\em
Dipartimento di Fisica Teorica,
Universit\`a degli Studi di Torino\\
Via Pietro Giuria 1,
10125 Torino,
Italy
}
\\[2mm]
{\large F. Piccinini}\\
{\em INFN - Sezione di Pavia,\\
    Via Bassi 6, 27100 Pavia,
Italy}
\\[2mm]
{\large R. Pittau}\\
{\em Departamento de F\'{\i}sica Te\'orica y del Cosmos
        and Centro Andaluz de F\'{\i}sica de Part\'{\i}culas 
        Elementares (CAFPE),
        Universidad de Granada, E-18071 Granada, Spain}
\\[2mm]
{\large J. Rathsman}\\
{\em Department of Physics and Astronomy, Uppsala University, \\
PO Box 516, 751 20 Uppsala, Sweden}
\\[1.5mm]
\end{center}
\begin{abstract}
{\noindent\small 
At the Large Hadron Collider, we prove the 
feasibility to detect pair production of 
the lightest CP-even Higgs boson $h$ of a
Type II 2-Higgs Doublet Model
through the process
$q\bar q^{(')} \to V {hh}$ (Higgs-strahlung, $V=W^\pm,Z$),
in presence of two $h\to
 b\bar b$ decays.
We also show that, through such production and decay
channels, one has direct access to the following Higgs self-couplings,
thus enabling one to distinguish between a standard and the
Supersymmetric version of the above model: 
$\lambda_{Hhh}$ -- which constrains the form of the Higgs potential
-- as well as 
$\lambda_{W^\pm H^\mp h}$ and
$\lambda_{Z A h}$ -- which are required by gauge invariance.
Unfortunately, such claims cannot be extended to
the Minimal Supersymmetric Standard Model, 
where the extraction of the same signals is impossible.
}
  
\end{abstract}


\section{Introduction}
\label{sec:intro}
\noindent
The ability to access Higgs self-couplings would enable one 
to constrain the form of the Higgs potential responsible for Electro-Weak
Symmetry Breaking (EWSB) and, in particular, to assess whether such a potential
corresponds to the one embodied in the Standard Model (SM) or indeed 
in extended versions of it. Lately, amongst the scenarios with an 
enlarged Higgs sector, much attention has been devoted to a   
(CP-conserving) Type II 2-Higgs Doublet Model (2HDM),  
possibly in presence of minimal 
Supersymmetry (SUSY) -- the 
combination of the two yielding the so-called Minimal 
Supersymmetric Standard Model (MSSM)
\cite{guide}. In a generic 2HDM and in the MSSM, of the
initial eight degrees of freedom pertaining to the two complex Higgs
doublets, only five survive as real particles upon EWSB, 
labelled as $h, H$, $A$
(the first two are CP-even or `scalars' (with $M_h<M_H$) whereas the third is CP-odd or
`pseudoscalar') and $H^\pm$, as three degrees of freedom 
are absorbed into the definition
of the longitudinal polarisation for the gauge bosons $Z$ and $W^\pm$,
upon their mass generation after EWSB. This thus makes four additional 
Higgs particle
states with respect to the SM, wherein only one complex Higgs doublet
is allowed which, after EWSB, generates a lone CP-even (scalar) Higgs boson
(henceforth also denoted by $h$). 

The detection of a $h$ state, independently of the underlying model
(amongst those mentioned),
is guaranteed at the Large Hadron Collider (LHC) \cite{Abdel,ATLAS,CMS}, 
over the theoretically allowed mass intervals. However, notice that
while the Higgs boson mass $M_h$ is a free 
parameter in both the SM and a generic 2HDM, 
so that its upper limit (of about 700 GeV)
is only dictated by unitarity requirements, in the MSSM the SUSY
relations
interlinking gauge and Higgs couplings imply an absolute 
upper limit
on the mass of the lightest Higgs boson $M_h$ of about 140 GeV
(see the second paper in \cite{Abdel} and references therein). Furthermore,
there is a lower limit from LEP, of about 115 GeV, applicable directly to
$M_h$ in the SM but also adoptable in both a generic 2HDM and
the MSSM over large areas of their 
parameter spaces. Finally, the Tevatron has put an exclusion limit on $M_h$ in the SM in the range 
162-166 GeV \cite{Aaltonen:2010yv}.

The CP-conserving 2HDM, with soft $Z_2$ symmetry breaking, is identified uniquely at tree level 
by seven independent parameters (as will be made clear below): 
$M_{h}$, $M_{H}$, $M_{A}$, $M_{H^\pm}$, $\tan\beta$ (the ratio between the
vacuum expectation values of the two Higgs doublets), 
$\alpha$ (the mixing angle between
the two CP-even neutral Higgs states) and
$\lambda_5=(\Lambda_5-\Lambda_6)/2$, 
one of the parameters of the Higgs potential as defined in \cite{THDM} 
(see also \cite{Fawzi}). In the case of the MSSM, at tree level,
only two parameters are needed, $\tan\beta$
and any of the Higgs masses (henceforth $M_A$). 
(Recall that,
 in the SM, $M_h$ is the only free parameter in the Higgs potential.)

An intriguing situation that could arise at the LHC is then the following.
That only a light Higgs state, with mass below 140 GeV, is found and it is 
impossible to establish whether it belongs to the SM, a generic 2HDM or indeed
the MSSM. In fact, this  happens in the so-called
`decoupling region' of the extended scenarios, i.e.,
when $M_{H},M_{A},M_{H^\pm}\gg M_{h}$ 
(and for suitable
choices of the additional free inputs in the generic
2HDM), quite independently of  $\tan\beta$, where - for the same mass - 
the $h$ couplings
to ordinary matter in the SM are the same as those
in both the 2HDM and MSSM\footnote{In the MSSM, one
also has to presume that the sparticle states are very heavy, so that the existence
of a SUSY scenario is not obviously manifest through their detection.}. 
Even in these conditions, however, it has been proved that one could
possibly establish the presence of an extended Higgs sector by determining
the size of the trilinear Higgs self-coupling $\lambda_{hhh}$ 
\cite{HHH}, by looking at the Vector Boson Fusion (VBF) process,
$q q^{(')}\to q q^{(')} hh$, and possibly also 
distinguish between a generic 2HDM and the MSSM by isolating
the $\lambda_{Hhh}$ trilinear Higgs self-coupling in the same 
VBF channel \cite{HHH_MSSM-2HDM}. This
can be achieved by looking at decays of the two light Higgs states 
produced into bottom-quark pairs.

In general though, this distinction between a generic 2HDM and the MSSM is not
always possible. Not
even the discovery of also 
the $A$ and $H^\pm$ states may say the conclusive word 
in this respect. In fact, despite there exist well established spectra among 
the four different masses in the MSSM
(dictated indeed by SUSY for fixed, say, $M_{h}$ and $\tan\beta$),
it may well be possible that the additional 2HDM parameters arrange
themselves to produce an identical mass pattern. However, such a degeneracy
between the two models would not typically
persist if one were able to also measure other Higgs-gauge couplings, chiefly
 $\lambda_{W^\pm H^\mp h}$ and/or $\lambda_{Z A h}$, 
which are both proportional to $\cos(\beta-\alpha)$ times gauge couplings. 
Even though the form of the Higgs-gauge coupling is specified by
reasons of gauge invariance, the actual size can be different in the MSSM and
in a generic 2HDM since in the latter both $\alpha$ and $\beta$ are independent
parameters.  Hence, the measurement of at least one of these 
Higgs-gauge couplings would constitute a confirmation,
or otherwise, of the MSSM
relations if one knew $M_{h}$ (and possibly the heavier masses, $M_H$, $M_A$
plus $M_{H^\pm}$) and $\tan\beta$ but not $\alpha$.

These Higgs-gauge couplings are accessible via the double Higgs-strahlung 
process, that we intend to study here\footnote{In principle, they can
also be accessed in VBF: see the last four graphs of Fig.~1 
in \cite{HHH_MSSM-2HDM}. In practise, we had verified in
that paper that the contribution due to those diagrams was not accessible,
the main reason being that they are not resonant in VBF, unlike 
the case of Higgs-strahlung.}. 
Our approach is very conservative, as 
we make the assumption that only one parameter is known, 
$M_{h}$, measured for some value below 140 GeV, as may well happen at the LHC 
after only a $h$ resonance is detected. We further imply that all SUSY
states are much heavier than ordinary particles (with the possible exception
of the lightest SUSY particle, which may well
escape detection) and not yet detected
by the time the $h$ resonance is isolated
and our advocated studies are pursued. In particular,
in this paper, we show that,
by resorting to double Higgs-strahlung, it is possible
to extract both $\lambda_{W^\pm H^\mp h}$ 
and $\lambda_{Z A h}$ in a generic
2HDM whilst in the MSSM neither of them is accessible\footnote{Similar 
conclusions also hold at linear colliders, as shown in 
Ref.~\cite{Ferrera:2007sp}.}. 

Therefore, in 
conjunction with the measurement of
{$\lambda_{Hhh}$}, accessible in VBF \cite{HHH_MSSM-2HDM} and also in the
process considered here (as we shall show), one will in principle be
able to disentangle between the 2HDM and the MSSM, even if only a light
Higgs resonance is found after earlier LHC data and no SUSY states are 
visible throughout.  However, all this phenomenology requires 
that enough luminosity can be accumulated at the LHC, such that
the Super-LHC (SLHC) option \cite{SLHC} becomes needed to perform
such kind of studies.

\begin{figure}[!ht]
\vspace*{5.0cm}
  ~\hskip1.0cm\epsfig{file=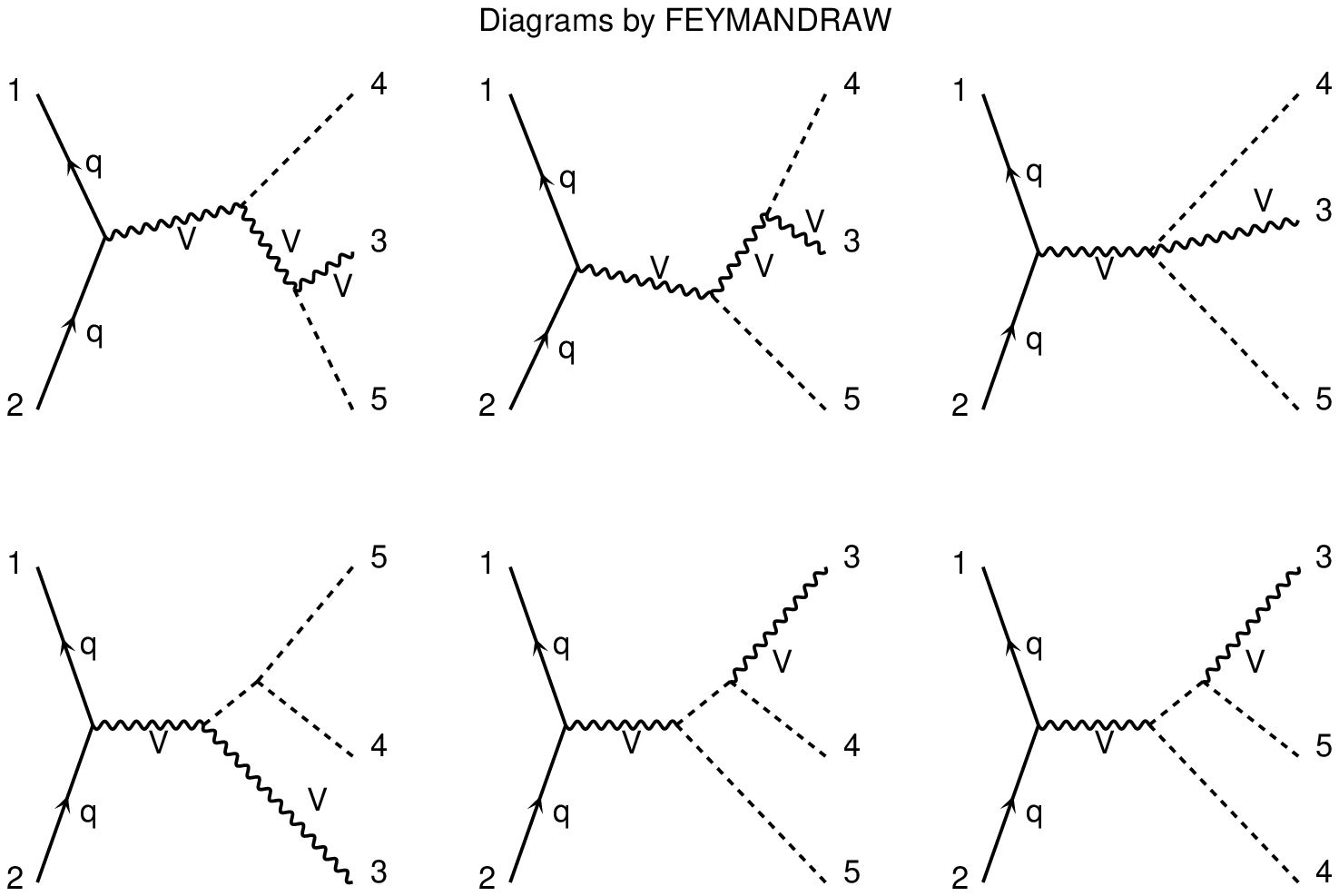,height=18cm,angle=0}
\vspace*{-10cm}
\caption{Feynman diagrams for $q_1\bar q^{(')}_2\to V_3h_4h_5$.}
\label{fig:Vgraphs}
\end{figure}

We will illustrate 
how we have come to this conclusion, i.e., after investigating 
the production of lightest Higgs boson pairs in the Higgs-strahlung
process, namely \cite{VHH,VHHAbdel}:
\bea\label{proc}
q\bar q^{(')} \to V {hh},
\eea
with
$V=W^\pm~{\rm or}~Z$  
and $q^{(')}$ referring to any possible (anti)quark
flavour. The relevant Feynman diagrams corresponding to this
channel in both
the 2HDM and MSSM considered here can be 
found in Fig.~\ref{fig:Vgraphs}. (Herein, notice the three graphs at the bottom,
where the $\lambda_{hhh}$ or  $\lambda_{Hhh}$ and  $\lambda_{W^\pm H^\mp h}$ or $\lambda_{Z A h}$ couplings 
are located.)
In our selection analysis, we 
will resort to the extraction of two
$h\to b\bar b$ resonances, in presence of both hadronic and leptonic
$V$ decays, i.e., we will study the following signatures:
\begin{enumerate}
\item `four $b$-quark jets and two light-quark jets', emerging
from (\ref{proc}) if $W^\pm,Z\to $~jet-jet;
\item `four $b$-quark jets and a lepton pair', emerging
from (\ref{proc})  if $Z\to \ell^+\ell^-$;
\item `four $b$-quark jets, one lepton and missing energy', emerging
from (\ref{proc})  if $W^\pm\to \ell^\pm\nu_\ell$.
\end{enumerate}
We will limit ourselves to the case $\ell=e,\mu$ and we will discuss the
procedure to tag the $b$-jets.

Our paper is organised as follows. In the next section, we outline
the computational procedure. Sect.~3 presents our numerical results
and discusses these. Sect.~4 contains our conclusions.

\section{Calculation}
\label{sec:calcul}

We have assumed $\sqrt s=14$ TeV for the LHC energy throughout. 
Our numerical results are obtained by setting the renormalisation and 
factorisation scales to $2M_{h}$
for the signal while for the QCD
background we have used the average jet transverse momentum 
($p_T^2=\sum_{1}^{n}p_{Tj}^2/n$).
Both Higgs processes and noise were 
estimated by using the Parton Distribution Function (PDF) set MRST99(COR01)
\cite{hepdata}. All signal and background calculations  
 were based on exact tree-level Matrix
Elements (MEs) using either the ALPGEN program \cite{ALPGEN} or the
HELAS subroutines \cite{HELAS}. Furthermore, 
VEGAS \cite{VEGAS} or Metropolis
\cite{Metropolis} were used for the multi-dimensional
integrations over the phase space. 
As for numerical input values of SM parameters,
we adopted the ALPGEN defaults. 

Concerning the MSSM setup, the two independent tree-level parameters that
we adopt are $M_A$ and $\tan\beta$. Through higher orders, 
we have considered the so called `Maximal
Mixing' scenario ($X_t=A_t-\mu/\tan\beta =\sqrt6 M_{\rm{SUSY}}$) \cite{MaxMix}, 
wherein 
we have chosen for the relevant SUSY input parameters: 
$\mu=200$ GeV, $A_b=0$, with 
$M_{\rm{SUSY}}=5$ TeV, the latter -- as already intimated --
implying a sufficiently heavy scale for all sparticle masses, so
that these are not accessible at the LHC and
no significant interplay between the SUSY and Higgs sectors of the model
can take place\footnote{The only 
possible exception in this mass hierarchy would be the
Lightest Supersymmetric Particle (LSP), as intimated already, whose mass may well be smaller
than the lightest Higgs mass values that we will be considering. However,
we have verified that invisible $h$ decays (including the one
into two LSPs) have negligible decay rates.}. Masses and couplings
within the MSSM have been obtained by using the {HDECAY} program \cite{HDECAY}.
The $\tan\beta$ and $M_A$ values adopted as representative of the MSSM
parameter space were: $\tan\beta=3$ and 40 for $M_A=160$, 200 and 240
GeV (yielding $M_{h,H}=108(178)$ GeV, $M_{h,H}=112(212)$ GeV
and $M_{h,H}=114(249)$ GeV for the low $\tan\beta$ value plus
$M_{h,H}=129(160)$ GeV, $M_{h,H}=129(200)$ GeV
and $M_{h,H}=129(240)$ GeV for the high one, respectively). 

Before giving the details of the 2HDM setup we are using, let us recall the
most general CP-conserving 2HDM scalar potential which is ($Z_2$) symmetric under 
$\Phi_{1(2)}\to-(+)\Phi_{1(2)}$ up to softly breaking
dimension-2 terms (thereby allowing for loop-induced flavour 
changing neutral currents)~\cite{guide},
\begin{eqnarray}\label{eq:potential}
V &=& 
m_{11}^2\Phi_1^\dagger\Phi_1
+
m_{22}^2\Phi_2^\dagger\Phi_2
-
\left\{m_{12}^2\Phi_1^\dagger\Phi_2 + h.c.\right\}
+
\frac{1}{2}\lambda_1\left(\Phi_1^\dagger\Phi_1\right)^2
+
\frac{1}{2}\lambda_2\left(\Phi_2^\dagger\Phi_2\right)^2
+
\nonumber \\  &&
+
\lambda_3\left(\Phi_1^\dagger\Phi_1\right)\left(\Phi_2^\dagger\Phi_2\right)
+
\lambda_4\left(\Phi_1^\dagger\Phi_2\right)\left(\Phi_2^\dagger\Phi_1\right)
+
\left\{ \frac{1}{2}\lambda_5 \left(\Phi_1^\dagger\Phi_2\right)^2  + h.c.\right\}.
\end{eqnarray}
In the following,
the parameters $m_{11}$, $m_{22}$, $m_{12}$, $\lambda_1$, $\lambda_2$, 
$\lambda_3$ and $\lambda_4$ are replaced by
$v$, $M_{h}$, $M_{H}$, $M_{A}$, $M_{H^\pm}$, $\beta$ and $\alpha$ 
(with $v$ fixed). 
Hence, as intimated already, the CP-conserving 2HDM potential is parameterised by 
seven free parameters. 
Notice that 
from the scalar potential all the different Higgs couplings needed for our study
can easily be obtained. (See~\cite{THDM,Fawzi} for a 
complete compilation of couplings in a general CP-conserving 2HDM.) In addition, since we are 
considering a so called Type II 2HDM the Yukawa couplings are also fixed by $\tan\beta$ in the same 
way as in the MSSM.

In our 2HDM, we will
fix $M_{h}$ and $M_{H}$ to values similar to the ones found in the
MSSM scenario we are considering, by adopting
three different setups:
\begin{enumerate}
\item
$M_h=115$ GeV, $M_{H}=250$ GeV, $M_{A}=220$ GeV, $M_{H^\pm}=300$ GeV,
\item
$M_h=115$ GeV, $M_{H}=300$ GeV, $M_{A}=305$ GeV, $M_{H^\pm}=310$ GeV,
\item
$M_h=115$ GeV, $M_{H}=400$ GeV, $M_{A}=405$ GeV, $M_{H^\pm}=410$ GeV,
\end{enumerate}
where in the first scenario the masses have been chosen such that $M_H > 2M_h$, $M_A> M_h +M_Z$ 
and $M_{H^\pm}$ is above the limit given by measurements of $b \to s \gamma$ 
(for recent limits see \cite{Mahmoudi:2009zx,Deschamps:2009rh} and references therein).

We always scan over the remaining parameters
in the ranges
\begin{eqnarray*}
&-\pi/2<\alpha<\pi/2,&\\
&-4\pi<\lambda_5<4\pi,&\\
&0<\tan\beta<50. &
\end{eqnarray*}

\begin{figure}[!ht]
\center
\epsfig{file=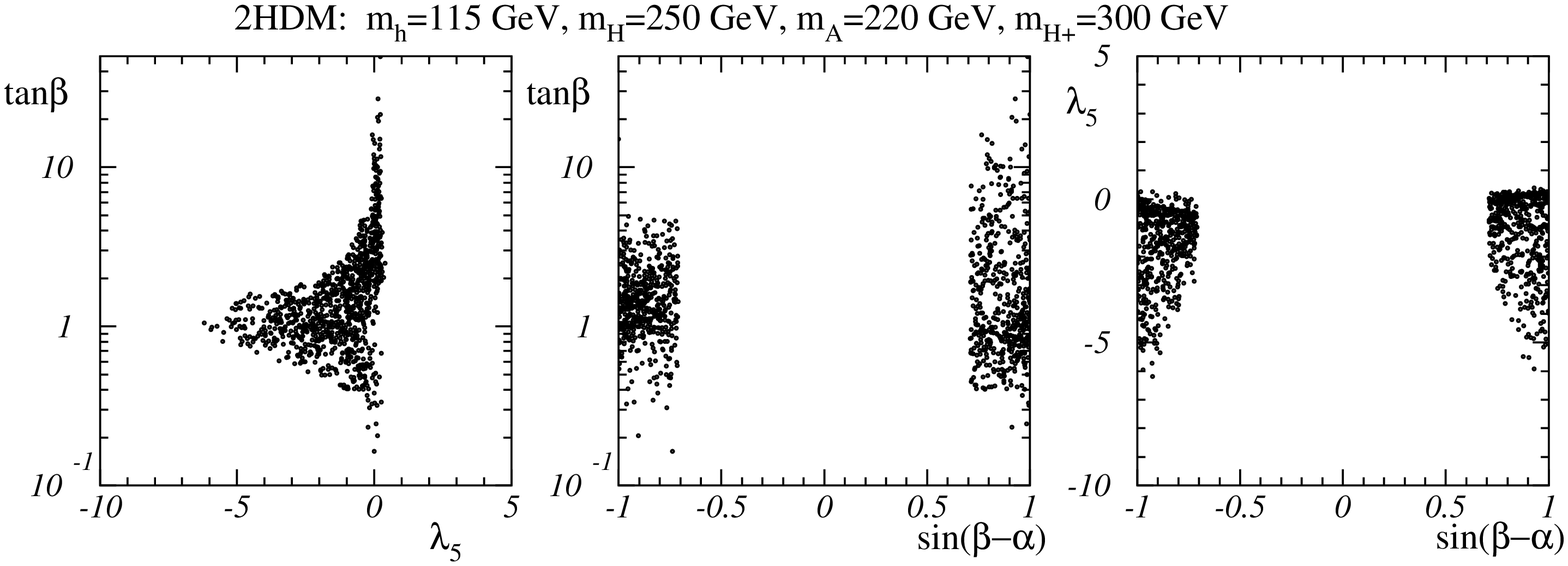,height=6cm}
\epsfig{file=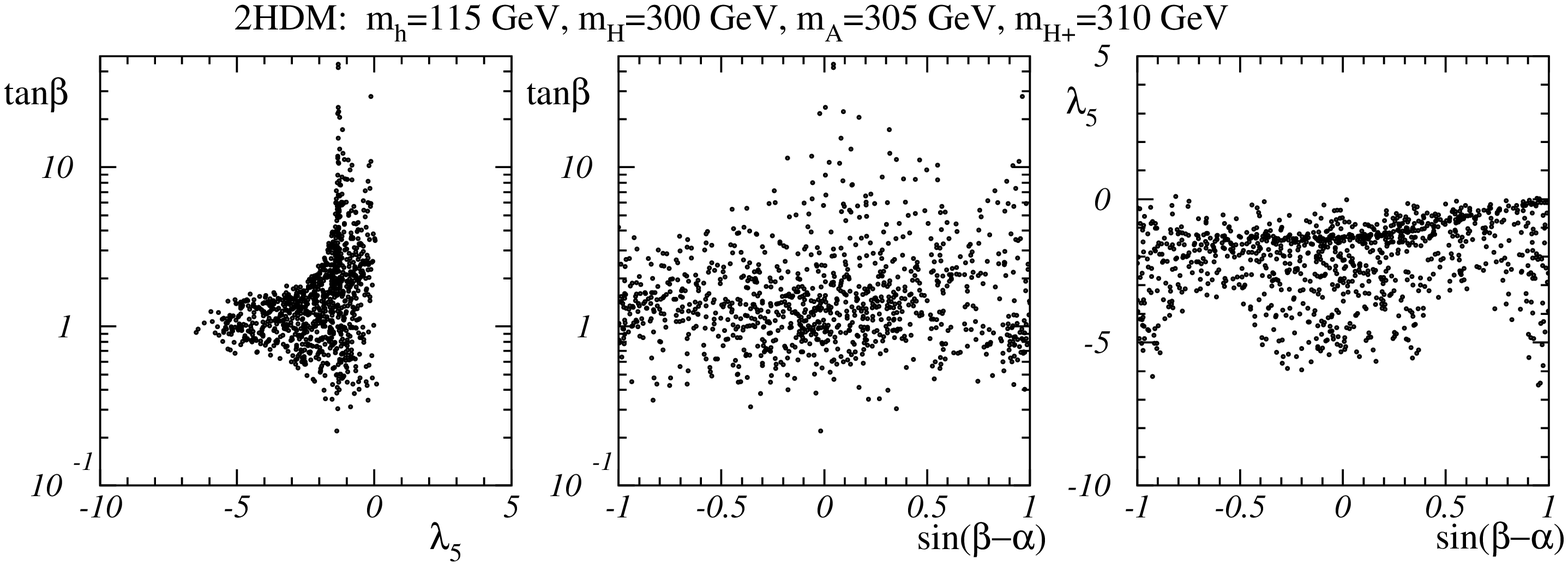,height=6cm}
\epsfig{file=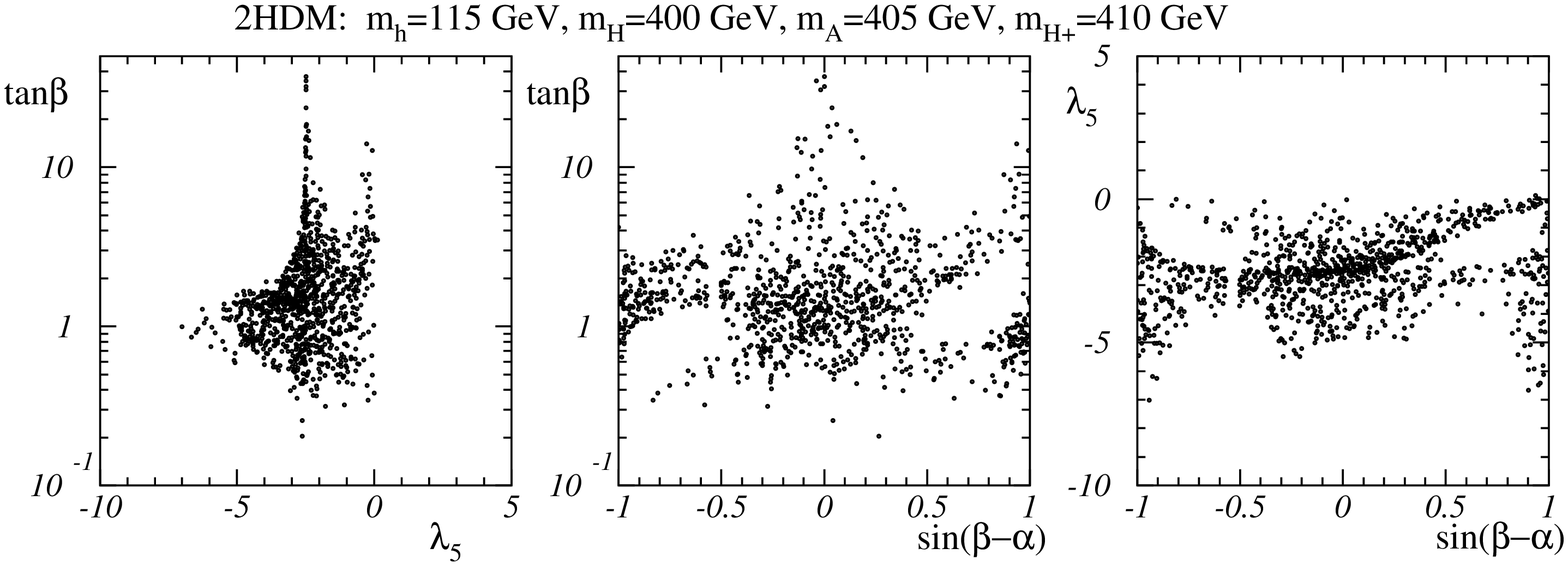,height=6cm}
\caption{The projection of the allowed points in the 2HDM parameter space 
on three different planes, $\lambda_5$--$\tan\beta$, 
$\sin(\beta-\alpha)$--$\tan\beta$ and $\sin(\beta-\alpha)$--$\lambda_5$, in the 
three different mass scenarios: 
(top) $M_{H}=250$ GeV, $M_{h}=115$ GeV, $M_{A}=220$ GeV and $M_{H^\pm}=300$ GeV,
(middle) 
$M_{H}=300$ GeV, $M_{h}=115$ GeV, $M_{A}=305$ GeV and $M_{H^\pm}=310$ GeV,
(bottom) $M_{H}=400$ GeV, $M_{h}=115$ GeV, $M_{A}=405$ GeV and $M_{H^\pm}=410$ GeV. 
The projections have been obtained with a sample of 1000 random points in the
allowed 2HDM parameter space.}
\label{fig:allowed2HDM}
\end{figure}

In order to accept a point from the scan we also check 
that the following conditions are fulfilled: the potential is bounded
from below, 
the $\lambda_i$'s comply with the unitarity 
constraints of \cite{Akeroyd:2000wc} and yield a contribution 
to $|\Delta\rho| < 10^{-3}$.
In short, the unitarity constraints amount to putting limits on the 
eigenvalues of the $S$ matrices for the scattering of various combinations of Higgs and
EW gauge bosons. We have followed the normal 
procedure~\cite{guide} of requiring the $J=0$ partial waves ($a_0$) of 
the different scattering processes to fulfill  $|{\rm Re}(a_0)|<1/2$, which 
corresponds to applying the condition that the eigenvalues\footnote{Here, 
$Z_2$ refers to the $Z_2$ symmetry, $Y$ is the hypercharge, and 
$\vec{\sigma}$ is the total weak isospin.} 
$\Lambda_{Y\sigma\pm}^{Z_2}$ of the scattering matrices (or more precisely 
$16 \pi S$) fulfill $|\Lambda_{Y\sigma\pm}^{Z_2}| < 8 \pi$~\cite{Ginzburg:2005dt}. 
In other words we allow parameter space points all the way up to
the tree-level unitarity constraint $|{\rm Re}(a_0)|<1/2$. 
The spectrum of masses, couplings and decay rates
in our 2HDM is similar to the one exploited 
in Ref.~\cite{HHH_MSSM-2HDM}\footnote{For a variety of
studies of Higgs pair production at the LHC, primarily through gluon-gluon fusion,
see Refs.~\cite{Glover:1987nx}--\cite{Arhrib:2009hc}. Recent studies at 
linear and $\gamma \gamma$ colliders can
be found in Refs.~\cite{LopezVal:2009qy},~\cite{Hodgkinson:2009uj} and 
\cite{Bernal:2009rk}.}, 
obtained by using a modification of {HDECAY} \cite{HDECAY}
(consistent with a similar manipulation of the program used in
Ref.~\cite{BRs}). 
For each accepted point in the scan the partial decay rates 
for the different Higgs bosons are then calculated using 
such {HDECAY} modification
and also taking possible additional (i.e.,
2HDM specific) partial widths of the 
${H}$, ${A}$, and ${H^\pm}$ bosons into account.
We have also cross-checked the results by using the 2HDMC program~\cite{2HDMC} 
to perform the same calculations. The results of the above discussed 
constraints on the parameter space are shown in 
Fig.~\ref{fig:allowed2HDM}, where the projections of the allowed parameters 
on three different planes, $\lambda_5$--$\tan\beta$, 
$\sin(\beta-\alpha)$--$\tan\beta$ and 
$\sin(\beta-\alpha)$--$\lambda_5$, with a sample of 1000 
randomly chosen points are plotted. It is worth noting that the 
allowed region in the plane $\lambda_5$--$\tan\beta$ complies with 
the constraints found in Refs.~\cite{LopezVal:2009qy,LopezVal:2010vk}, 
where the neutral Higgs-pair production and Higgs-strahlung 
at linear colliders within the general 
2HDM have been investigated at the one-loop level. 

While the parameter dependence of the MSSM Higgs sector renders
the computation of the tree-level MSSM cross-sections rather straightforward
(as the latter depends on two parameters only, $M_{A}$ and $\tan\beta$),
the task becomes much more time-consuming in the context of the 2HDM.
In order to calculate the cross-sections in this scenario, the latter are written as a
combination of couplings and kinematic factors in the following way ($n=5$):
\begin{equation}
\sigma_{\rm tot} = \int \left| \sum_{i=1}^{n} g_i M_i 
\right|^2 d {\rm{LIPS}} = 
\sum_{i=1}^{n}\sum_{j=i}^{n} g_ig_j\sigma_{ij},
\label{eqn:sigma_sum}
\end{equation}
where all the explicit 
dependence on $\alpha$, $\beta$, $\lambda_{Hhh}$ and 
$\lambda_{hhh}$ is contained in the couplings $g_i$: 
$ g_1= \sin^2(\beta-\alpha)$, 
$ g_2 = \cos^2(\beta-\alpha) $, 
$ g_3 = \cos(\beta-\alpha)\lambda_{Hhh}$, 
$ g_4 = \sin(\beta-\alpha)\lambda_{hhh}$, and 
$ g_5 = 1$,
whereas the 
dependence on masses and other couplings is in the factors 
\begin{equation}
\sigma_{ij}=\frac{1}{1+\delta_{ij}}\int \left( M_i^\dagger M_j + 
M_j^\dagger M_i \right) d {\rm{LIPS}}.
\end{equation}
Note that the sum over subamplitudes $M_i$ also contains all interference 
terms and that colour factors etc.\ are included properly. 
The $\sigma_{ij}$ are then calculated numerically for fixed masses. 
We can then get the cross-section in an arbitrary parameter space point by
multiplying the kinematic factors with the appropriate couplings. However,
there is a slight complication since the kinematic factor for the
$H \to hh$ contribution depends on the width $\Gamma_{H}$ if 
there is a $s$-channel resonance and the width in turn depends on the couplings.
In this case the kinematic factor scales as $1/\Gamma_{H}$ which is 
accounted for by assuming a fixed value for the width when the kinematic factor
is calculated and then rescaling the result with the true width when calculating
the contribution to the cross-section. 
The same type of procedure is also applied for the 
$A \to Zh$ and $H^\pm \to W^\pm h$ resonant contributions. 

Notice that such a 
procedure, after the summation in eq.~(\ref{eqn:sigma_sum}), amounts to the
exact (i.e., complete and gauge invariant\footnote{Apart from subleading effects of
${\cal O}(\Gamma/M)$, where $M$ and $\Gamma$ are the mass and width, respectively,
of a resonant particle. }) calculation of the cross section for process (\ref{proc}).

\section{Results}
\label{sec:results}

\subsection{Inclusive Signal Results}
\label{subsec:inclusive}

In this section, after a preliminary analysis of the Higgs mass and coupling spectra
in the MSSM and a general Type II 2HDM, we will start our numerical analysis by investigating the model 
parameter dependence of the Higgs pair production process in (\ref{proc})
at fully inclusive level, assuming on-shell production of gauge and Higgs
bosons, followed by the decay of the former in all possible final states
(of quarks, leptons and neutrinos, as appropriate) and of the latter into
two $b\bar b$ pairs, with the integration over the  phase space being performed
with no kinematical restrictions. This will be followed by an analysis of the
production and decay processes pertaining to the Higgs signals at fully differential
level, in presence of detector acceptance cuts and kinematical selection constraints.
Finally, we will compare the yield of the signals to that of the corresponding
backgrounds and perform a dedicate signal-to-background study including an
optimisation of the cuts in order to enhance the overall signal significance. 
We will treat the MSSM and 2HDM in two separate subsections.
 
\subsubsection{MSSM}
\label{subsubsec:MSSM}

\begin{figure}[!ht]
  ~\hskip0.0cm\epsfig{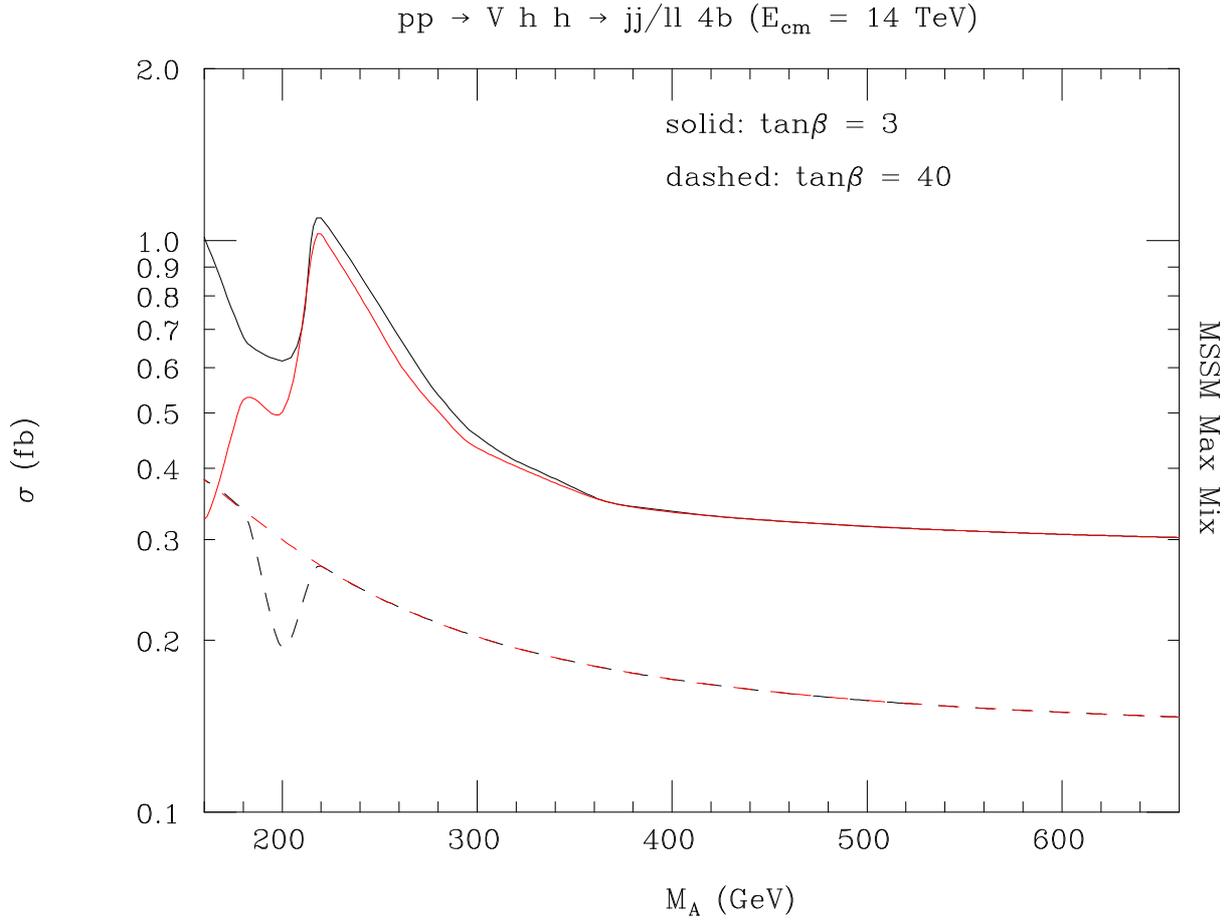}
\caption{The inclusive cross-sections  (as defined in the text) for
Higgs-strahlung in (\ref{proc}), followed by
$hh\to b\bar bb\bar b$  and $V\to$ `anything' decays, as a function of 
 the CP-odd Higgs boson  mass, $M_A$, for two choices of $\tan\beta$, 
assuming the MSSM in Maximal Mixing configuration. The red lines 
correspond to the
case in which we have removed from the computation the last two Feynman diagrams
in the previous figure.}
\label{fig:scan_hhv_MSSM_MaxMix}
\end{figure}

As representative of the low and high $\tan\beta$ regime, we will use in the remainder
the values of 3 and 40. We have instead treated $M_{A}$ as a continuous parameter,
varying between 100 and 700 GeV or so\footnote{Values of $M_{A}$ below 90 GeV or so
are actually excluded by LEP for the lower $\tan\beta$ value: see \cite{HiggsLEP2}.}. 
Fig.~\ref{fig:scan_hhv_MSSM_MaxMix}
presents the fully inclusive MSSM cross-section for the process
of interest, as defined in (\ref{proc}), followed by 
two $h\to b \bar b$ decays and of those of the $W^\pm$ and $Z$ bosons
into anything. 
In the same figure, we also show the rates corresponding to the
case in which we have removed from the list of diagrams the last two in 
Fig.~\ref{fig:Vgraphs}\footnote{Notice in 
fact that the two sets of diagrams peak in different
phase space regions, so that, modulo the aforementioned ${\cal O}(\Gamma/M)$ effects,
they can safely be separated.},
in order to individuate the origin of possible 
dynamic structures. In fact, the difference between the red
and black lines illustrate the effects of the diagrams
involving the $\lambda_{W^\pm H^\mp h}$ and
$\lambda_{Z A h}$ couplings, which can amount to a resonant
contribution (at low $\tan\beta$) or a constructive interference
(at large $\tan\beta$).

The displayed rates however correspond to the ideal situation
in which all final state objects (quarks and gauge bosons) are 
detected with unit efficiency and the detector coverage extends to their
entire phase space, so that they only serve as a guidance in rating the phenomenological
relevance of the processes discussed.

A more realistic analysis is in order, which we have performed as follows.
In all channels, the four $b$-jets emerging from the decay of the $hh$ pair are accepted
according to the following criteria:
\begin{equation}\label{precuts1}
p_T^b>30~{\rm{GeV}},\qquad 
\vert \eta^b \vert < 2.5,\qquad
\Delta R_{bb} > 0.7.
\end{equation}
Their tagging efficiency is taken as $\epsilon_{b}=50\% $ for each $ b $ satisfying
these requirements, $ \epsilon_{b}=0 $ otherwise\footnote{Here and in the remainder,
the label $b$ refers to jets that are $b$-tagged while $j$ to any jet (even those
originating from $b$-quarks)
which is not. As probability for a gluon or light-quark
jet to fake a $b$ jet we take $R=1/100$.}. We assume no $b$-jet charge determination. In addition,
to enforce the reconstruction of the two Higgs
bosons, we require all such $b$'s in the event to be tagged 
and that at least one out of the three possible double 
pairings of $b$-jets satisfies the following mass preselection:
\be 
\label{eq:mhchi2}
(m_{b_1,b_2}-M_h)^2+(m_{b_3,b_4}-M_h)^2 < 2~\sigma_m^2, 
\ee 
where $\sigma_m= 0.12~ M_h$ whereas $b_1 b_2$ and $b_3 b_4$ refer, hereafter,  to the two pairs of
$b$-jets best reconstructing the Higgs masses with $b_1$ being the $b$-jet with highest $p_T$. (As intimated, 
we will be working
under the assumption that a light Higgs state $h$ is found already.)

As for tagging the gauge bosons produced via the Higgs-strahlung 
process (\ref{proc}), there are two possibilities, depending on whether the gauge boson
decays are  `leptonic' or `hadronic'. In the former case,
the additional cuts are
\begin{equation}\label{precuts2}
 p_T^{\ell,{\rm{miss}}} >  20~{\rm{GeV}},\qquad
 |\eta^{\ell,{\rm{miss}}}| <  2.5,\qquad
\Delta R_{b\ell,b{\rm{miss}}} > 0.4,\qquad
 |M_{\ell^+\ell^-} -M_Z| <  5 ~{\rm{GeV}}
\end{equation}                                                           
(with $\ell=e,\mu$ and wherein the neutrino transverse momentum
is equated to the missing one, $p_T^{{\rm{miss}}}$, while the
longitudinal one is reconstructed by assuming $M_{\ell\nu_\ell}\equiv
M_{W^\pm}$ and choosing the solution with smallest magnitude).
In the latter case, we use
\begin{equation}\label{precuts3}
p_T^j>20~{\rm{GeV}},\qquad 
\vert \eta^j \vert < 2.5,\qquad
\Delta R_{bj,jj} > 0.7,\qquad
 70 ~{\rm GeV} < M_{jj} < 100 ~{\rm GeV},
\end{equation}
where $M_{jj}$ is the invariant mass of the two non-$b$-jets in the event.

In our investigation of hadronic
final states, we will
assume that $b$-quark jets are distinguishable from light-quark and gluon
ones and, as already mentioned, neglect considering $b$-jet charge determination. In the
case of leptonic ones, we will limit ourselves to the
case of electrons and muons (collectively denoted
by the symbol $\ell$).
Finite calorimeter resolution
has been emulated through a Gaussian smearing in transverse momentum,
$p_T$, with $(\sigma(p_T)/p_T)^2=(0.60/\sqrt{p_T})^2 +(0.04)^2$ for all
jets and $(\sigma(p_T)/p_T)^2=(0.12/\sqrt{p_T})^2 +(0.01)^2$ for
leptons. The corresponding missing transverse momentum,
$p_T^{\mathrm{miss}}$, was reconstructed from the vector sum of
the visible momenta after resolution smearing. Also,
in our parton level analysis, we
have identified jets with the
partons from which they originate and applied all cuts directly to the
latter, since parton shower and hadronisation effects were not included in our
study.
 
Tab.~\ref{tab:lambda} shows the rates of the signal
after the implementation of the
acceptance and preselection cuts in eqs.~(\ref{precuts1})--(\ref{precuts3})
(hereafter, also referred to as `primary cuts'). Notice
that $b$-tagging efficiencies
are not taken into account in this table. In fact, for the `$4b$-jet' tagging 
option that we are advocating, one should multiply the
numbers in Tab.~\ref{tab:lambda} by $\epsilon_b^4$, that is, 1/16. 
 
The conclusion here is rather straightforward: given the smallness of the
inclusive signal rates, the scope of extracting $hh\to 4b$ signals in the
MSSM from
the double Higgs-strahlung process (\ref{proc}) is basically nil. Therefore, in our 
forthcoming signal-to-background analysis, we will not pursue the MSSM case. 

\begin{table}[t]
\begin{center}
\begin{tabular}{|c|c|}
\hline
$M_A$ (GeV)  & $\sigma(q\bar q^{(')}\to Vhh)$ (fb) \\
\hline
\hline
160   & 0.028(0.034)  \\
200   & 0.019(0.037)   \\
240   & 0.022(0.053)   \\
\hline
\multicolumn{2}{|c|}{$\tan\beta=3$} \\ 
\hline
\hline
160   & 0.013(0.031)  \\
200   & 0.011(0.024)  \\
240   & 0.0087(0.020) \\
\hline
\multicolumn{2}{|c|}{$\tan\beta=40$} \\ 
\hline
\end{tabular}
\end{center}
\caption{Cross-sections for  Higgs-strahlung after Higgs/gauge boson decays (relevant Branching Ratios (BRs) are all included)
and the acceptance and preselection cuts defined
in (\ref{precuts1})--(\ref{precuts3}), for two choices of $\tan\beta$ and a selection
of $M_{A}$ values, assuming the MSSM
in Maximal Mixing configuration. Here,  
the numbers outside(inside) brackets refer to the `leptonic'(`hadronic') decay channel
of the gauge boson as described in the text. No $b$-tagging efficiencies are
included here.   
}
\label{tab:lambda}
\end{table}

\subsubsection{2HDM}
\label{subsec:2HDM}

\begin{figure}[!ht]
\center
\epsfig{file=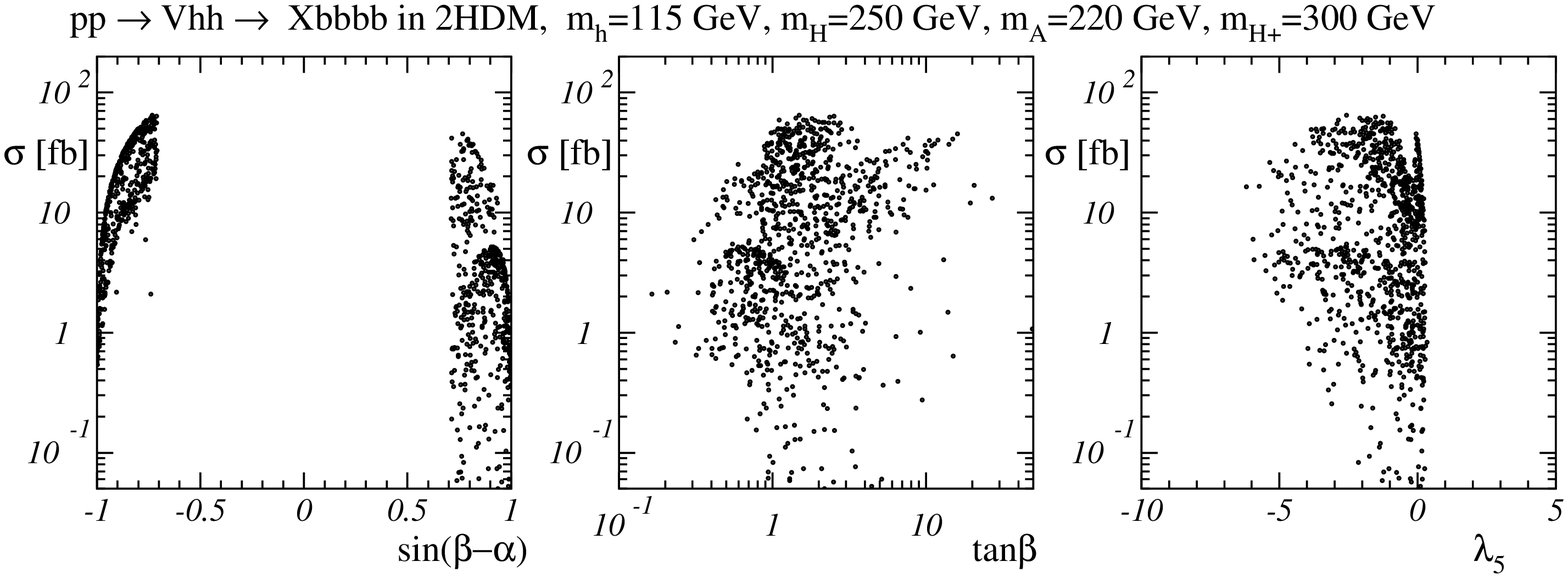,height=6cm}
\epsfig{file=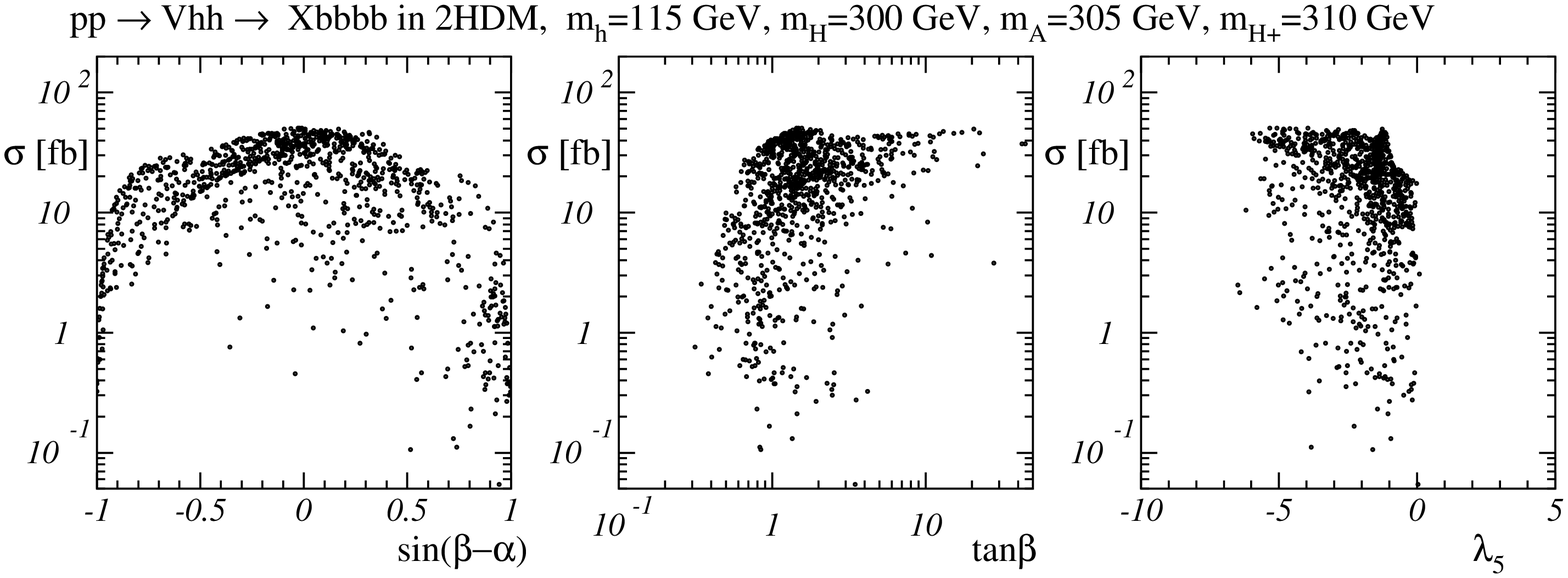,height=6cm}
\epsfig{file=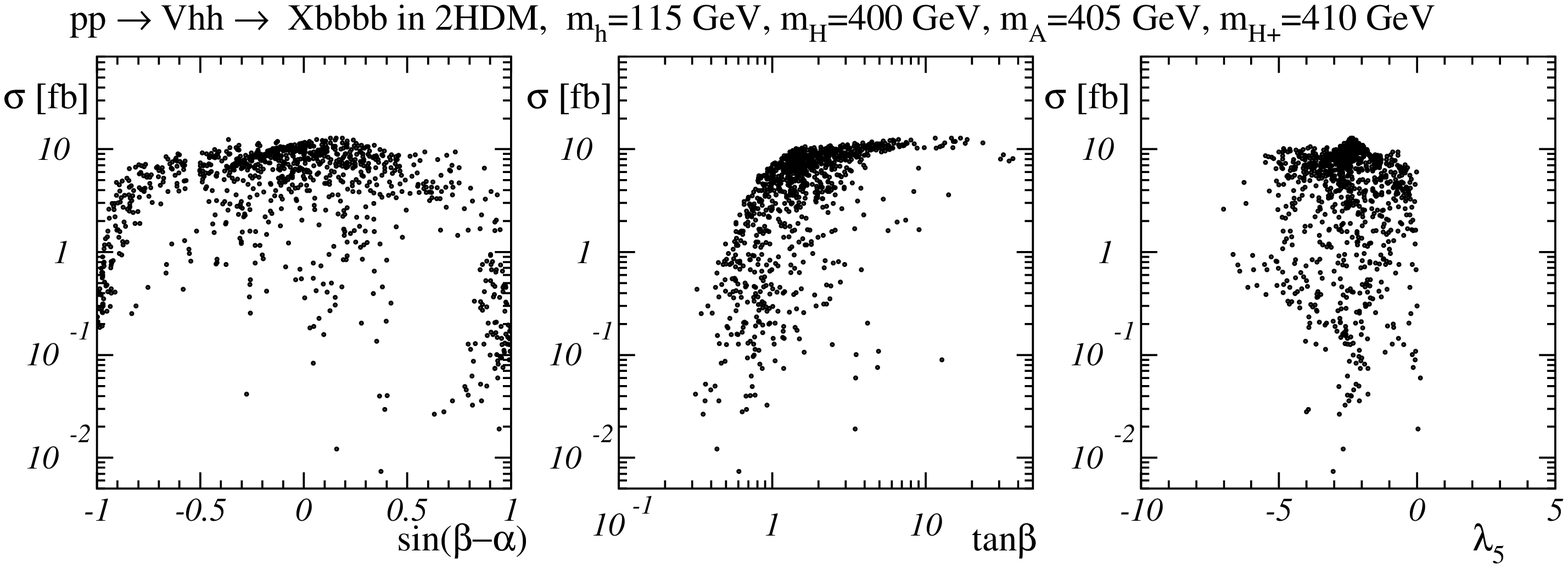,height=6cm}
\caption{The inclusive cross-section  in the 2HDM 
(as defined in the text)  for Higgs-strahlung in (\ref{proc}), 
followed by $hh\to b\bar bb\bar b$  and $V\to $ `anything' decays 
in the 
three different 
mass scenarios: 
(top) $M_{H}=250$ GeV, $M_{h}=115$ GeV, $M_{A}=220$ GeV and $M_{H^\pm}=300$ GeV,
(middle) 
$M_{H}=300$ GeV, $M_{h}=115$ GeV, $M_{A}=305$ GeV and $M_{H^\pm}=310$ GeV,
(bottom) $M_{H}=400$ GeV, $M_{h}=115$ GeV, $M_{A}=405$ GeV and $M_{H^\pm}=410$ GeV,
as a function of the remaining three parameters, for 1000 random points in the
allowed 2HDM parameter space.}
\label{fig:sigma_hhv_2hdm_notag_nocut}
\end{figure}

As already alluded to earlier the parameter space of the general CP-conserving
Type II 2HDM we are considering here is quite large as it depends
on seven independent parameters. 
In order to get a feeling for the dependence of
the signal cross-section of process (\ref{proc}) upon them, we present in
Fig.~\ref{fig:sigma_hhv_2hdm_notag_nocut} the results of our
selected mass scenarios where we scan the allowed parameter space over 1000 
randomly chosen points. 

Comparing with the cross-sections obtained in the MSSM we notice that in the 
2HDM, see  Fig.~\ref{fig:sigma_hhv_2hdm_notag_nocut}, 
they can be up to two orders of magnitude larger. 
The main differences are due to the fact that:
\begin{itemize}
\item
the trilinear Higgs couplings are not restricted in size by SUSY relations;
\item
the different parameters can vary independently of each other.
\end{itemize}
At the same time the kinematic factors will be the same for a given set of
masses and widths of the different Higgs bosons. Therefore, in those cases 
many features of the signal, such as differential distributions, 
will be similar to the MSSM even though the
normalisation can be completely different.
We also see 
from Fig.~\ref{fig:sigma_hhv_2hdm_notag_nocut}
that there is
quite a strong dependence on the masses of the Higgs bosons, with the maximal
cross-section essentially decreasing by one order of magnitude as
the heavy Higgs boson masses are increased from the scenario with lowest masses considered to
 the one with the highest masses.

\begin{figure}[!ht]
\center
\epsfig{file=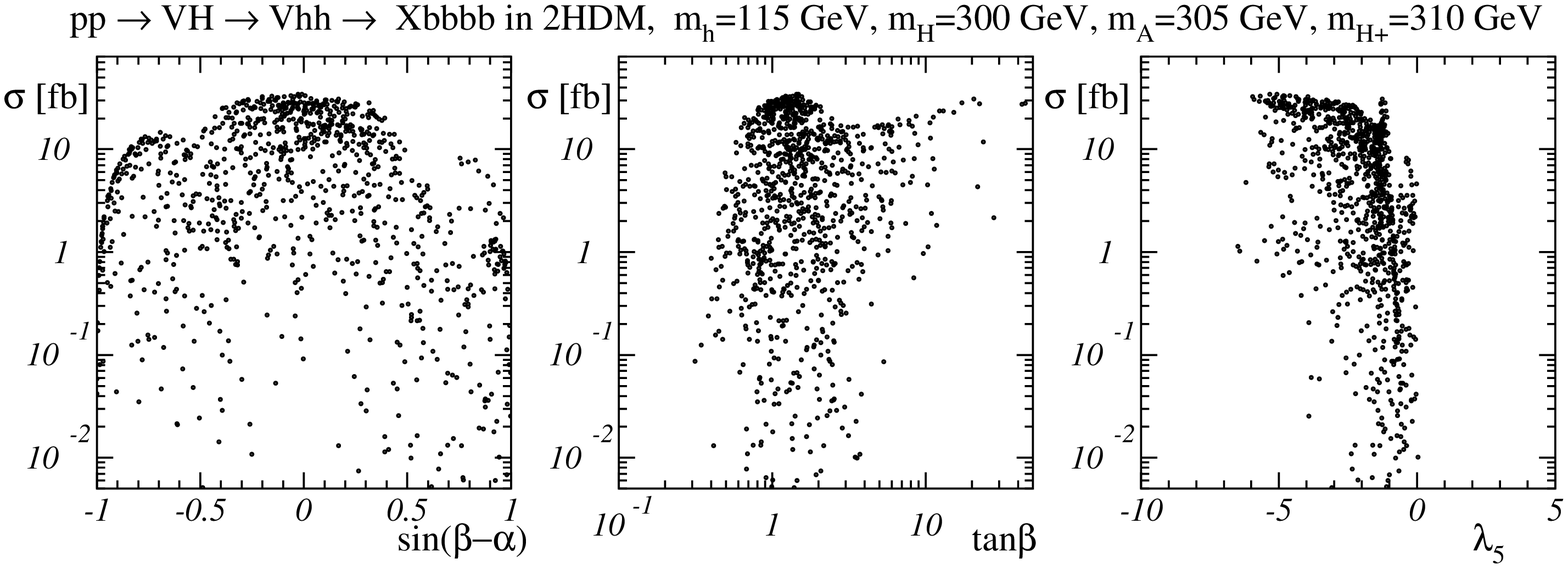,height=6cm}
\epsfig{file=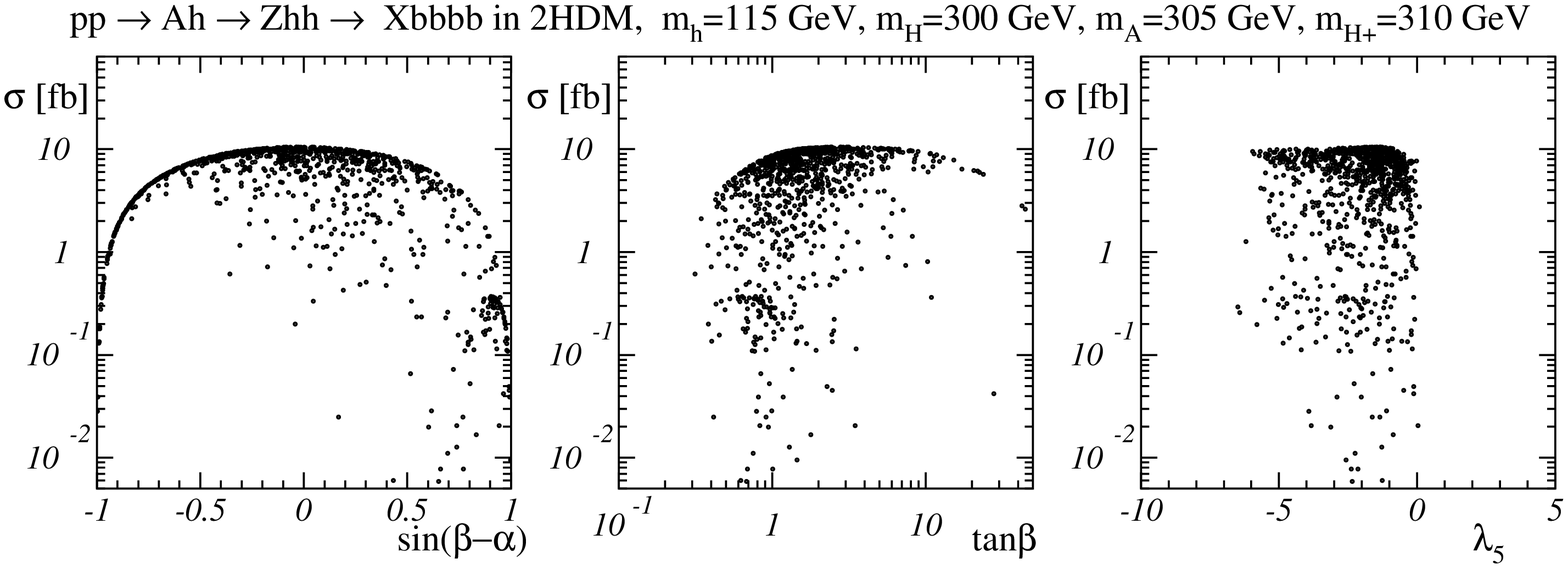,height=6cm}
\epsfig{file=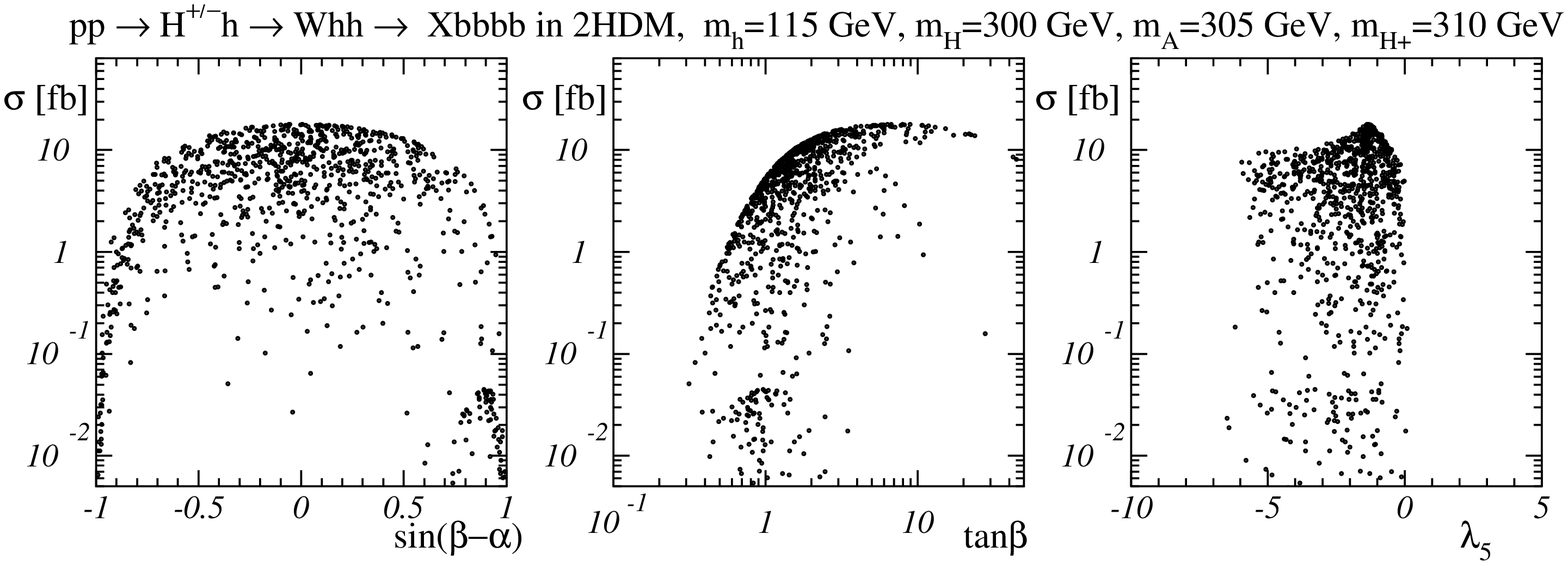,height=6cm}
\caption{The contributions to the inclusive cross-section  in the 2HDM 
(as defined in the text)  for Higgs-strahlung in (\ref{proc}), 
followed by $hh\to b\bar bb\bar b$  and $V\to $ `anything' decays 
from the: 
(top)  $H \to hh$,
(middle)  $A \to Zh$, and
(bottom)  $H^\pm \to W^\pm h$,
channels respectively
in the 
mass scenario: 
$M_{H}=300$ GeV, $M_{h}=115$ GeV, $M_{A}=305$ GeV and $M_{H^\pm}=310$ GeV,
as a function of the remaining three parameters, for 1000 random points in the
allowed 2HDM parameter space.}
\label{fig:sigma_hhv_2hdm_notag_nocut_rescon}
\end{figure}

As it is clear from  
Fig.~\ref{fig:sigma_hhv_2hdm_notag_nocut_rescon}, the largest 
inclusive cross-sections are obtained from the resonant channel  $H \to hh$, 
whereas the contributions from the $A \to Zh$ and
$H^\pm \to W^\pm h$ channels at the most reach about 10--20 fb. It is also clear
from the figures that, even though there are slight differences in the parametric
dependencies of the three different resonant contributions, the overall picture is
very similar in all cases.

Finally, we have also verified, though not shown, that the most promising signals are the ones where the 
$W^\pm$ decays leptonically, which is
sensitive to both the $H^\pm \to W^\pm h$  and  $H \to hh$ resonant enhancements,
and possibly also the leptonic decay of the $Z$, which is   sensitive
to possible $A \to Zh$ and $H\to hh$ resonances, whereas it is evident that the
hadronic decay channels will not be relevant phenomenologically since they are
only a factor $\sim$ 2--5 larger than the leptonic ones,
yet burdened by overwhelming QCD backgrounds\footnote{Hence, we will
not consider hadronic signatures any further in this study.}.

\subsection{Signal-to-Background Differential Analysis}
\label{subsec:S-to-B}

The backgrounds that we have considered are as follows.
In the case of process (\ref{proc}) for $V=W^\pm$, they are:
\begin{enumerate}
\item $Wb\bar bb\bar b$, irreducible (i.e., it has exactly the same particle content as the signal);
\item $t\bar t$, with two mistags (i.e., when two light-quark jets are erroneously classified
as $b$-jets);
\item $Wb\bar bjj$, with two mistags;
\item $Wjjjj$,  with four mistags;
\item $Whb\bar b\to Wb\bar bb\bar b$,  irreducible and co-resonant (i.e., it resonates at the $h$ mass,
albeit only once, unlike the signal, which does so twice);
\item $Whjj\to Wb\bar bjj$, co-resonant with two mistags.
\end{enumerate}
In the case of process (\ref{proc}) for $V=Z$, they are:

\begin{enumerate}
\item $Zb\bar bb\bar b$, irreducible;
\item $Zb\bar bjj$, with two mistags;
\item $Zjjjj$,  with four mistags;
\item $Zhb\bar b\to Zb\bar bb\bar b$,  irreducible and co-resonant;
\item $Zhjj\to Zb\bar bjj$, co-resonant with two mistags.
\end{enumerate}

In all cases both QCD and EW channels were considered, as appropriate. 
The former by exploiting 
ALPGEN and the latter the HELAS subroutines (including all 2HDM background channels to the
signal but not the interferences). We have however verified that, already after the
acceptance and preselection cuts, the last two channels (5. and 6. for $V=W^\pm$
plus 4. and 5. for $V=Z$) 
are negligible
in both cases, so that 
we have ignored them in the reminder of the analysis. In fact, also the backgrounds 4.\ for $V=W^\pm$
and 3.\ for $V=Z$ are very small so that they could be neglected in first approximation. We have however included
 them 
in our simulation. The backgrounds 3.\ for $V=W^\pm$
and 2.\ for $V=Z$ are similar both in magnitude and shape to the corresponding irreducible ones (i.e., 1.\ for 
both  $V=W^\pm$ and $V=Z$) after taking into account the appropriate weights, including the $b$-tagging/rejection 
factors discussed previously, although for clarity we do not plot them. Further, notice that background 2.\ for 
$V=W^\pm$ is 
significant before the final selection cuts, so that we have included it in our simulations.

We start by investigating the differential structure of process (\ref{proc})
in the attempt to extract the contributions that are most sensitive to the
$\lambda_{Hhh}$,
$\lambda_{W^\pm H^\mp h}=\cos(\beta-\alpha)$, and 
$\lambda_{ZAh}=\cos(\beta-\alpha)$ 
couplings, respectively. 
(Unfortunately, we can anticipate already that our studies revealed impossible
to extract the $h\to hh$ component of process (\ref{proc}).) 
To this end, we isolate in  (\ref{eqn:sigma_sum})
the corresponding ``resonant" contributions, more specifically $\sigma_{33}$ for $\lambda_{Hhh}$ and 
$\sigma_{22}$ for $\lambda_{W^\pm H^\mp h}$ and $\lambda_{ZAh}$\footnote{We have verified,
through BRS tests, that, although such contributions are not separately
gauge invariant, gauge violating effects always occur at the permille level over
the kinematic regions we are interested in (again, modulo the usual
${\cal O}(\Gamma/M$) effects).}. Note that this corresponds to the approximation 
that the respective final states are $WH$ or $ZH$ and $H^\pm h$ or $Ah$, with the consequent decays $H \to hh$, 
$ H^\pm \to W^\pm h$ and $A \to Z h$, for the two different processes at hand in (\ref{proc}).
We have then identified, for the three different mass scenarios 1.--3.\
introduced in Sect.~2, those differential spectra that show significant
differences from the yield of all the backgrounds above.

\begin{figure}[!ht]
\center
\epsfig{file=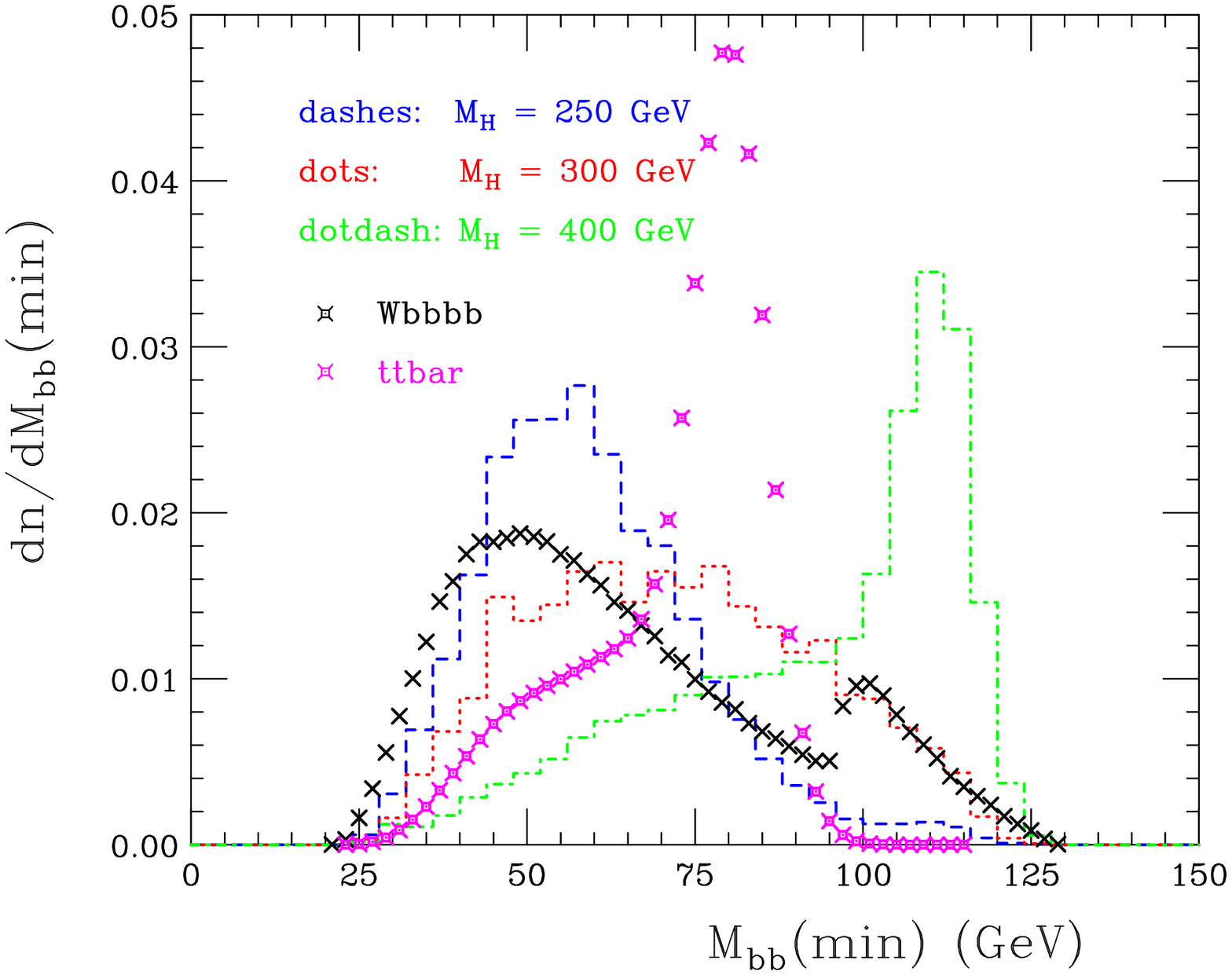, height=8cm, angle=90}
\epsfig{file=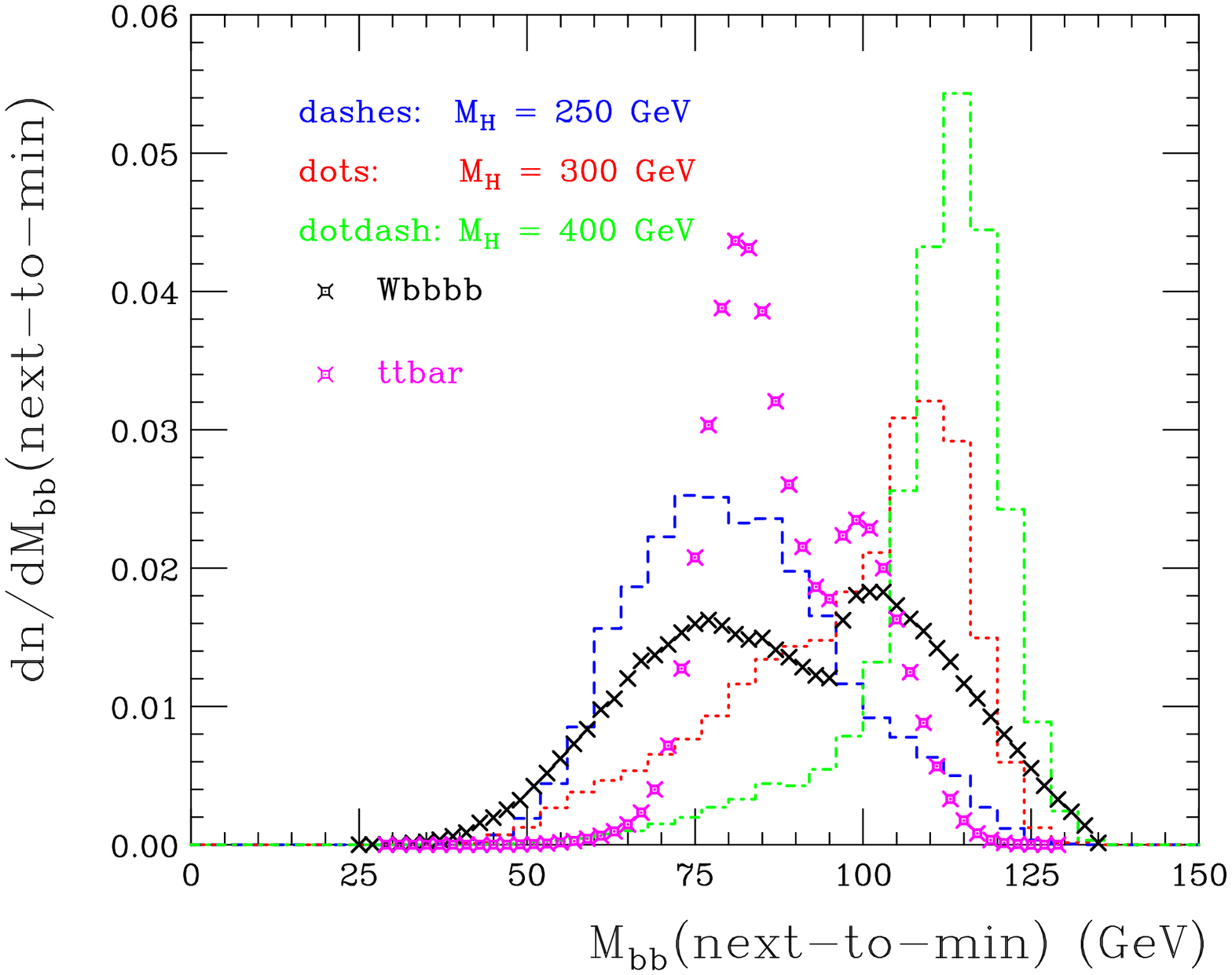, height=8cm, angle=90}
\epsfig{file=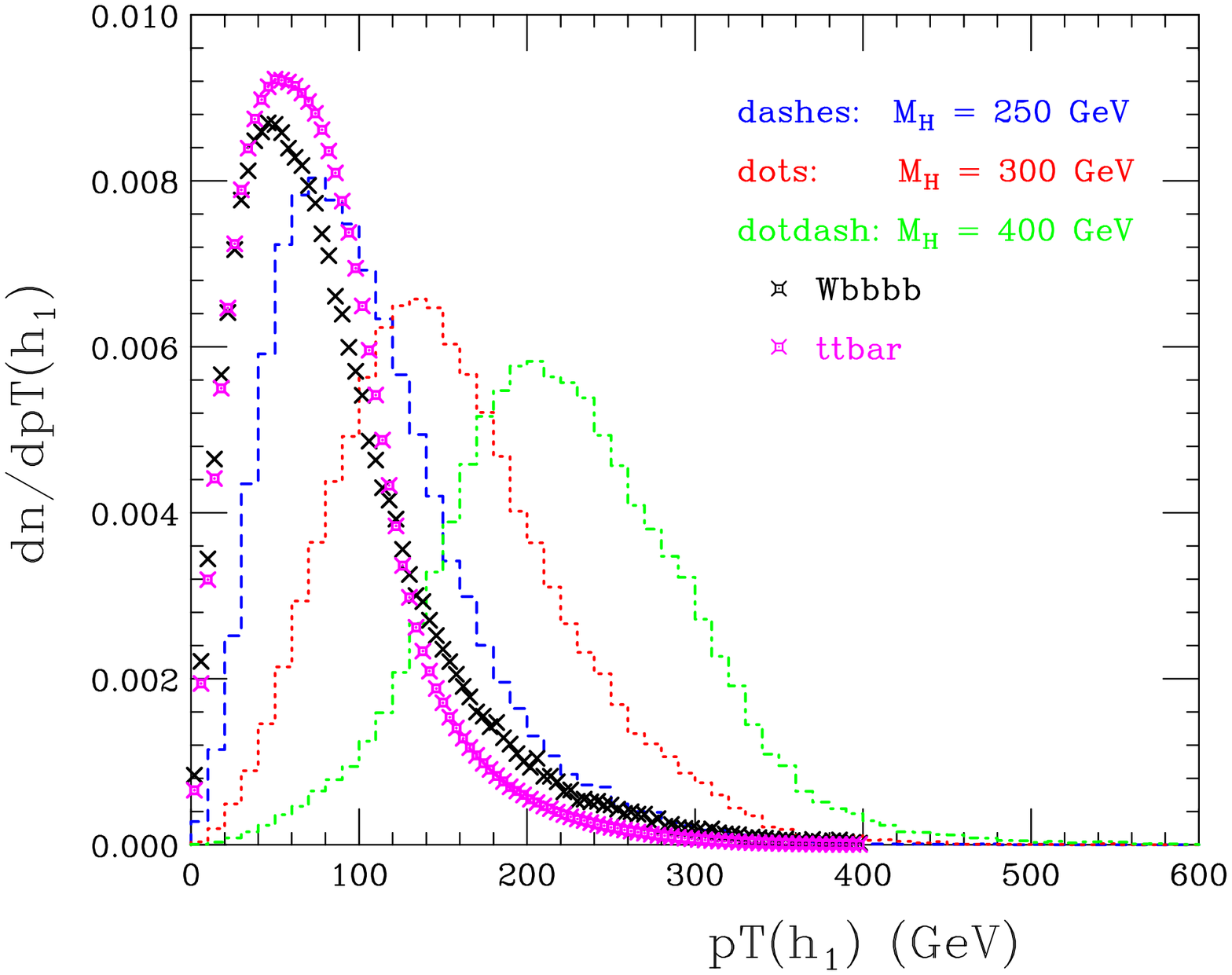, height=8cm, angle=90}
\epsfig{file=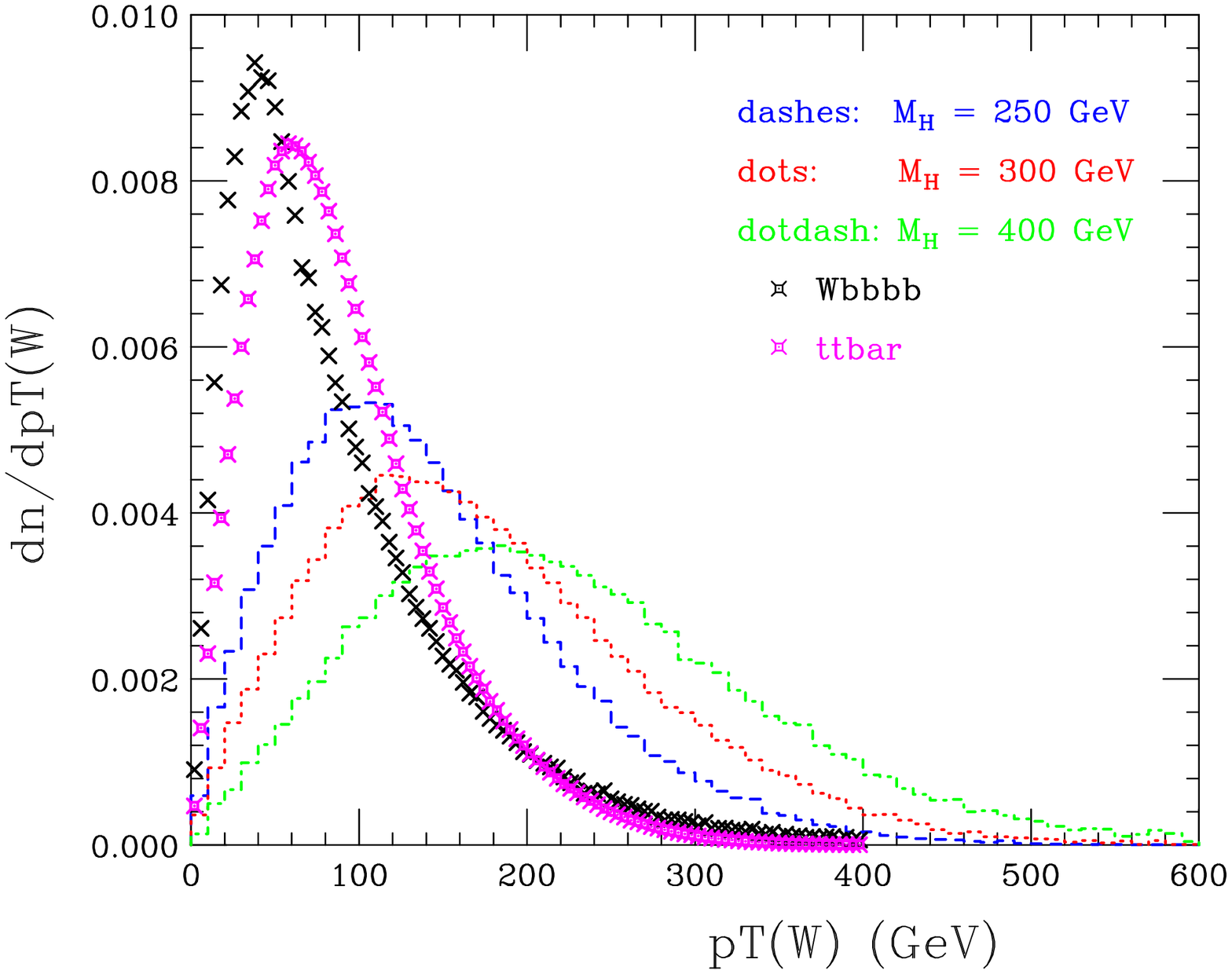, height=8cm, angle=90}

\caption{Differential distributions (normalised to unit area) 
of the  $H \to h h$ contribution to Higgs-strahlung (solid lines) 
after the
primary cuts of eqs.~(\ref{precuts1})--(\ref{precuts3}) in the 2HDM 
(as defined in the text)
followed by $hh\to b\bar bb\bar b$  and $W^\pm\to \ell \nu_\ell$ decays 
in the different mass scenarios indicated. Background
spectra from the $Wbbbb$ and $t\bar{t}$ processes are also given (crosses). 
The distributions are as follows:
the minimum $bb$ invariant mass (top left);
the next-to-minimum $bb$ invariant mass (top right);
the transverse momentum  of the reconstructed Higgs boson containing
the $b$-jet with largest transverse momentum (bottom left);
the transverse momentum of the reconstructed $W^\pm$ (bottom right).
\label{fig:WHkin} }
\end{figure}

\begin{figure}[!ht]
\center
\epsfig{file=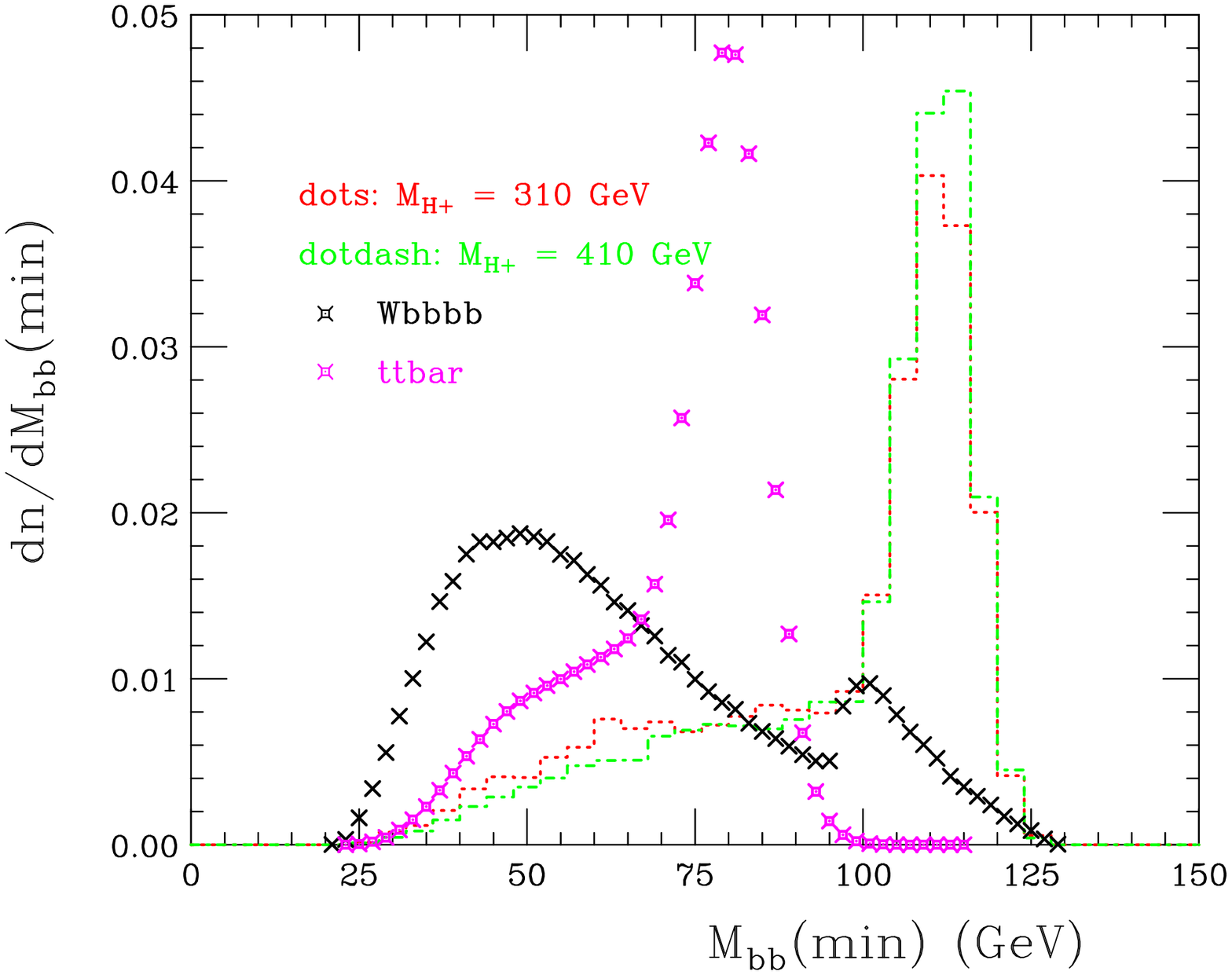, height=8cm, angle=90}
\epsfig{file=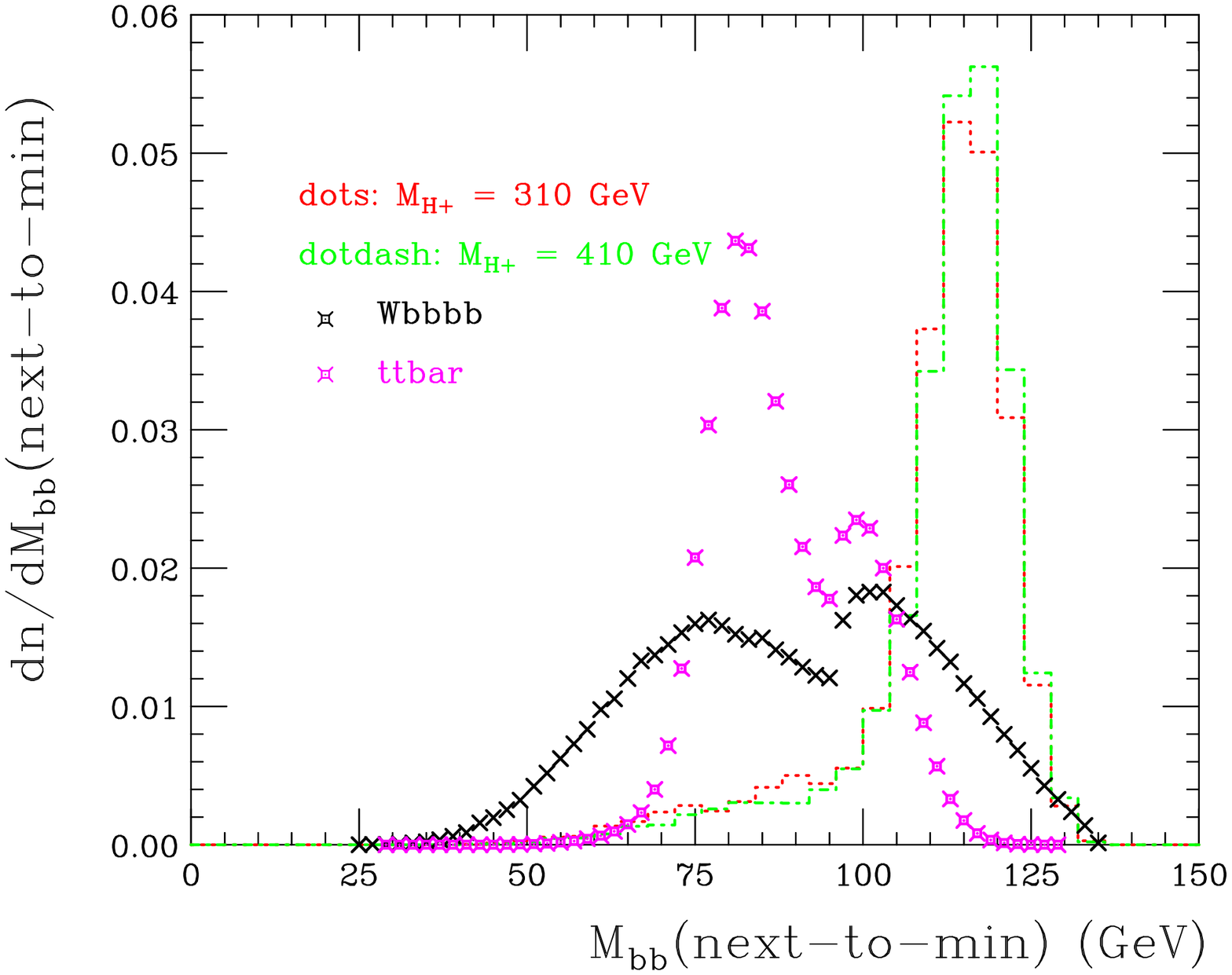, height=8cm, angle=90}
\epsfig{file=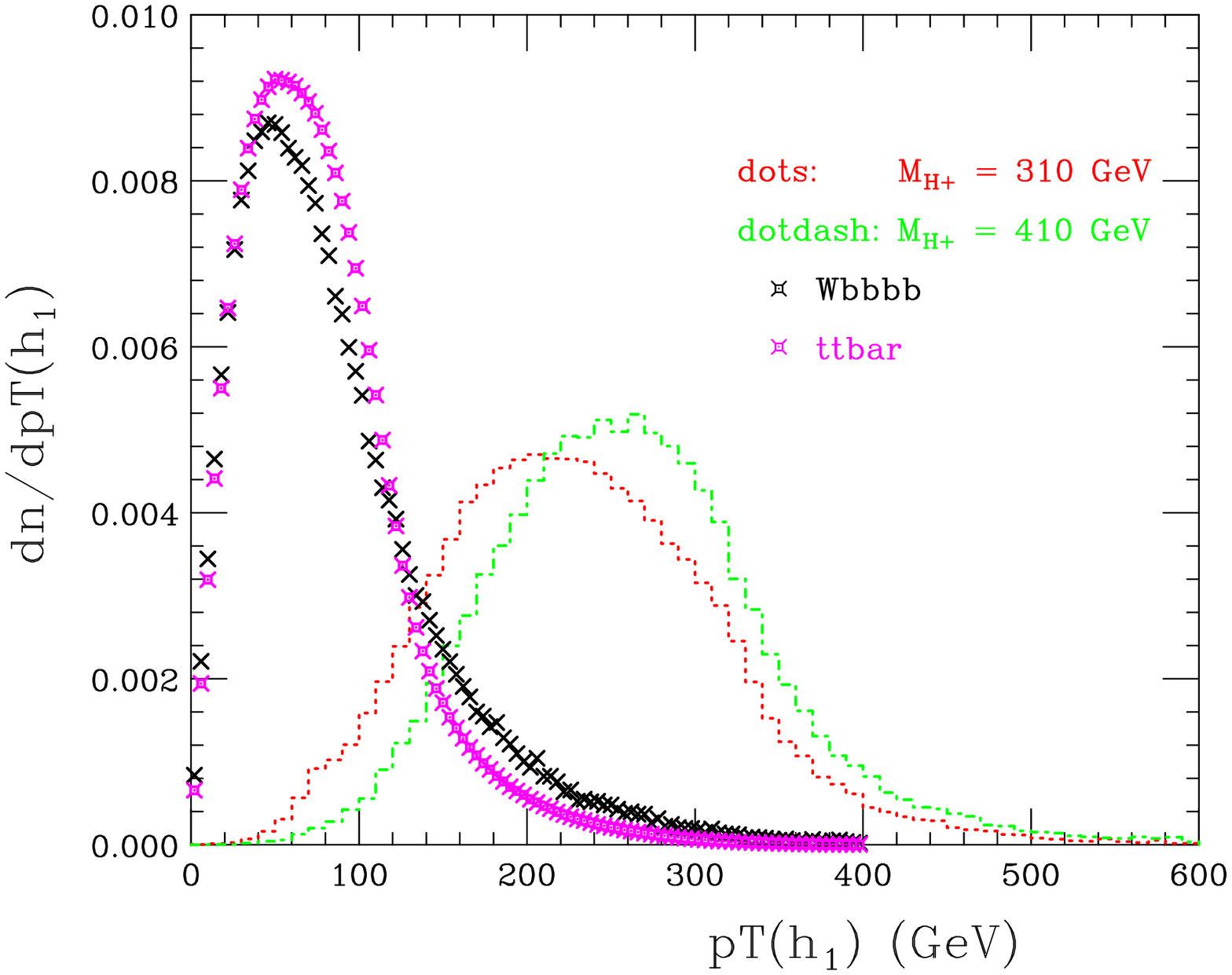, height=8cm, angle=90}
\epsfig{file=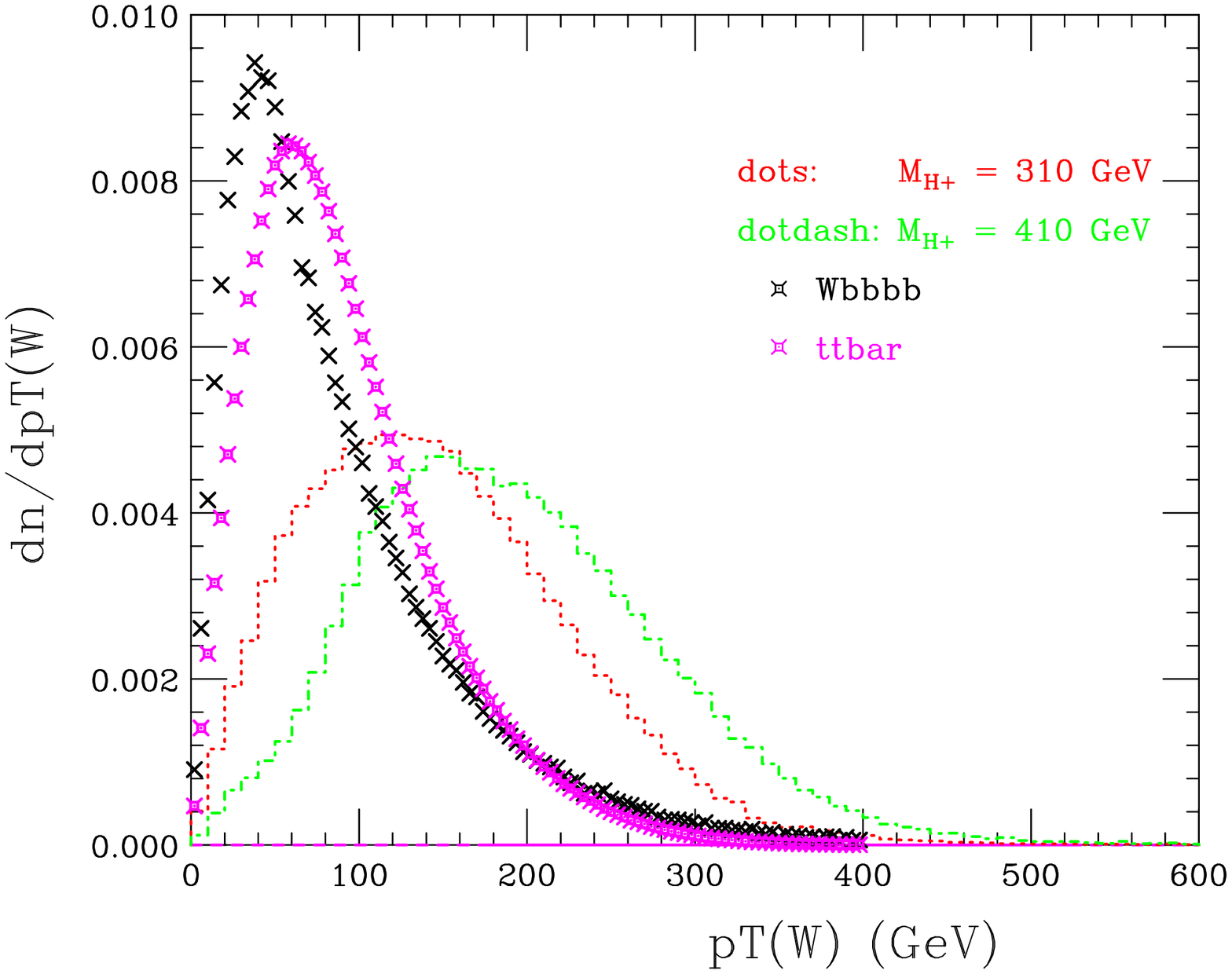, height=8cm, angle=90}
\caption{Differential distributions (normalised to unit area) 
of the  $H^\pm \to h W^\pm$ contribution to Higgs-strahlung  (solid lines) 
after the
primary cuts of eqs.~(\ref{precuts1})--(\ref{precuts3}) in the 2HDM 
(as defined in the text)
followed by $hh\to b\bar bb\bar b$  and $W^\pm\to \ell \nu_\ell$ decays 
in the different mass scenarios indicated. Background
spectra from the $Wbbbb$ and $t\bar{t}$ processes are also given (crosses). 
The distributions are as follows:
the minimum $bb$ invariant mass (top left);
the next-to-minimum $bb$ invariant mass (top right);
the transverse momentum  of the reconstructed Higgs boson containing
the $b$-jet with largest transverse momentum (bottom left);
the transverse momentum of the reconstructed $W^\pm$ (bottom right).
\label{fig:WHpmkin} }
\end{figure}

\begin{figure}[!ht]
\center
\epsfig{file=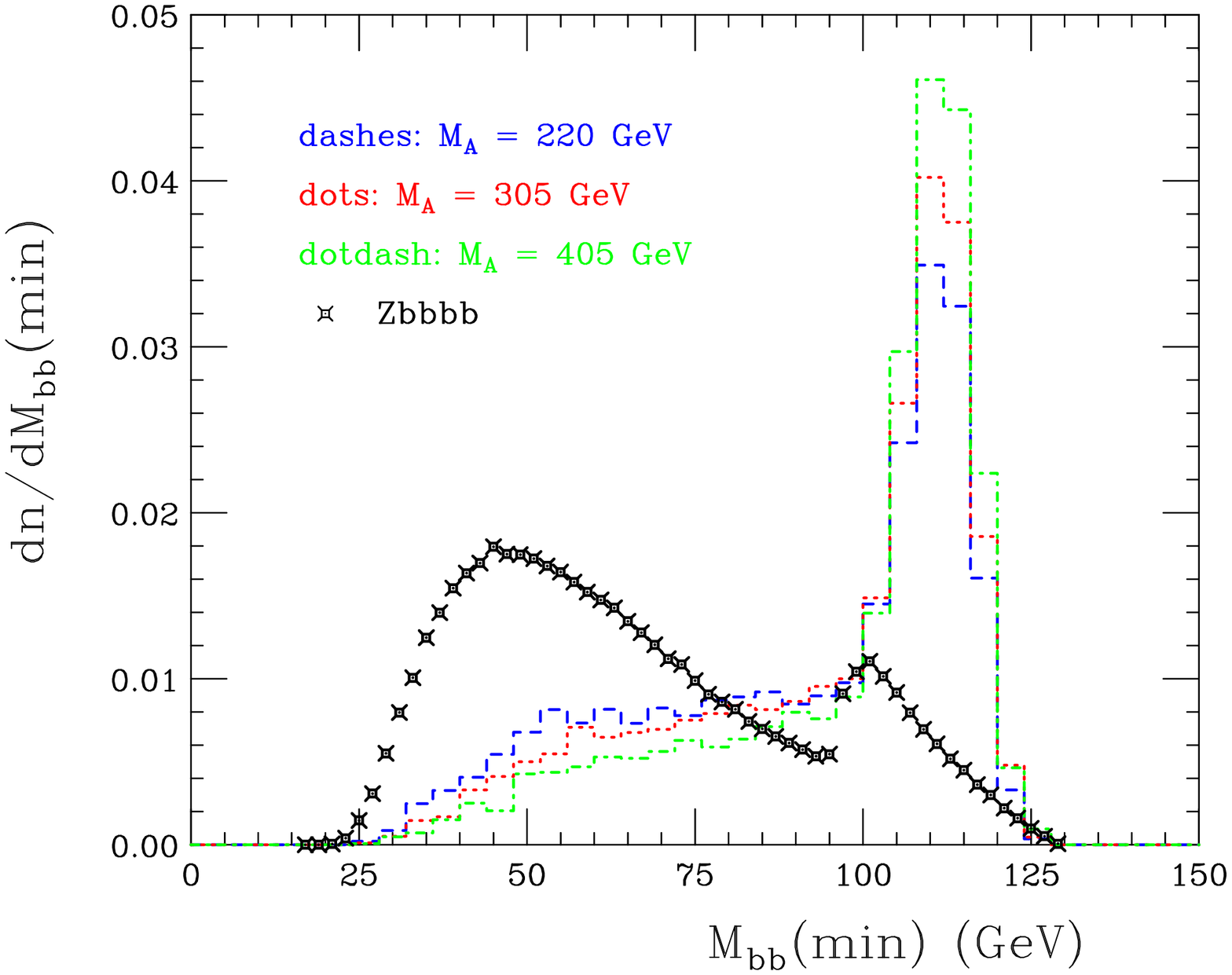, height=8cm, angle=90}
\epsfig{file=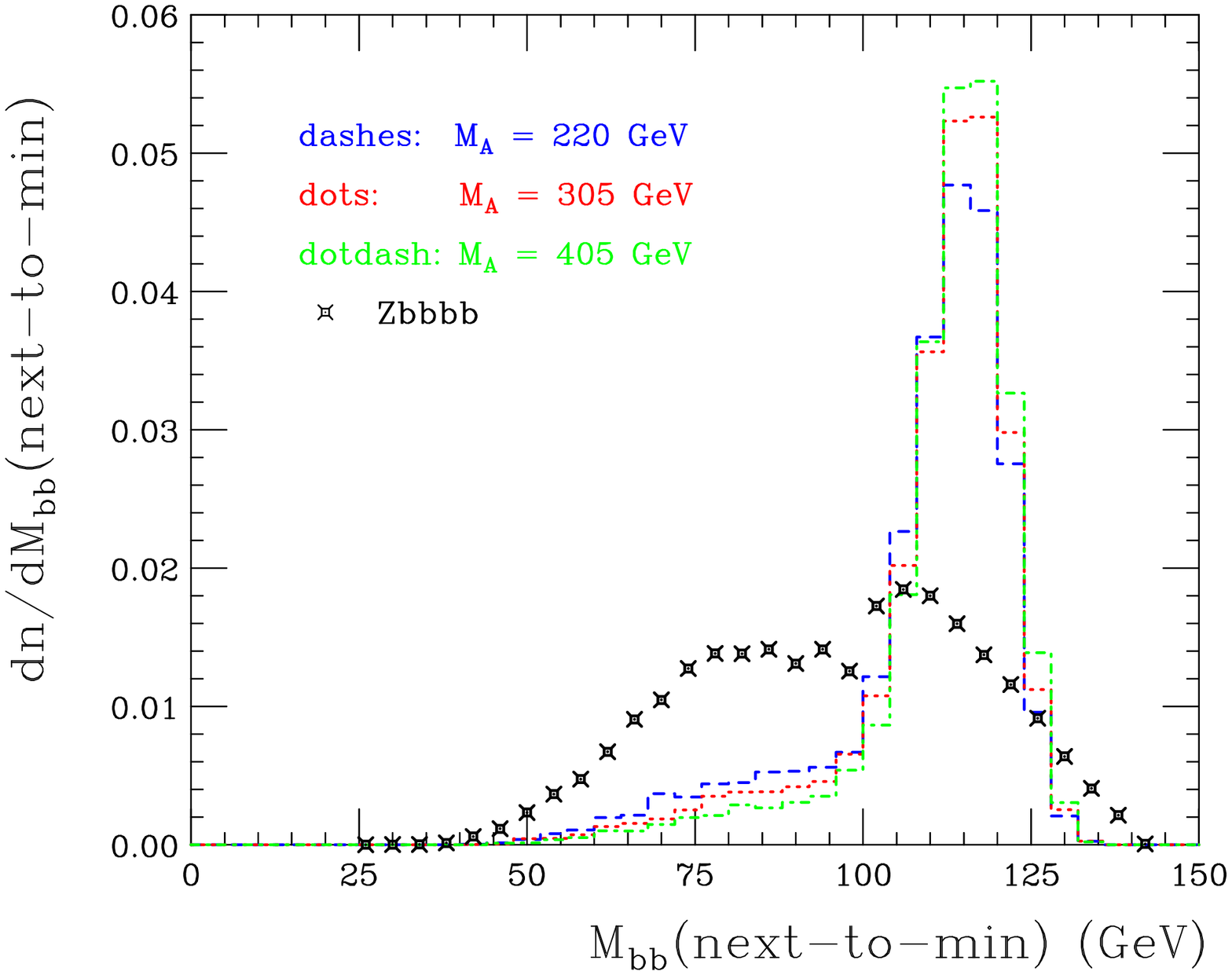, height=8cm, angle=90}
\epsfig{file=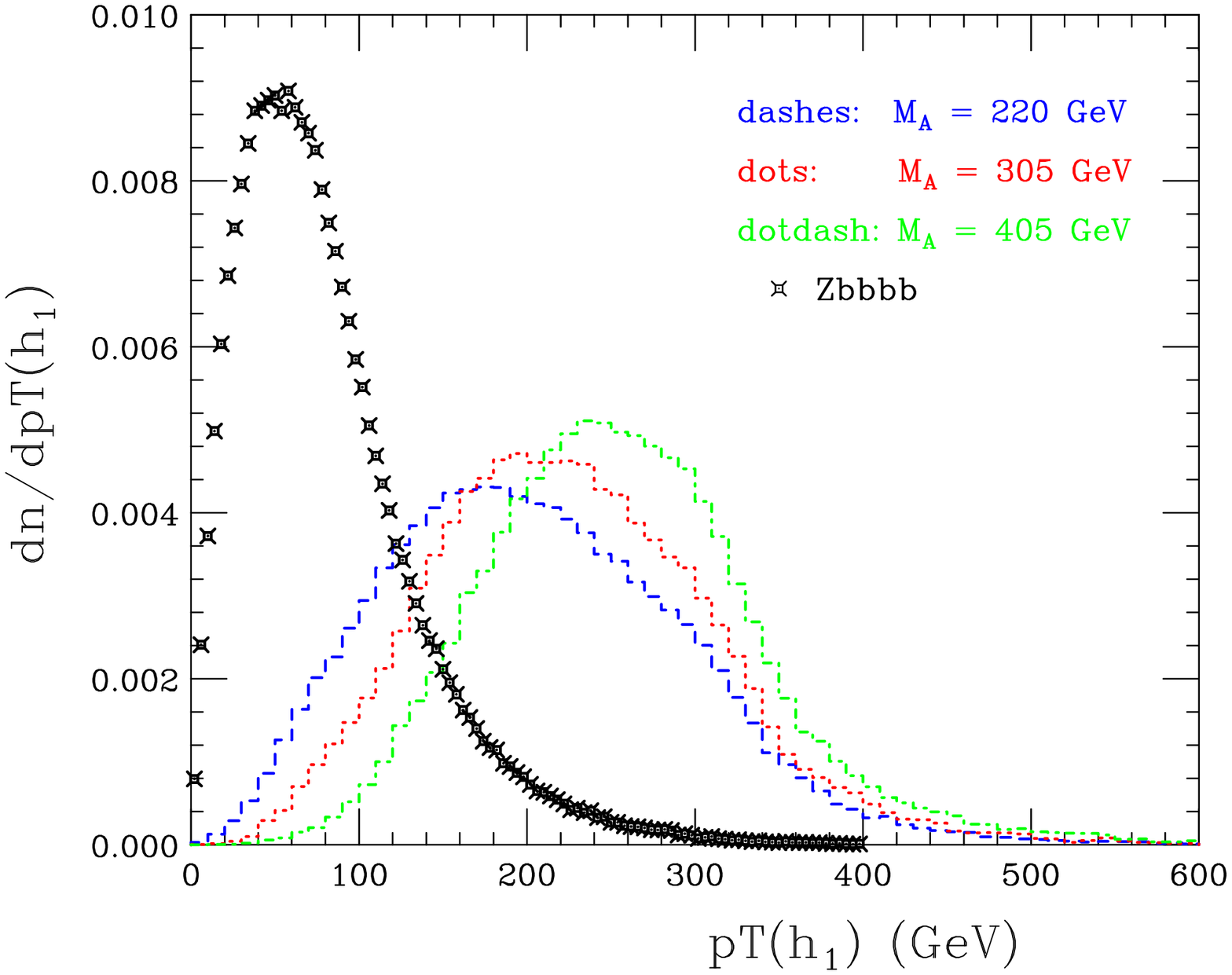, height=8cm, angle=90}
\epsfig{file=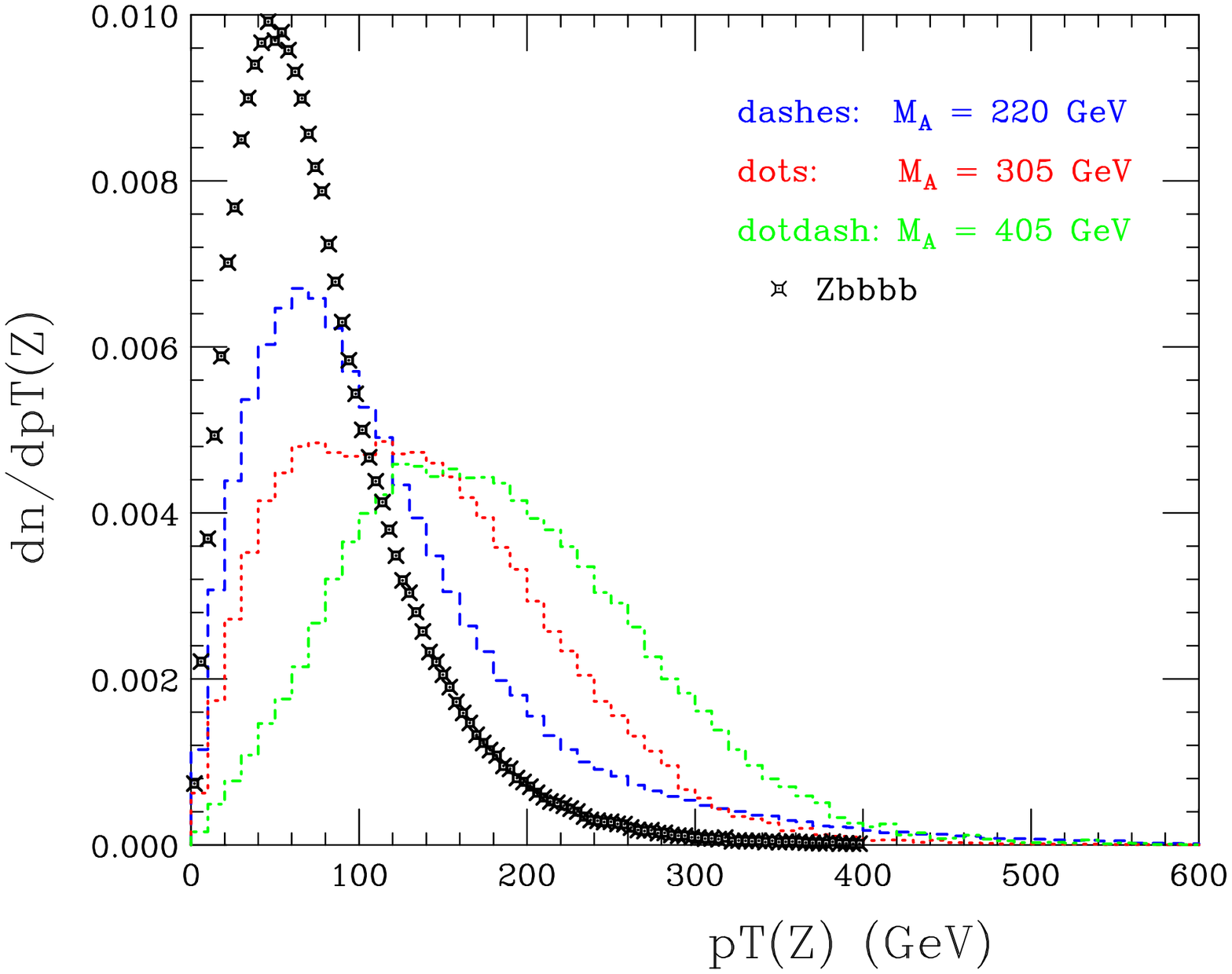, height=8cm, angle=90}
\caption{Differential distributions (normalised to unit area) 
of the  $A \to h Z$ contribution to Higgs-strahlung  (solid lines) 
after the
primary cuts of eqs.~(\ref{precuts1})--(\ref{precuts3}) in the 2HDM 
(as defined in the text)
followed by $hh\to b\bar bb\bar b$  and $Z\to \ell \ell$ decays 
in the different mass scenarios indicated. Background
spectra from the dominant irreducible background $Zbbbb$ are also given (crosses). 
The distributions are as follows:
the minimum $bb$ invariant mass (top left);
the next-to-minimum $bb$ invariant mass (top right);
the transverse momentum  of the reconstructed Higgs boson containing
the $b$-jet with largest transverse momentum (bottom left);
the transverse momentum of the reconstructed $Z$ (bottom right).
\label{fig:ZAkin}}
\end{figure}

\begin{figure}[!ht]
\center
\epsfig{file=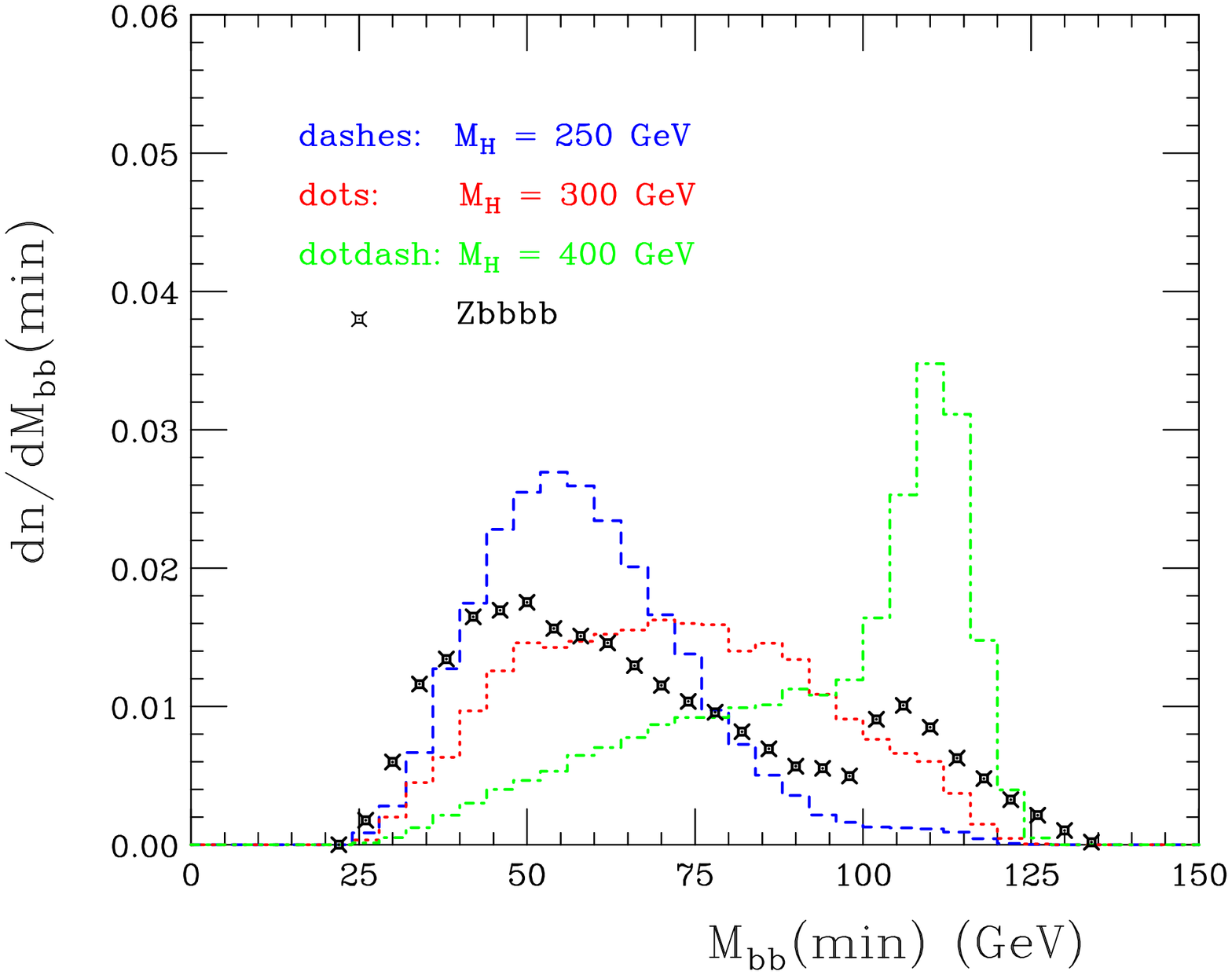, height=8cm, angle=90}
\epsfig{file=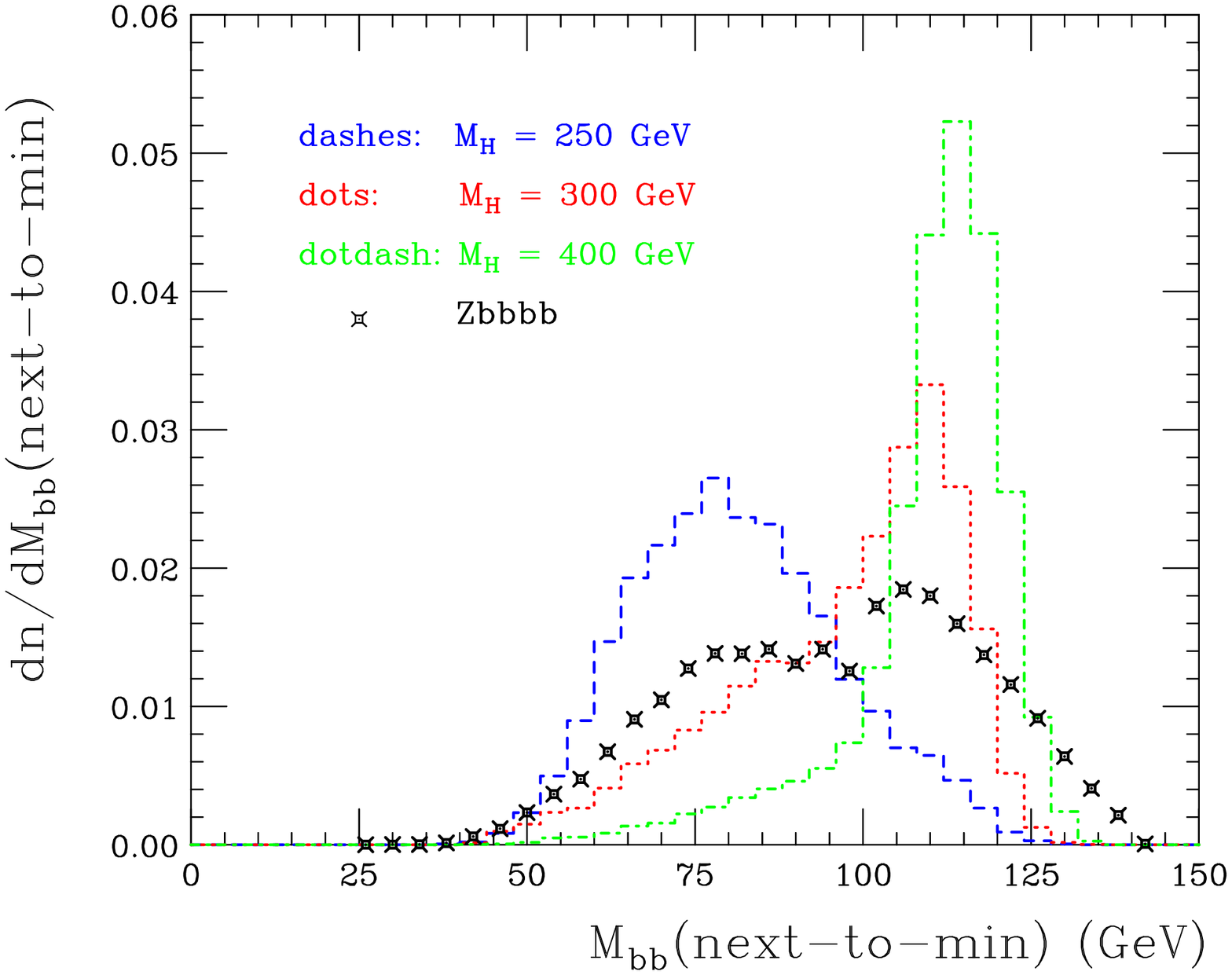, height=8cm, angle=90}
\epsfig{file=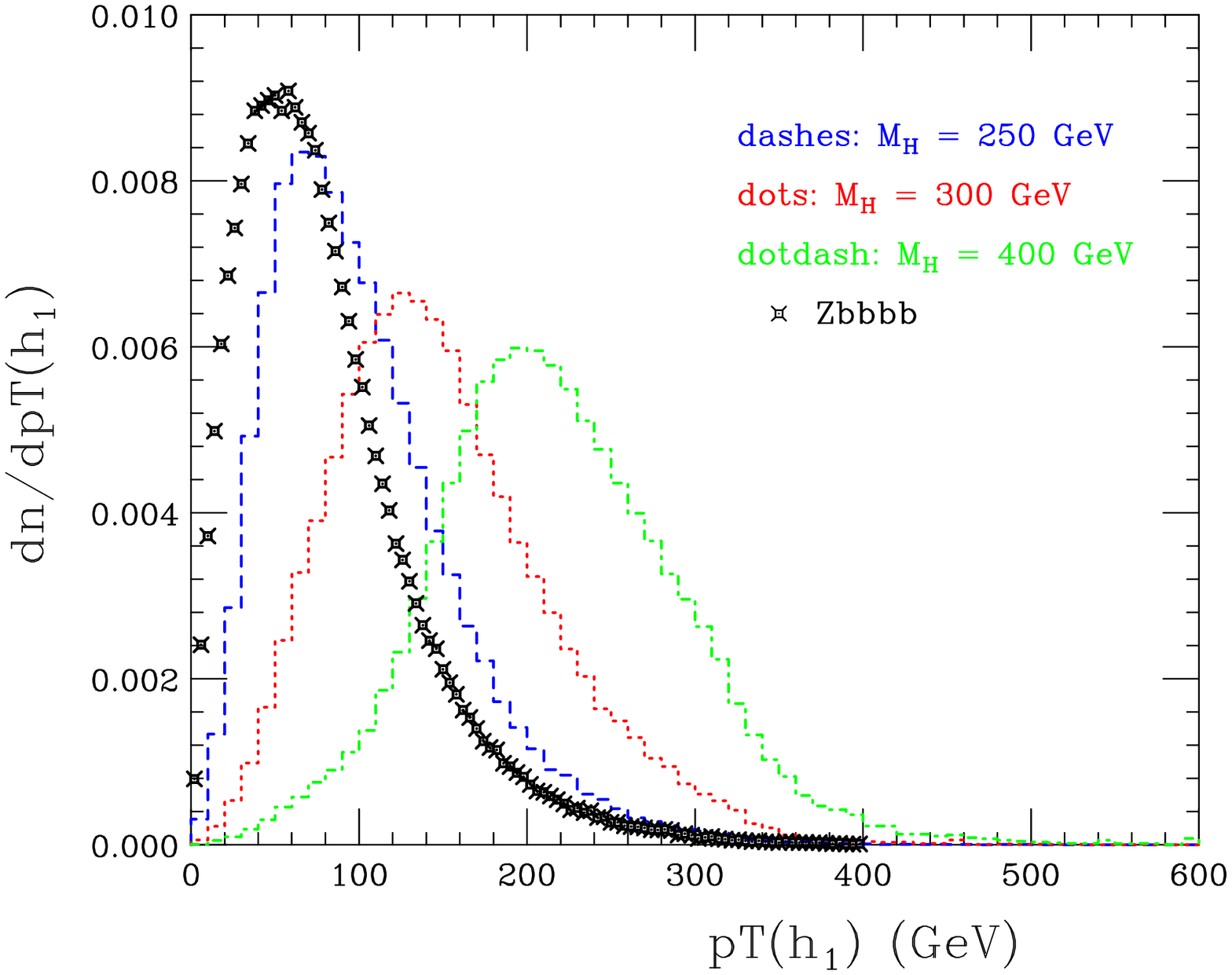, height=8cm, angle=90}
\epsfig{file=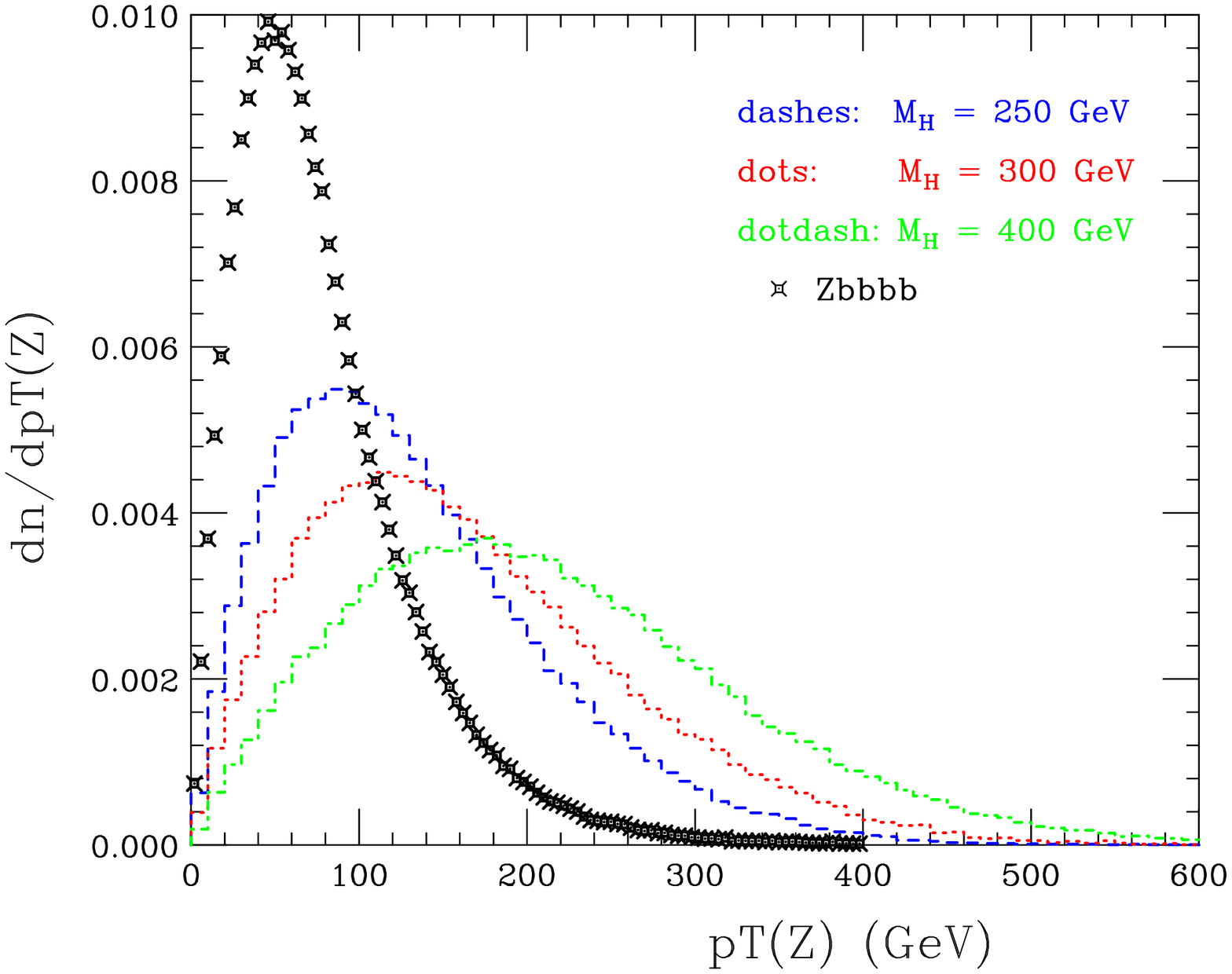, height=8cm, angle=90}
\caption{Differential distributions (normalised to unit area) 
of the  $H \to h h$ contribution to Higgs-strahlung  (solid lines) 
after the
primary cuts of eqs.~(\ref{precuts1})--(\ref{precuts3}) in the 2HDM 
(as defined in the text)
followed by $hh\to b\bar bb\bar b$  and $Z\to \ell \ell$ decays 
in the different mass scenarios indicated. Background
spectra from the dominant irreducible background $Zbbbb$ are also given (crosses).
The distributions are as follows:
the minimum $bb$ invariant mass (top left);
the next-to-minimum $bb$ invariant mass (top right);
the transverse momentum  of the reconstructed Higgs boson containing
the $b$-jet with largest transverse momentum (bottom left);
the transverse momentum of the reconstructed $Z$ (bottom right).
\label{fig:ZHkin}}
\end{figure}

In Figs.~\ref{fig:WHkin}--\ref{fig:ZHkin}
the resulting distributions\footnote{Notice that the normalisation in these figures
is to unit area.} after the primary cuts in 
eqs.~(\ref{precuts1})--(\ref{precuts3}) for the leptonic $W^\pm$ and $Z$ 
decays are given in the 2HDM with heavy Higgs masses in the range from $\approx 200$ GeV to $\approx 400$ GeV, limited to the contributions from the resonances, 
$H \to hh$ (with the $H$ produced in association with a $W^\pm$) in Fig.~\ref{fig:WHkin},
$H^\pm\to hW^\pm$ in Fig.~\ref{fig:WHpmkin},
$A \to Zh$ in Fig.~\ref{fig:ZAkin}  
plus $H \to hh$ (with the $H$ produced in association with a $Z$) in Fig.~\ref{fig:ZHkin}. Note that in  Fig.~\ref{fig:WHpmkin}, as well as in the following Figs.~\ref{fig:scan_optcon_W_2hdm} and \ref{fig:final} and Tab.~\ref{tab:best2HDM}, only the two mass scenarios with $M_{H^\pm}=310$, $410$ GeV are shown since the results for the $M_{H^\pm}=300$ GeV scenario is very similar to the one for  $M_{H^\pm}=310$ GeV.
The
plotted observables are as follows: $M_{bb}^{\min}$ -- the minimal mass of any pair of $b$-jets, 
$M_{bb}^{\rm nmin}$ -- the next-to-minimal mass of any pair of $b$-jets, 
$p_{T}^{h_1}$ -- the transverse momentum of the reconstructed $h$ containing the $b$-jet with 
highest transverse momentum, 
$p_{T}^{W}$($p_{T}^{Z}$) -- the transverse momentum of the reconstructed $W$($Z$) boson. In general 
terms, in all cases, the larger the Higgs masses involved in the decay
the more effective the use of the described observables for the isolation of the signals.

The shape of the plotted distributions, after accounting for correlations amongst the kinematic variables,
leads to the following choice of ``optimised cuts'' in order to enhance the signals of interest compared to 
the backgrounds trying to isolate the couplings $\lambda_{Hhh}$,
$\lambda_{W^\pm H^\mp h}$ and $\lambda_{ZAh}$

\begin{itemize}
\item $\lambda_{Hhh}$ from $q\bar q^{'} \to HW$ with $H \to hh$ 
\begin{equation}\label{optcuts1}
M_{bb}^{\min}>100~{\rm{GeV}},\qquad 
M_{bb}^{\rm nmin}>100~{\rm{GeV}},\qquad 
p_{T}^{h_1}>0~{\rm{GeV}},\qquad 
p_{T}^{W}>100~{\rm{GeV}} 
\end{equation}
\item $\lambda_{W^\pm H^\mp h}$ from $q\bar q^{'} \to H^\pm h$ with $H^\pm\to hW^\pm$ 
\begin{equation}\label{optcuts2}
M_{bb}^{\min}>100~{\rm{GeV}},\qquad 
M_{bb}^{\rm nmin}>100~{\rm{GeV}},\qquad 
p_{T}^{h_1}>100~{\rm{GeV}},\qquad 
p_{T}^{W}>0~{\rm{GeV}} 
\end{equation}

\item $\lambda_{ZAh}$ from $q\bar q \to A h$ with $A\to hZ$ 
\begin{equation}\label{optcuts3}
M_{bb}^{\min}>70~{\rm{GeV}},\qquad 
M_{bb}^{\rm nmin}>100~{\rm{GeV}},\qquad 
p_{T}^{h_1}>100~{\rm{GeV}},\qquad 
p_{T}^{Z}>0~{\rm{GeV}} 
\end{equation}

\item $\lambda_{Hhh}$ from $q\bar q \to HZ$ with $H \to hh$ 
\begin{equation}\label{optcuts4}
M_{bb}^{\min}>40~{\rm{GeV}},\qquad 
M_{bb}^{\rm nmin}>70~{\rm{GeV}},\qquad 
p_{T}^{h_1}>0~{\rm{GeV}},\qquad 
p_{T}^{Z}>100~{\rm{GeV}} 
\end{equation}
\end{itemize}

The resulting cross-sections for the relevant backgrounds after applying these optimised cuts 
are given in Tab.~\ref{tab:sigma2HDM}. After taking into account the $b$-tagging ($\epsilon_b=0.5$) 
and mis-tagging ($R=0.01$) factors, this means that only the $Vbbbb$ and $Vbbjj$ backgrounds are important, 
although in our numerical analysis we have kept all the discussed backgrounds.

\begin{table}[t]
\begin{center}
\begin{tabular}{|c|c|c|c|c|}
\hline
cut-comb & $\sigma(pp \to Wb\bar bb\bar b)$ & $\sigma(pp \to Wb\bar bjj)$& $\sigma(pp \to Wjjjj)$ & $\sigma(pp \to t\bar t \to Wb\bar bjj)$ \\
 & (fb) & (fb) &  (fb) &  (fb)\\
\hline
(\ref{optcuts1})   & $9.2 \cdot 10^{-4}$   & $1.4$   & $82$   & $0.32$    \\
(\ref{optcuts2})   & $1.8 \cdot 10^{-3}$   & $2.9$   & $1.6 \cdot 10^{2}$   & $0.59$    \\
\hline
\hline
cut-comb & $\sigma(pp \to Zb\bar bb\bar b)$ & $\sigma(pp \to Zb\bar bjj)$& $\sigma(pp \to Zjjjj)$ &   \\
 & (fb) & (fb) &  (fb) & \\
\hline
(\ref{optcuts3})   & $3.9 \cdot 10^{-4}$   & $1.1$   & $19$   &      \\
(\ref{optcuts4})   & $4.0 \cdot 10^{-4}$   & $1.2 $   & $20 $   &      \\
\hline
\end{tabular}
\end{center}
\caption{Cross-sections  for the most important background processes after the optimised cuts defined in
 (\ref{optcuts1})--(\ref{optcuts4}), respectively, have been applied in addition to the acceptance and preselection cuts defined
in (\ref{precuts1})--(\ref{precuts3}). Note that $b$-tagging ($\epsilon_b=0.5$) and mis-tagging ($R=0.01$) factors have not been applied.
}
\label{tab:sigma2HDM}
\end{table}

In the next step of our analysis, we have scanned again the 2HDM parameter space in our three mass scenarios to 
search for points where, in presence of the above optimised cuts, one
can establish a resonant signal at the $5\sigma$ level. Since the number of expected background events ($B$) 
is very small we have used Poisson statistics in way similar to the one suggested in \cite{Bityukov} and 
calculated the number of signal events $S$ that gives a probability for detecting $S+B$ or more events, 
given $B$ expected ones, to be less than 
$2.8 \cdot 10^{-7}$ after $\int{\cal L}dt=3000$ fb$^{-1}$ of integrated luminosity, thus corresponding to 
a 5$\sigma$ significance. The required signal cross-sections obtained in this way are given in 
Tab.~\ref{tab:significance2HDM}.

The results of the scans are given in Figs.~\ref{fig:scan_optcon_W_2hdm} and \ref{fig:scan_optcon_Z_2hdm}.
From looking at the latter, 
there is obviously scope 
in finding a 5$\sigma$ signal in almost all cases, for both the charged and neutral current 
version of process (\ref{proc})\footnote{The only notable exception is the $H^\pm\to hW^\pm$ resonance
for $M_H=250$ GeV, $M_h=115$ GeV, $M_A=220$ GeV and $M_{H^\pm}=300$ GeV. This is due to the cut 
$M_{bb}^{\min}>100~{\rm{GeV}}$,
which is necessary against the $t\bar t$ background but at the same time kills the signal in this case.}, as shown in
Figs.~\ref{fig:scan_optcon_W_2hdm} and \ref{fig:scan_optcon_Z_2hdm}, respectively.

\begin{table}[t]
\begin{center}
\begin{tabular}{|c|c|c|c|}
\hline
cut-comb & $B$ & $S$ & $\sigma(q\bar q^{'} \to Whh \to \ell \nu_\ell b \bar b b \bar b )$ (fb) \\
\hline
(\ref{optcuts1})   &  0.30  & 6.1   & 0.033 \\
(\ref{optcuts2})   &  0.59  & 7.3   & 0.041 \\
\hline
\hline
cut-comb & $B$ & $S$ & $\sigma(q\bar q \to Zhh \to \ell \ell b \bar b b \bar b)$ (fb) \\
\hline
(\ref{optcuts3})   &  0.16  & 5.1   & 0.027 \\
(\ref{optcuts4})   &  0.16  & 5.1   & 0.027 \\
\hline
\end{tabular}
\end{center}
\caption{The number of background $(B)$ events expected after $\int{\cal L}dt=3000$ fb$^{-1}$ of integrated luminosity when applying the optimised cuts
defined in (\ref{optcuts1})--(\ref{optcuts4})
(in addition to the acceptance and preselection cuts) and the number of signal ($S$) events that give a probability less than $2.8 \cdot 10^{-7}$ to observe $S+B$ events when expecting $B$ events using Poisson statistics. In addition the corresponding signal cross-sections are given (without any $b$-tagging efficiency).
}
\label{tab:significance2HDM}
\end{table}

As it is clear from the 
plots the cross-sections with a $H^\pm \to W^\pm h$ or a $A \to Z h$
resonance are typically dominating the total cross-sections after the optimised cuts 
whereas the $H \to hh$ resonance is typically not dominant even after applying them. The reason for 
this is that the width of the $A$ and $H^\pm$ are
typically small in the 2HDM scenarios we are considering and thus these resonances always
contribute significantly to the total cross-section. In contrast,
 the  $H \to hh$ is typically wide and it cannot be small at the same time as the $A$ or $H^\pm$ are wide.

Finally, Fig.~\ref{fig:final} shows the phenomenal level of purity of the resonant signal samples that
can be achieved, after the optimised cuts are implemented, for the ``best case scenarios'' in the 2HDM
 as given in Tab.~\ref{tab:best2HDM}. Note that here the signals have been calculated from the sum of all 
diagrams in (\ref{eqn:sigma_sum}) and thus all interferences are accounted for whereas the scenarios have 
been chosen to give the highest resonant cross-sections (the short-dashed lines in 
Figs.~\ref{fig:scan_optcon_W_2hdm} and \ref{fig:scan_optcon_Z_2hdm}).
The high purity is exemplified by the invariant masses that would enable the
reconstruction of the decaying Higgs state, i.e., the four $b$-jet invariant mass (for the case of $H\to hh$)
and the invariant mass of the Higgs $h$ not containing the highest transverse momentum $b$-jet with the 
lepton-lepton
(for the case $A\to hZ$) 
and lepton-neutrino   
(for the case $H^\pm\to hW^\pm$) pair. (Here, the $t\bar t$ background is not shown, as it has become totally
negligible.) 
As a final note we have investigated how far the different ``best case
scenarios''  in Table \ref{tab:best2HDM} are from the strong coupling
limit. By applying the coupled renormalisation group evolution equations
for the $\lambda$-couplings in Eq.~(\ref{eq:potential}) as well as the
$t$, $b$, $\tau$ Yukawa couplings and the gauge
couplings~\cite{Ferreira:2009jb}, we can determine how far a given point
in parameter space can be evolved in renormalisation scale before it
reaches the tree-level unitarity limit~\footnote{We would like thank
Oscar St{\aa}l for providing us with a code for solving the RGEs.}. This
in turn gives an indication at what scale one can expect to see
additional states and/or interactions (a.k.a. UV-completion). Applying
this procedure we find that for most of the ``best case scenarios'', the
tree-level unitarity limit is reached at a scale which is a factor 3-4
larger than the input scale. The only two exceptions are the two
scenarios with $\sin(\beta-\alpha)=-0.01$ in Table \ref{tab:best2HDM}
where this limit is reached only after evolving to a scale which is a
factor $15-30$ times larger than the input scale.    
\begin{table}[t]
\begin{center}
\begin{tabular}{|c|c|c|c|c|c|}
\hline
cut-comb (signal) & $\sin(\beta-\alpha)$ & $\lambda_5$ & $\tan\beta$  & $\lambda_{Hhh}$  & $\lambda_{hhh}$ \\
\hline
\hline
\multicolumn{6}{|c|}{$M_{H}=250$ GeV, $M_{A}=220$ GeV, $M_{H^\pm}=300$ GeV} \\ 
\hline
(\ref{optcuts1}) ($q\bar q^{'} \to HW \to hhW$)     & -0.82  & -4.4   & 1.1   & 1259 & 1749  \\  
(\ref{optcuts2}) ($q\bar q^{'} \to H^\pm h \to hhW$)&        &        &       &      &       \\ 
(\ref{optcuts3}) ($q\bar q^{} \to Ah \to hhZ$)      & -0.72  & -3.6   & 1.1   & 702  &  1805 \\ 
(\ref{optcuts4}) ($q\bar q^{} \to HZ \to hhZ$)      &  0.73  & -0.1   & 17.6  & 800  & -1862 \\  
\hline
\multicolumn{6}{|c|}{$M_{H}=300$ GeV, $M_{A}=305$ GeV, $M_{H^\pm}=310$ GeV} \\ 
\hline
(\ref{optcuts1}) ($q\bar q^{'} \to HW \to hhW$)     & -0.26  & -2.9   & 0.9   & -988  & 387  \\  
(\ref{optcuts2}) ($q\bar q^{'} \to H^\pm h \to hhW$)& -0.01  & -1.3   & 7.5   & -335  & -873 \\ 
(\ref{optcuts3}) ($q\bar q^{} \to Ah \to hhZ$)      & -0.01  & -1.4   & 2.9  & -397  &  161 \\ 
(\ref{optcuts4}) ($q\bar q^{} \to HZ \to hhZ$)      &  0.09  & -1.3   & 24  & -992  & -293 \\  
\hline
\multicolumn{6}{|c|}{$M_{H}=400$ GeV, $M_{A}=405$ GeV, $M_{H^\pm}=410$ GeV} \\ 
\hline
(\ref{optcuts1}) ($q\bar q^{'} \to HW \to hhW$)     &  0.73   & -0.7  & 5.2   & 867  & -1328 \\  
(\ref{optcuts2}) ($q\bar q^{'} \to H^\pm h \to hhW$)&  0.0034 & -2.5  & 7.5   & -671 &  68   \\ 
(\ref{optcuts3}) ($q\bar q^{} \to Ah \to hhZ$)      & -0.0012 & -2.5  & 6.4   & -651 &  58   \\ 
(\ref{optcuts4}) ($q\bar q^{} \to HZ \to hhZ$)      &  0.73   & -0.7  & 5.2   & 867  & -1328 \\  
\hline
\end{tabular}
\end{center}
\caption{The respective best case scenarios used for illustration in Fig.~\ref{fig:final}, after the implementation of the optimised cuts defined in (\ref{optcuts1})--(\ref{optcuts4}). Note that there is no entry for the $H^\pm$ resonant scenario for $M_{H^\pm}=300$ GeV since the results are very similar to the ones for 
$M_{H^\pm}=310$ GeV.
}
\label{tab:best2HDM}
\end{table}

\section{Conclusions}
\label{sec:summary}

In summary, we have shown the potential of a high luminosity LHC in extracting (resonant) decay signals of heavy Higgs states,
both charged ($H^\pm$) and neutral ($H$ and $A$), in a generic CP-conserving Type II 2HDM, using leptonic decays of
a gauge boson and $b$-quarks ones of the lightest Higgs state of the model ($h$), the latter produced in the Higgs-strahlung process
$q\bar q^{(')}\to Vhh$, with $V=W^\pm$ or $Z$ and $q^{(')}$ referring to any possible (anti)quark flavour. The expanse
of   parameter space of the enlarged Higgs model that can be covered is sizable, not fine-tuned and yields detectable and extremely
pure signals. This opens the prospect of accessing fundamental triple (Higgs and gauge) couplings of the model, notably
$\lambda_{Hhh}$ (but not $\lambda_{hhh}$) plus $\lambda_{W^\pm H^\mp h}$ and $\lambda_{Z A h}$, which would enable one to distinguish
between this Higgs scenario and alternative ones (such as the MSSM) even in presence of degenerate mass and coupling spectra between
the two. The possibility of extracting the latter and the precision at which this can be achieved, depending on the parameter
space regions considered, ought to be assessed in a full
detector environment, in presence of all necessary parton shower and hadronisation effects. However, the level of sophistication
of our study and the encouraging results obtained hint at the feasibility of these analyses. To confirm this, we 
make available
upon request all our codes.

\section*{Acknowledgments} 

SM thanks the Royal Society (London, UK) for partial financial support
during one of his visits to Uppsala and so does JR for a visit to Southampton.
RP acknowledges the financial support of MICINN under contract
FPA2008-02984 and of the RTN European Programme
MRTN-CT-2006-035505 (HEPTOOLS, Tools and Precision Calculations
for Physics Discoveries at Colliders). SM acknowledges the latter too
for partial funding. FP and MM thank the CERN Theory Unit for partial support. 
SM is financially supported in part by 
the scheme `Visiting Professor - Azione D - Atto Integrativo tra la 
Regione Piemonte e gli Atenei Piemontesi'.
RP, FP and MM acknowledge the financial support of the bilateral 
INFN/MICINN program ACI2009-1045 (Aspects of Higgs physics at the LHC). 

\def\pr#1 #2 #3 { {\rm Phys. Rev.} {\bf #1} (#2) #3}
\def\prd#1 #2 #3{ {\rm Phys. Rev. D} {\bf #1} (#2) #3}
\def\prl#1 #2 #3{ {\rm Phys. Rev. Lett.} {\bf #1} (#2) #3}
\def\plb#1 #2 #3{ {\rm Phys. Lett. B} {\bf #1} (#2) #3}
\def\npb#1 #2 #3{ {\rm Nucl. Phys. B} {\bf #1} (#2) #3}
\def\prp#1 #2 #3{ {\rm Phys. Rep.} {\bf #1} (#2) #3}
\def\zpc#1 #2 #3{ {\rm Z. Phys. C} {\bf #1} (#2) #3}
\def\epjc#1 #2 #3{ {\rm Eur. Phys. J. C} {\bf #1} (#2) #3}
\def\mpl#1 #2 #3{ {\rm Mod. Phys. Lett. A} {\bf #1} (#2) #3}
\def\ijmp#1 #2 #3{{\rm Int. J. Mod. Phys. A} {\bf #1} (#2) #3}
\def\ptp#1 #2 #3{ {\rm Prog. Theor. Phys.} {\bf #1} (#2) #3}
\def\jhep#1 #2 #3{ {\rm JHEP} {\bf #1} (#2) #3}
\def\jphg#1 #2 #3{ {\rm J. Phys. G} {\bf #1} (#2) #3}
\def\cpc#1 #2 #3{ {\rm Comp. Phys. Comm.} {\bf #1} (#2) #3} 

\clearpage

\clearpage

\clearpage

\clearpage

\clearpage\thispagestyle{empty}

\clearpage\thispagestyle{empty}

\clearpage\thispagestyle{empty}

\clearpage\thispagestyle{empty}

\clearpage\thispagestyle{empty}

\clearpage\thispagestyle{empty}\begin{figure}[!ht]
\center
\epsfig{file=scan_optcon_W85_2hdm_250.eps, height=6cm}
\epsfig{file=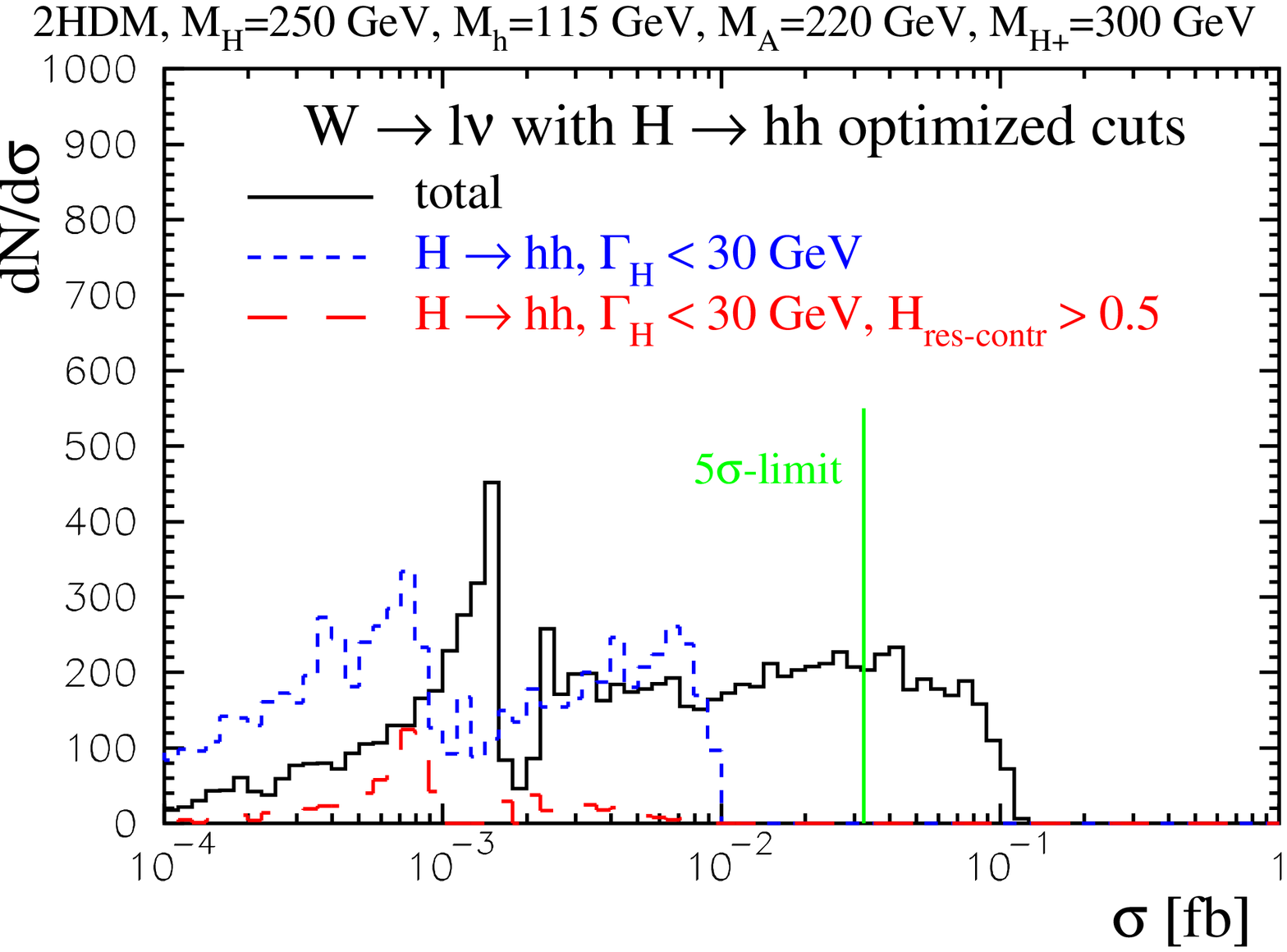, height=6cm}
\epsfig{file=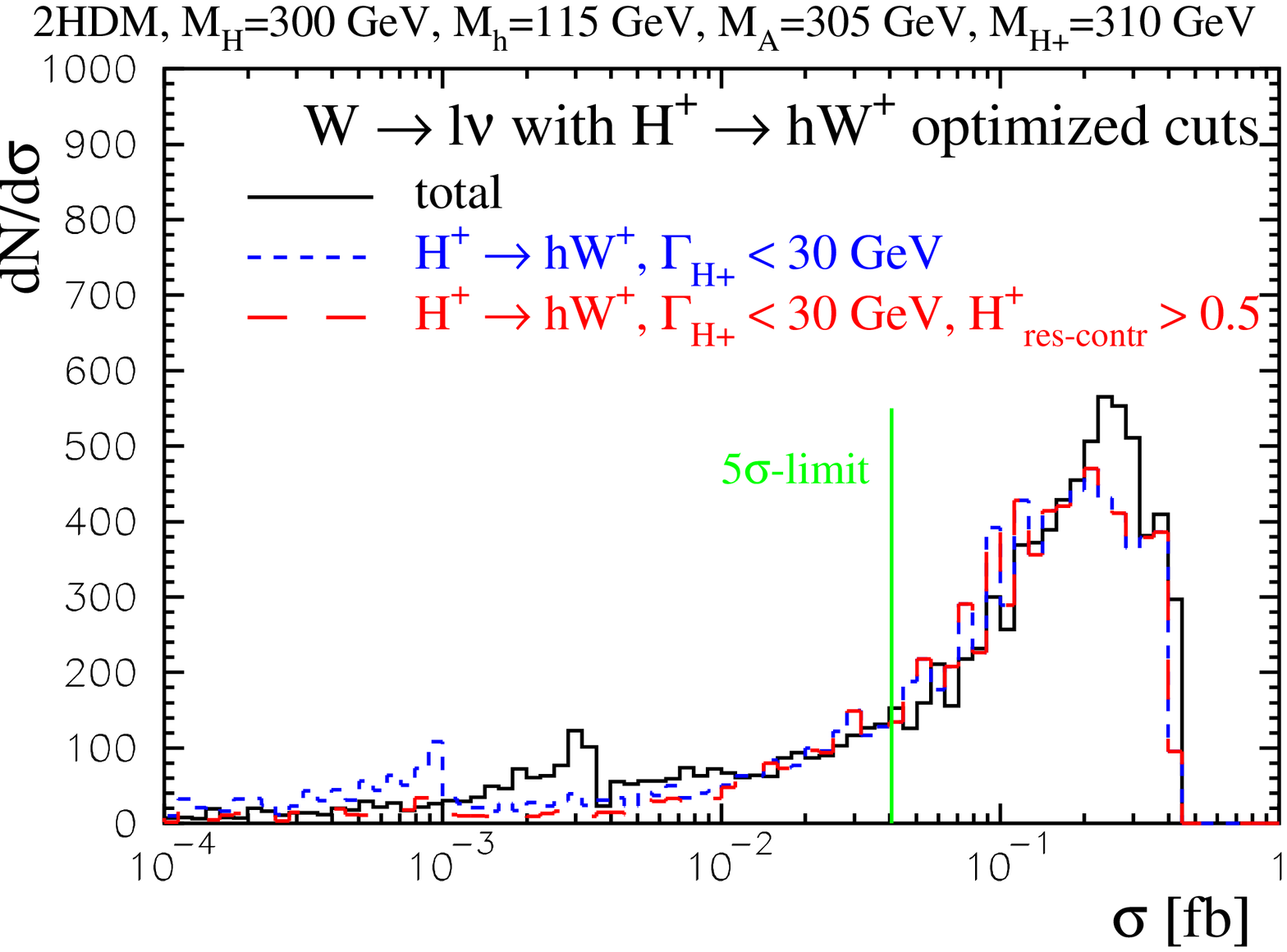, height=6cm}
\epsfig{file=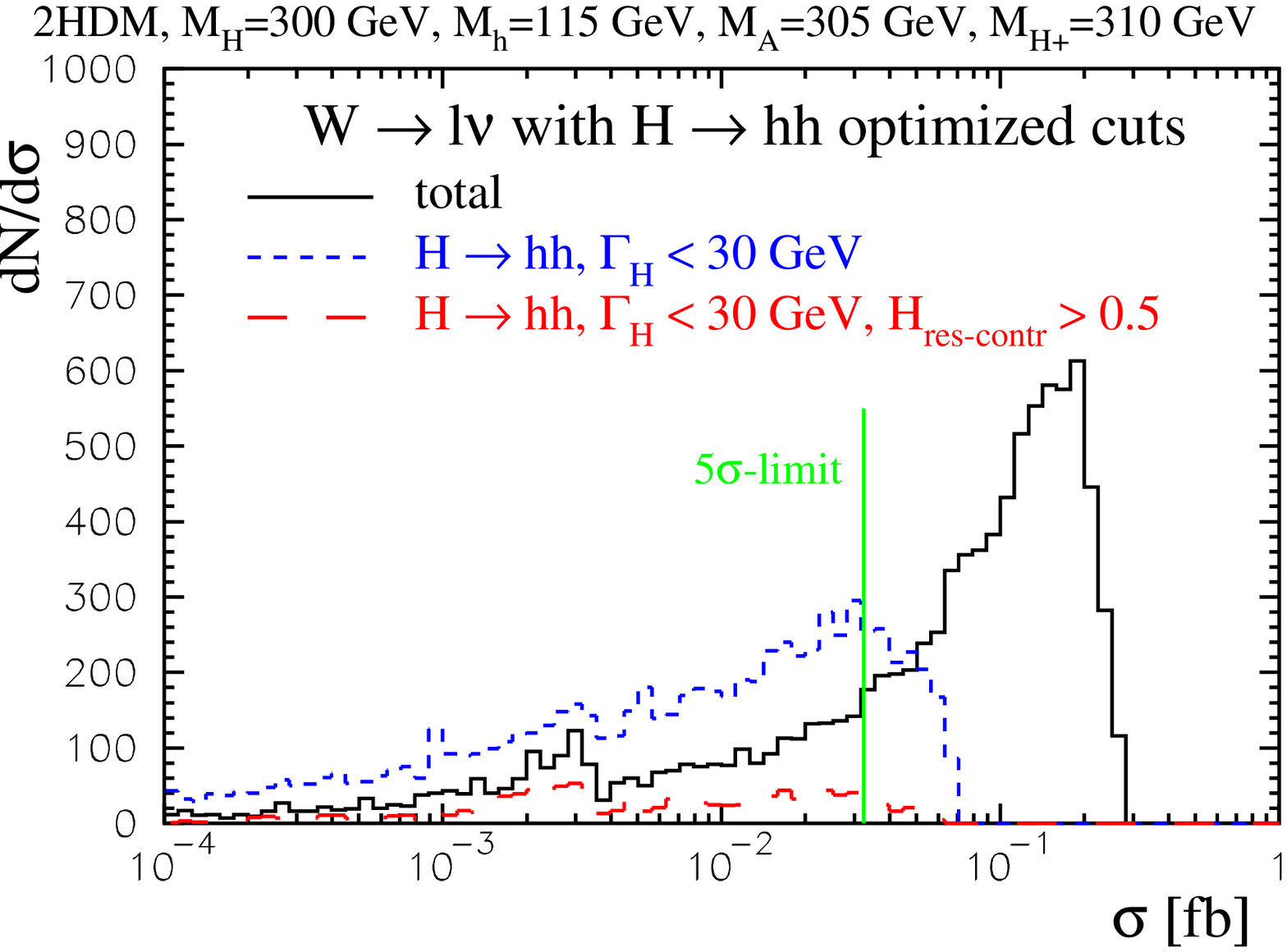, height=6cm}
\epsfig{file=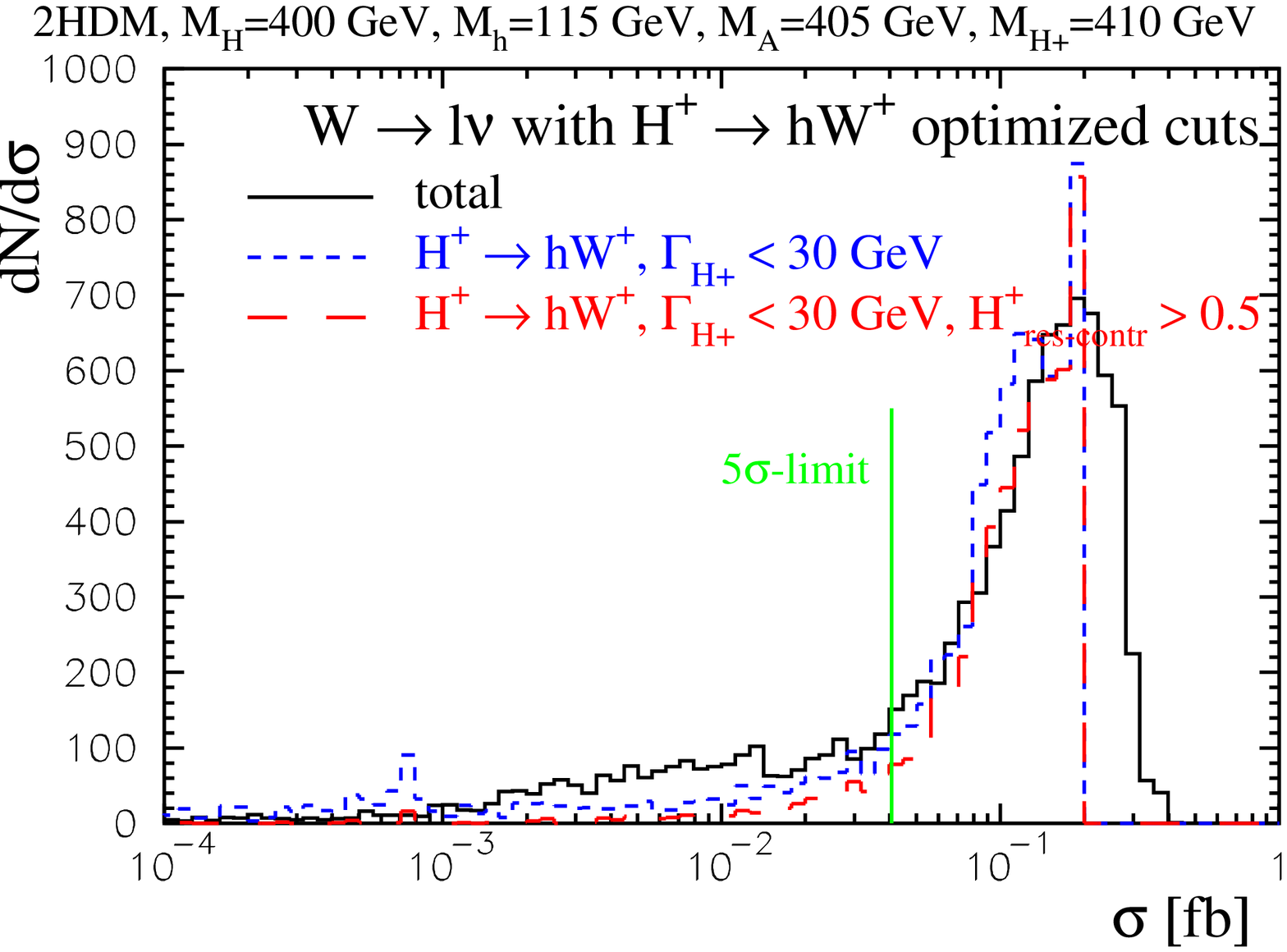, height=6cm}
\epsfig{file=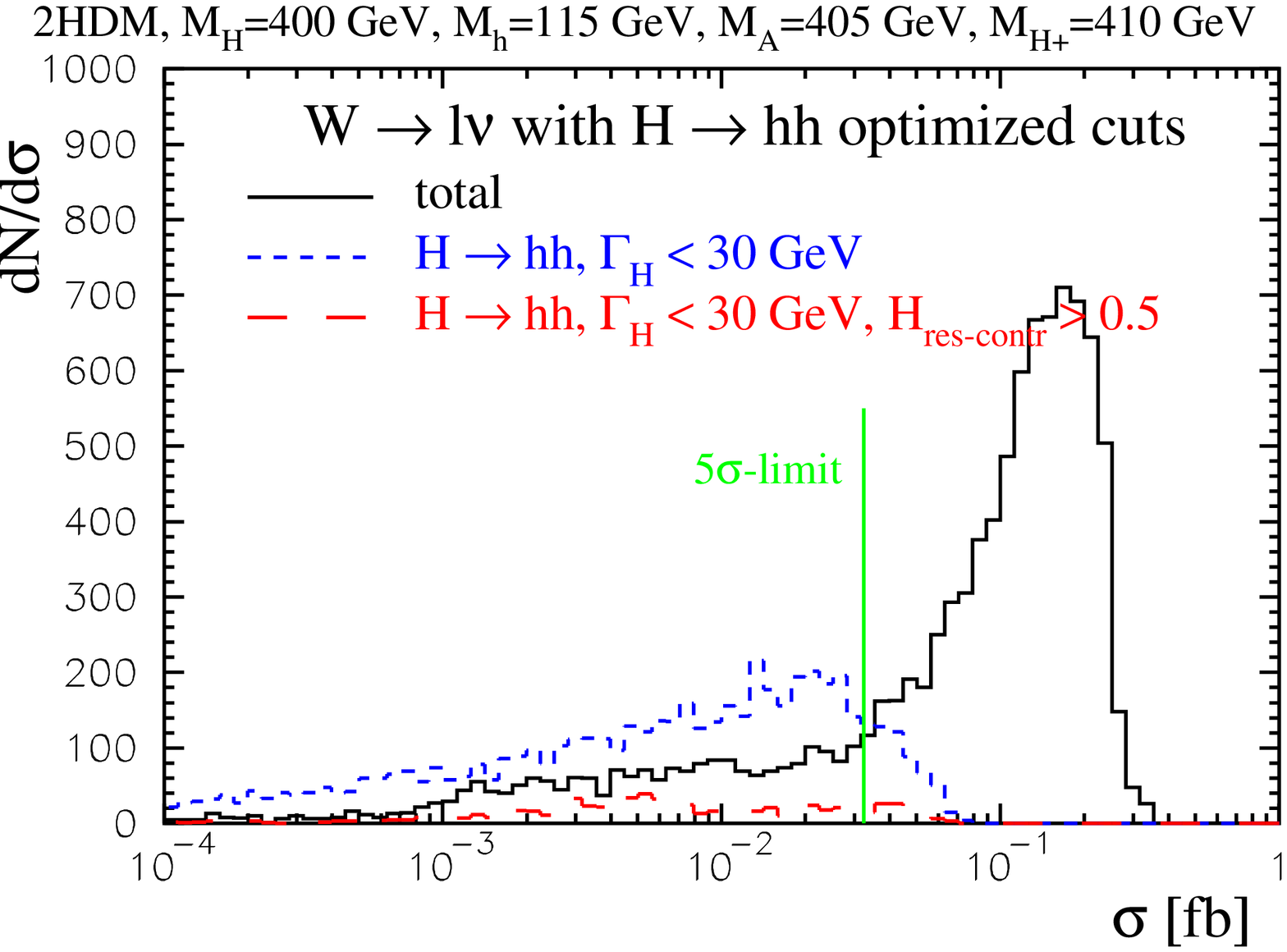, height=6cm}
\caption{
Distribution of resulting cross-sections 
in the 2HDM with leptonic $W^\pm$ decays when scanning over 10000 parameter 
space points for the different mass scenarios indicated.
The two columns show the resulting cross-sections after applying optimal cuts  
(as defined in the text) to select the $H^\pm \to h W^\pm$ (left) or $H \to hh$ (right) channels respectively.
The different lines show: the inclusive cross-section (solid), 
the cross-section corresponding to the $H^\pm \to h W^\pm$ (left) or $H \to hh$ (right) 
resonances (dashed blue), and the same cross-section when requiring that 
the contributions from the resonance is at least 50\% (red long dash).
 }
\label{fig:scan_optcon_W_2hdm}
\end{figure}

\clearpage\thispagestyle{empty}\begin{figure}[!ht]
\center
\epsfig{file=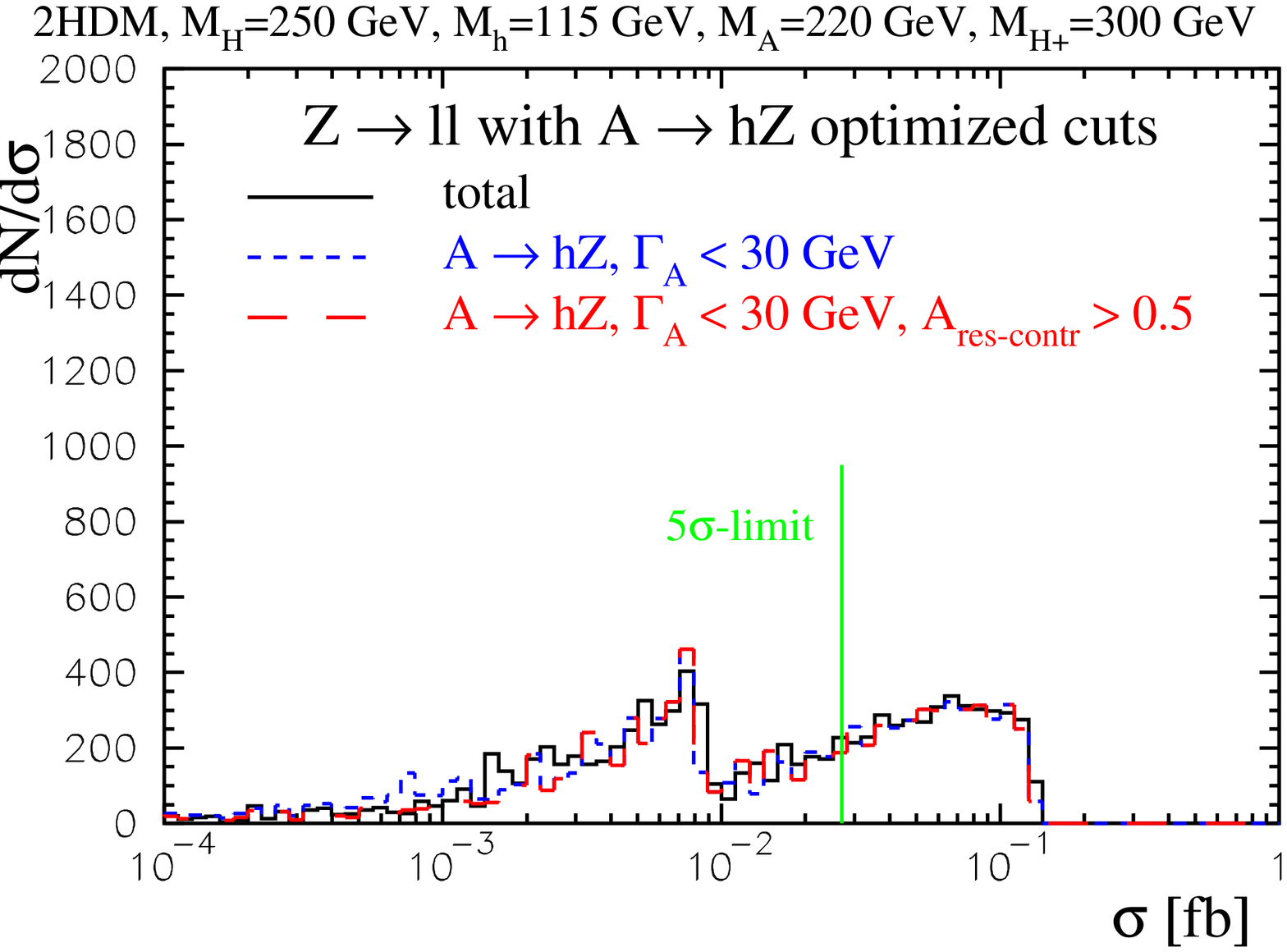, height=6cm}
\epsfig{file=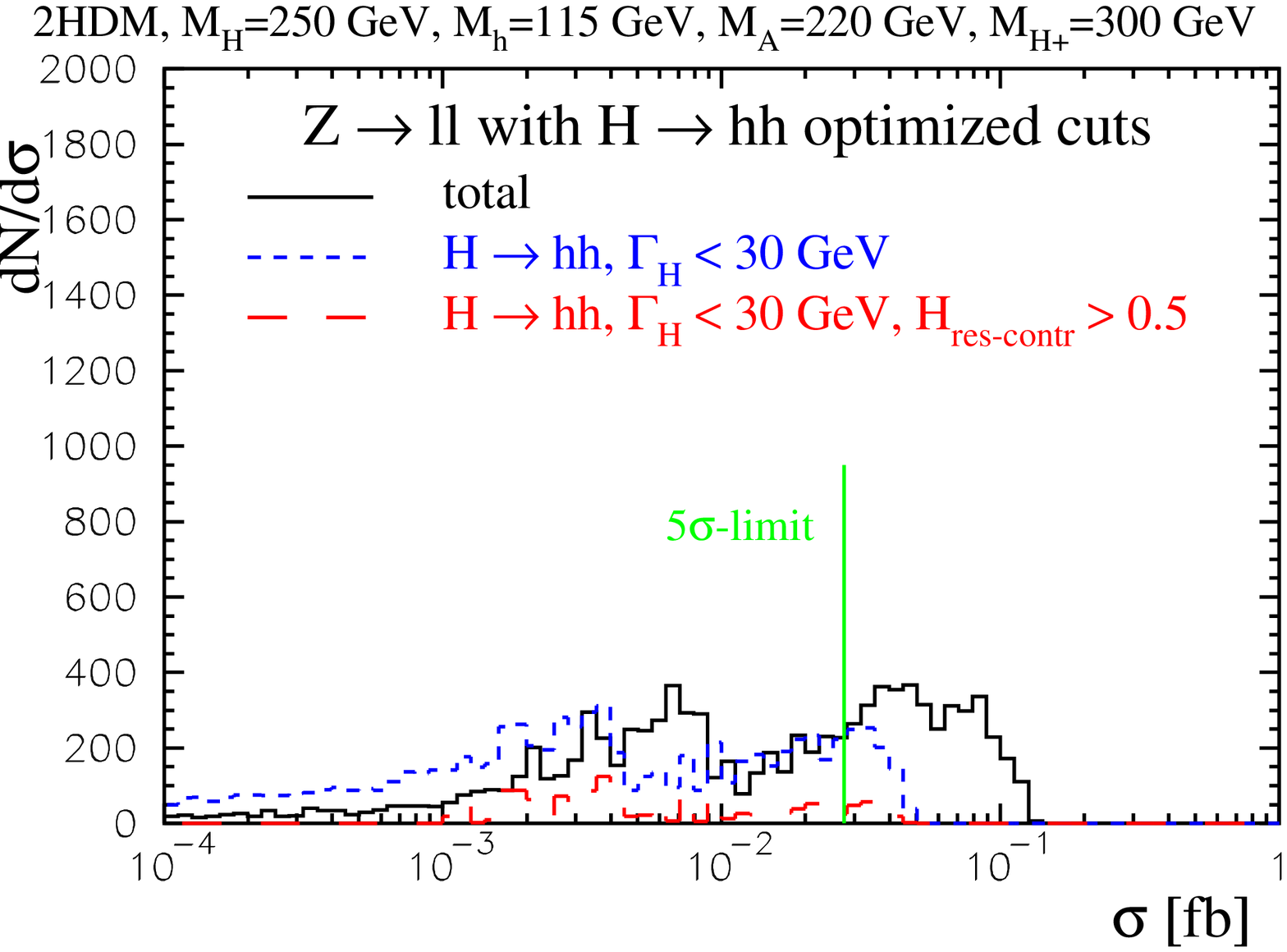, height=6cm}
\epsfig{file=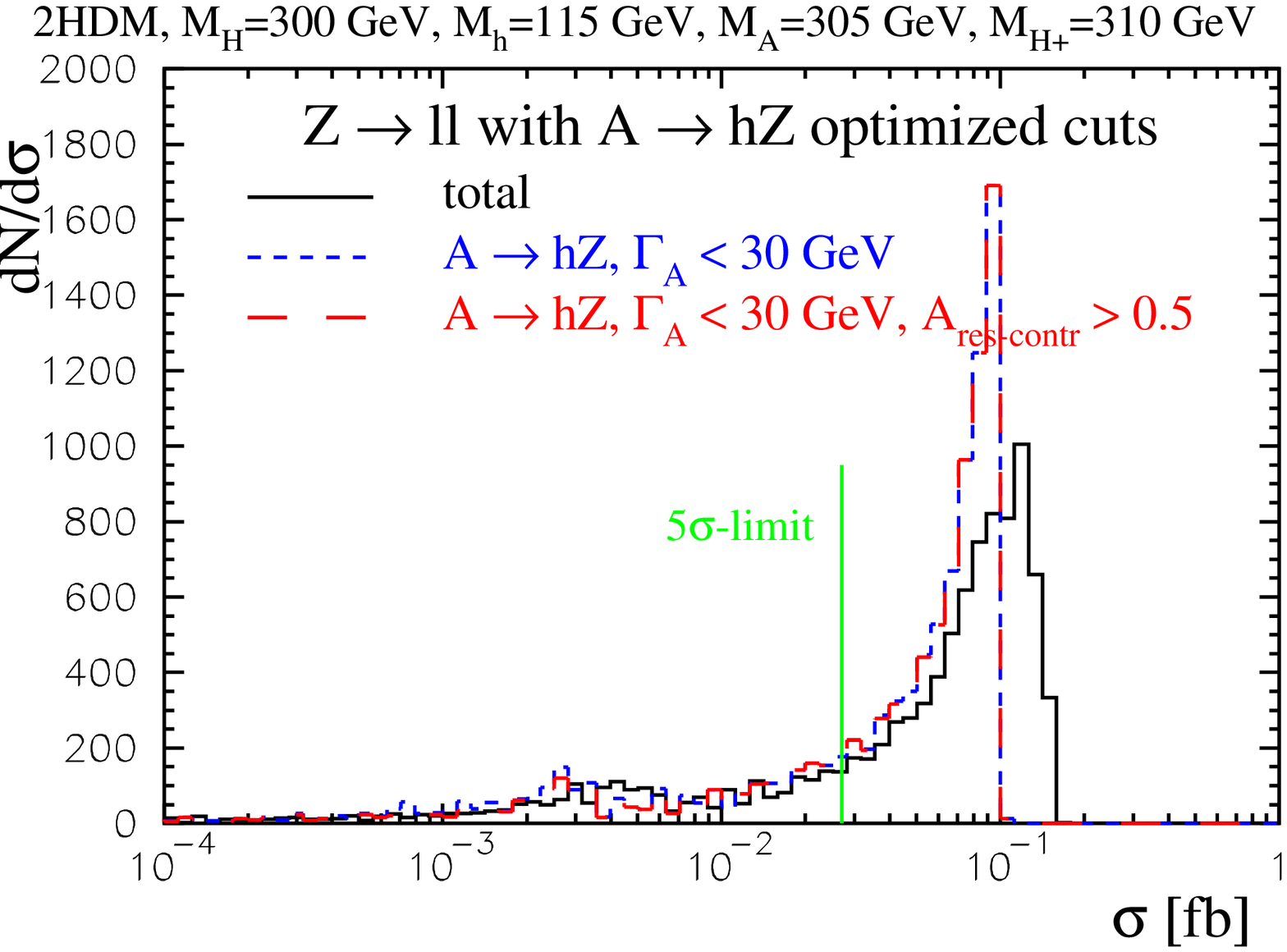, height=6cm}
\epsfig{file=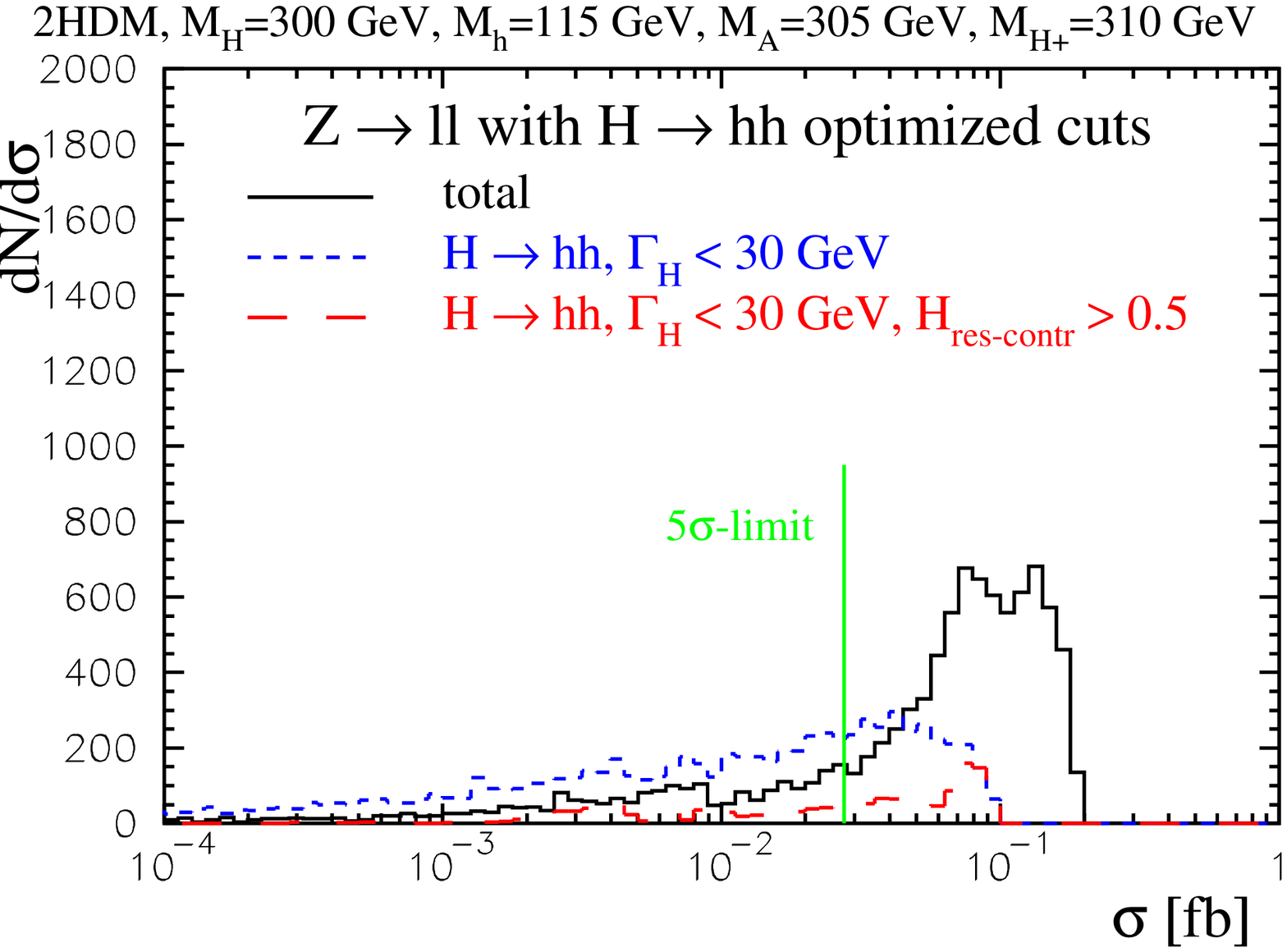, height=6cm}
\epsfig{file=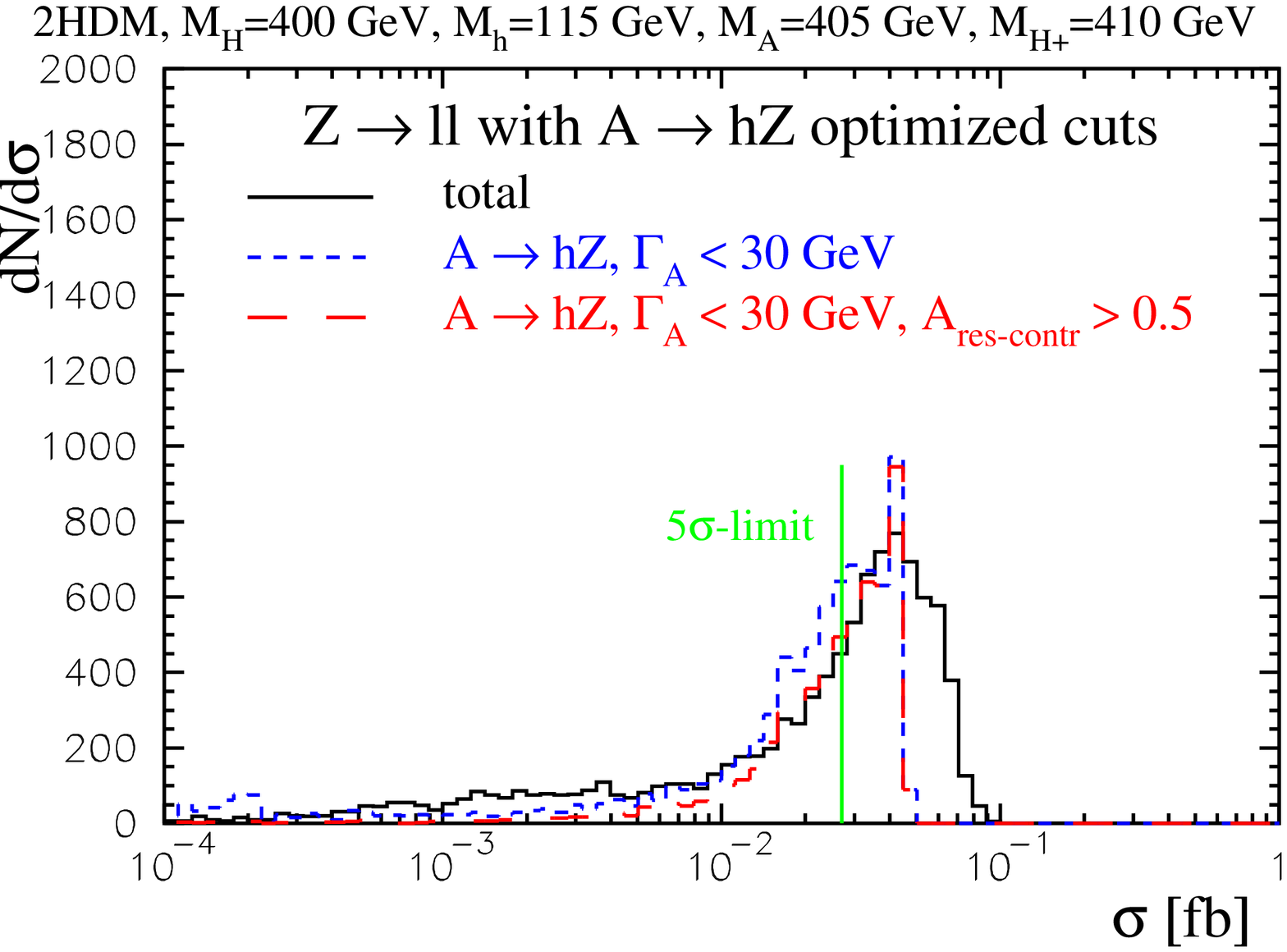, height=6cm}
\epsfig{file=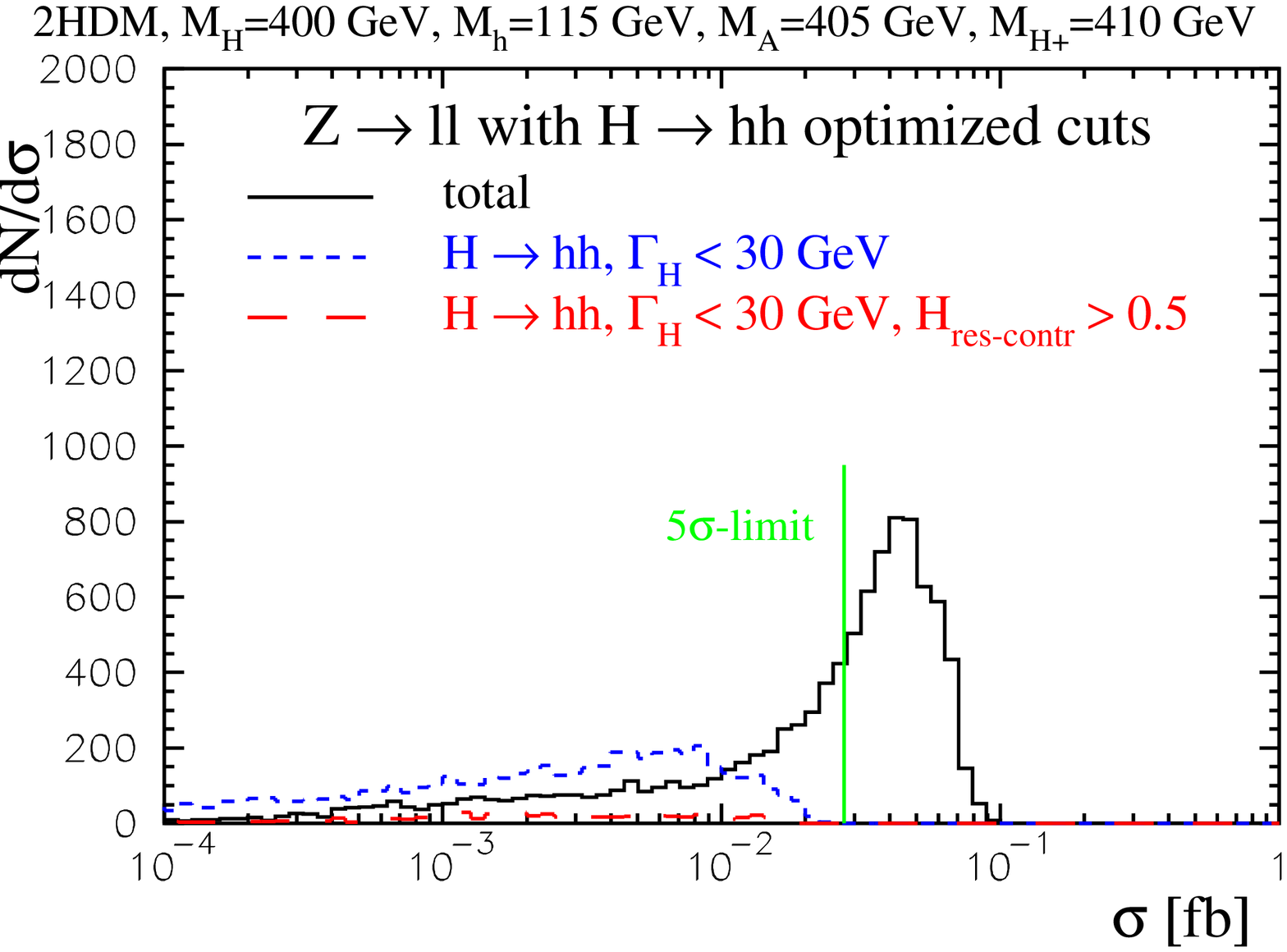, height=6cm}
\caption{
Distribution of resulting cross-sections 
in the 2HDM with leptonic $Z$ decays when scanning over 10000 parameter 
space points for the different mass scenarios indicated.
The two columns show the resulting cross-sections after applying optimal cuts  
(as defined in the text) to select the $A \to hZ$ (left) or $H \to hh$ (right) channels respectively.
The different lines show: the inclusive cross-section (solid), 
the cross-section corresponding to the $A \to hZ$ (left) or $H \to hh$ (right) 
resonances (dashed blue), and the same cross-section when requiring that 
the contributions from the resonance is at least 50\% (red long dash).
 }
\label{fig:scan_optcon_Z_2hdm}
\end{figure}

\clearpage\thispagestyle{empty}\begin{figure}[!ht]
\center
\epsfig{file=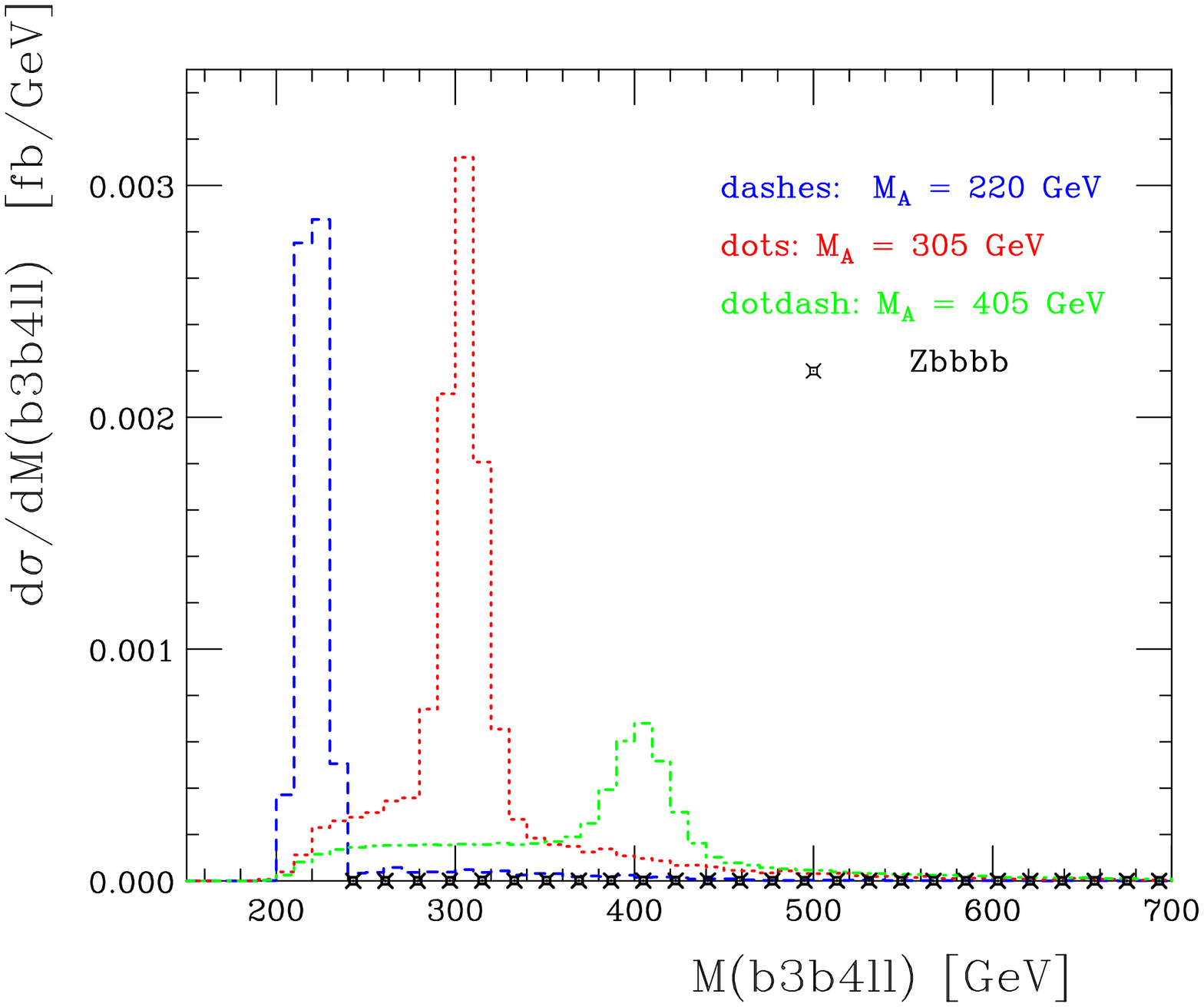, height=8cm, angle=90}
\epsfig{file=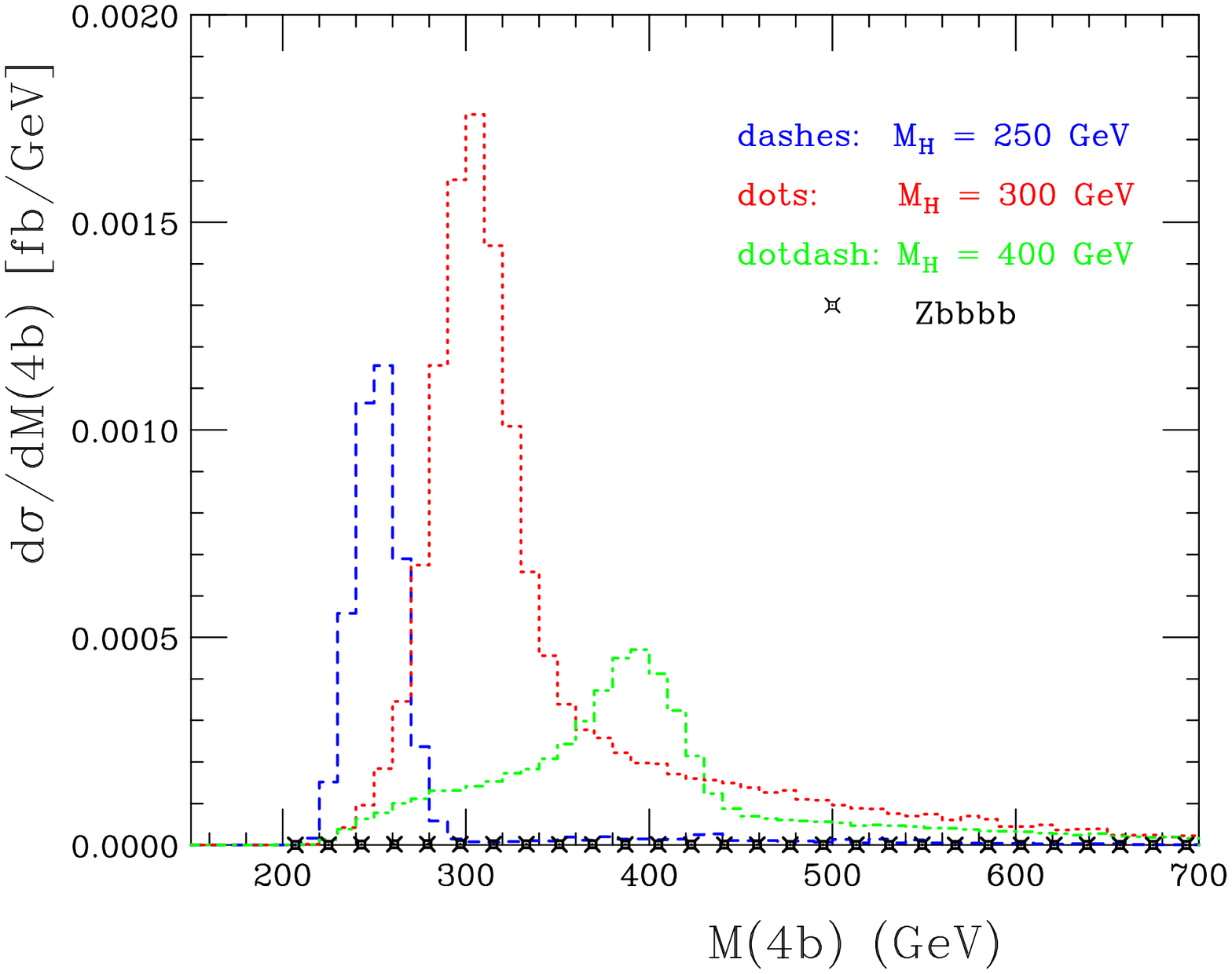, height=8cm, angle=90}
\epsfig{file=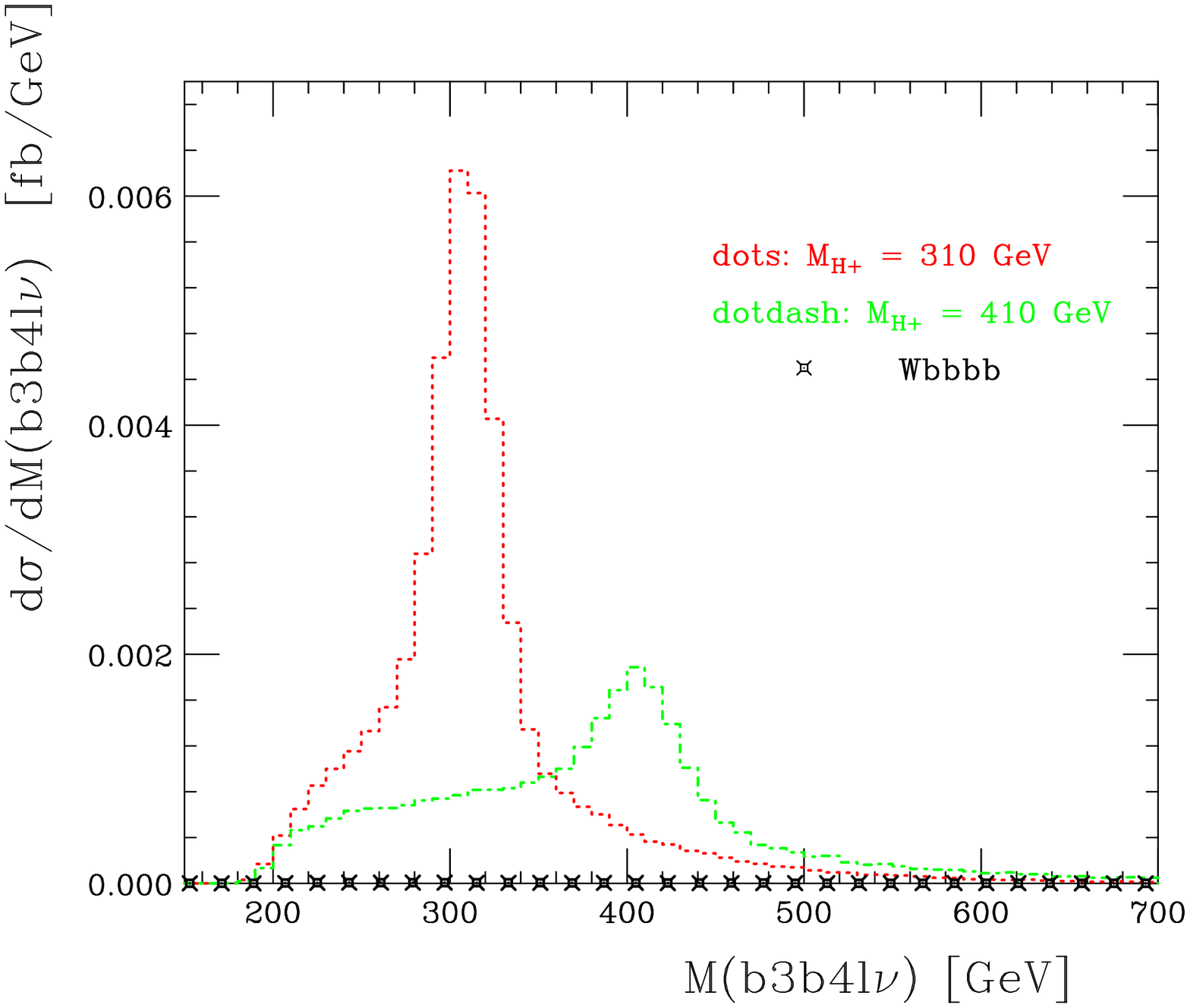, height=8cm, angle=90}
\epsfig{file=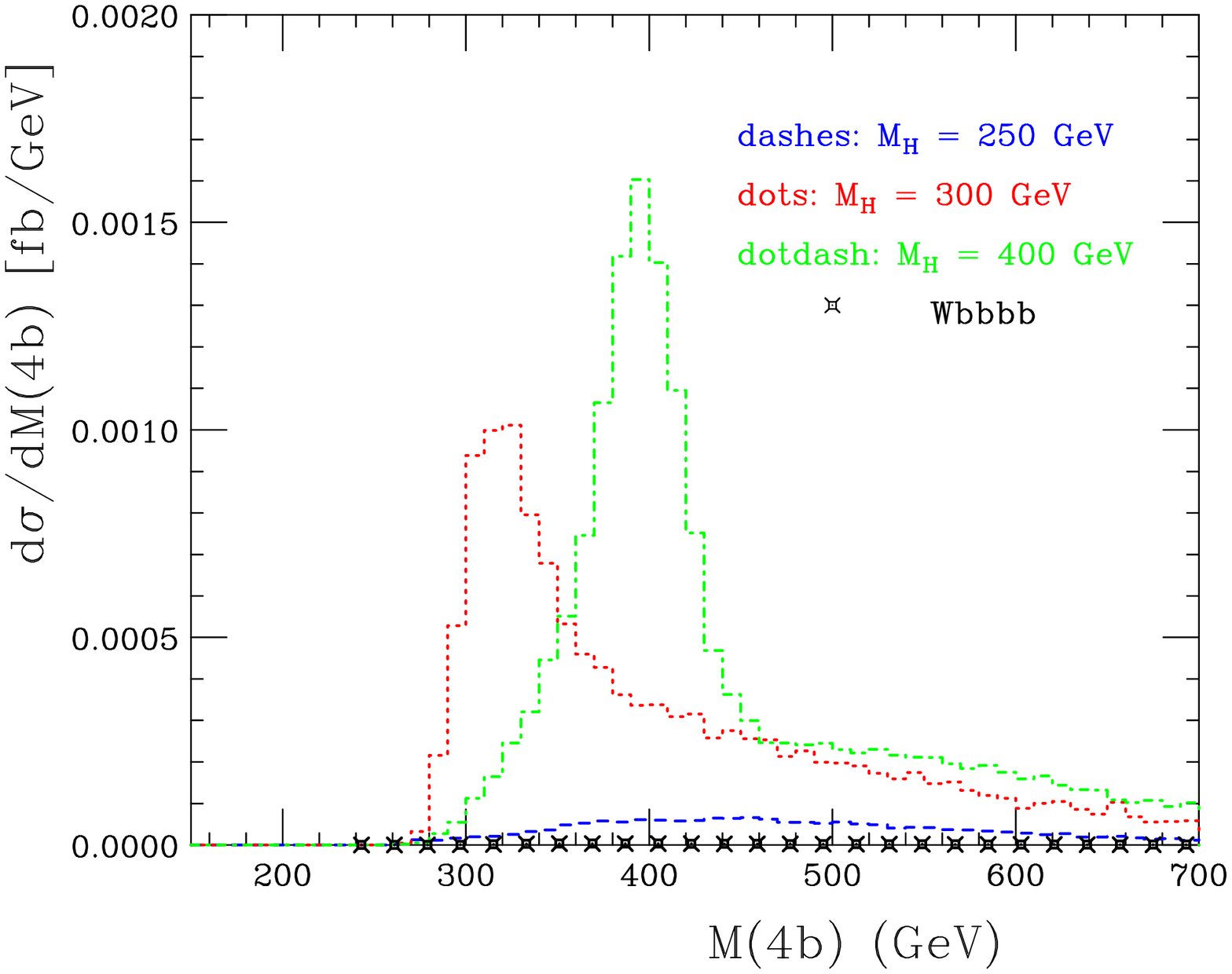, height=8cm, angle=90}
\caption{Differential distributions (normalised to the cross-sections in the respective ``best case scenarios") 
after the optimised cuts for the four different signals:
$pp \to Ah$ with  $A \to h Z$  (top left);
$pp \to ZH$ with $H \to h h$  (top right);
$pp \to H^\pm h$ with  $H^\pm \to h W^\pm$ (bottom left); 
$pp \to WH$ with $H \to h h$  (bottom right). 
The distributions are as follows:
 the invariant mass of the reconstructed $A$ 
boson using the second reconstructed $h$ (top left);
the invariant mass of the reconstructed $H$ boson (top right);
the invariant mass of the reconstructed $H^\pm$ 
boson using the second reconstructed $h$ (bottom left);
the invariant mass of the reconstructed $H$ boson (bottom right).
\label{fig:final}}
\end{figure}

\end{document}